\documentclass[11pt, a4paper, twosides]{Thesis} 
\usepackage{graphicx}
\usepackage[usenames,dvipsnames]{color}
\usepackage{calc}
\usepackage{fancyhdr}
\usepackage{ifthen}
\usepackage{setspace}
\usepackage{moaccopyright}
\usepackage{amsmath}
\usepackage{amsbsy}
\usepackage{bm}
\usepackage{multirow}
\graphicspath{{./Pictures/}} 

\newcommand{\tp}{\tilde{\psi}}
\newcommand{\tm}{\tilde{m}}
\newcommand{\tl}{\tilde{L}}
\newcommand{\tls}{\tilde{L}^2}
\newcommand{\te}{\tilde{\epsilon}}
\newcommand\be{\begin{equation}}
\newcommand\ee{\end{equation}}
\newcommand\bsym{\boldsymbol} 
\newcommand\nn{\notag \\ }

\renewcommand\headrulewidth{0pt}
  \renewcommand\footrulewidth{0pt}
  
\usepackage[titletoc]{appendix}

\fancypagestyle{myplain}
{ 
 \fancyhf{}
  \renewcommand\headrulewidth{0pt}
  \renewcommand\footrulewidth{0pt}
  \fancyfoot[C]{\thepage}
}

\fancypagestyle{myfancy}
{ \fancyhf{}
  \renewcommand\headrulewidth{0.5pt}
 \renewcommand\footrulewidth{0pt}
  \fancyfoot[C]{\thepage}
}

\usepackage[T1]{fontenc}
\usepackage[ansinew]{inputenc}
\usepackage{xcolor,graphicx,wallpaper}

\makeatletter
\let\orig@chapter\@chapter
\def\@chapter[#1]#2{\ifnum \c@secnumdepth >\m@ne
                       \if@mainmatter
                         \refstepcounter{chapter}%
                         \typeout{\@chapapp\space\thechapter.}%
                         \addcontentsline{toc}{chapter}%
                                   {Chapter~\protect\numberline{\thechapter.} #1}%
                       \else
                         \addcontentsline{toc}{chapter}{#1}%
                       \fi
                    \else
                      \addcontentsline{toc}{chapter}{#1}%
                    \fi
                    \chaptermark{#1}%
                    \addtocontents{lof}{\protect\addvspace{10\p@}}%
                    \addtocontents{lot}{\protect\addvspace{10\p@}}%
                    \if@twocolumn
                      \@topnewpage[\@makechapterhead{#2}]%
                    \else
                      \@makechapterhead{#2}%
                      \@afterheading
                    \fi}
\makeatother

\def\mypart#1#2#3{\par 
\pagestyle{plain}
\newpage\clearpage

\vspace*{2cm} 
\refstepcounter{part}
{\centering \textbf{\huge PART \thepart}\par}%
\vspace{1cm}
{\centering \linespread{2}\selectfont \textbf{\Huge #1}\par \linespread{1}\selectfont}%

\thispagestyle{empty}
\vspace{5cm}

\null
\vfill
#3
#2
\addcontentsline{toc}{part}{Part \Roman{part}:\; \textbf{#1}}
\pagestyle{myfancy}
}

\usepackage{moactitlepage}

\usepackage[Bjornstrup]{fncychap}

\usepackage{xcolor}
\colorlet{partbgcolor}{gray!30}
\colorlet{partnumcolor}{gray}
\colorlet{chapbgcolor}{gray!30}
\colorlet{chapnumcolor}{red}

\makeatletter
\renewcommand*{\@part}{}
\def\@part[#1]#2{%
  \ifnum \c@secnumdepth >-2\relax
    \refstepcounter{part}%
    \@maybeautodot\thepart%
    \addparttocentry{\thepart}{#1}%
  \else
    \addparttocentry{}{#1}%
  \fi
  \begingroup
    \setparsizes{\z@}{\z@}{\z@\@plus 1fil}\par@updaterelative
    \raggedpart
    \interlinepenalty \@M
    \normalfont\sectfont\nobreak
    \setlength\fboxsep{0pt}
    \colorbox{partbgcolor}{\rule{0pt}{40pt}%
    \makebox[\linewidth]{%
    \begin{minipage}{\dimexpr\linewidth+20pt\relax}
      \ifnum \c@secnumdepth >-2\relax
        \vskip-25pt
        \size@partnumber{\partformat}%
      \fi      %
      \vskip\baselineskip
      \hspace*{\dimexpr\myhi+10pt\relax}%
      \parbox{\dimexpr\linewidth-2\myhi-20pt\relax}{\raggedleft\LARGE#2\strut}%
      \hspace*{\myhi}\par\medskip%
    \end{minipage}%
      }%
    }%
    \partmark{#1}\par
  \endgroup
  \@endpart
}

\renewcommand\DOCH{%
  \settowidth{\py}{\CNoV\thechapter}
  \addtolength{\py}{-10pt}
  \fboxsep=0pt%
  \colorbox{chapbgcolor}{\rule{0pt}{40pt}\parbox[b]{\textwidth}{\hfill}}%
  \kern-\py\raise20pt%
  \hbox{\color{chapnumcolor}\CNoV\thechapter}\\%
}

\makeatother

\usepackage[square, numbers, comma, sort&compress]{natbib} 
\hypersetup{urlcolor=black, colorlinks=true} 
\title{\ttitle} 

\begin{document}








\pagenumbering{gobble}

\renewcommand{\moacdepartment}{Department of Physics}

\renewcommand{\moacdate}{August 2013}

\renewcommand{\moacsupervisors}{Prof. Clifford M. Will}

\renewcommand{\usemoaclogo}{false}






\moactitlepages



\frontmatter 

\setstretch{1.3} 

\fancyhead{} 
\rhead{\thepage} 
\lhead{} 

\pagestyle{fancy} 

\newcommand{\HRule}{\rule{\linewidth}{0.5mm}} 

\hypersetup{pdftitle={\ttitle}}
\hypersetup{pdfsubject=\subjectname}
\hypersetup{pdfauthor=\authornames}
\hypersetup{pdfkeywords=\keywordnames}

\clearpage
\pagestyle{myplain}
\null\vfill
\noindent
\begin{center}
{\textcopyright \: 2013, \moacauthor\\
All Rights Reserved}
\end{center}
\thispagestyle{empty}

\setcounter{tocdepth}{1}

\pagestyle{myfancy} 

\lhead{\emph{Contents}} 
\begingroup
\hypersetup{linkcolor=blue}
\tableofcontents
\endgroup 

\lhead{\emph{List of Figures}} 
\listoffigures 

\lhead{\emph{List of Tables}} 
\listoftables 
\clearpage

\clearpage 

\setstretch{1.5} 

\lhead{\emph{Abbreviations}} 
\listofsymbols{ll} 
{
\textbf{AGN} & \textbf{A}ctive \textbf{G}alactic \textbf{N}uclei  \\
\textbf{BT} & J.  \textbf{B}inney and S.  \textbf{T}remaine, Galactic Dynamics, second edition, 2008 \\
\textbf{CDM} & \textbf{C}old \textbf{D}ark \textbf{M}atter  \\
\textbf{CMB} & \textbf{C}osmic \textbf{M}icrowave \textbf{B}ackground  \\
\textbf{DM} & \textbf{D}ark \textbf{M}atter \\
\textbf{GC} & \textbf{G}alactic \textbf{C}enter \\
\textbf{GR} & \textbf{G}eneral \textbf{R}elativity \\
\textbf{GS} &  P. \textbf{G}ondolo and J. \textbf{S}ilk, Phys. Rev. Lett., 83: 1719-1722, 1999  \\
\textbf{MAMW} & D. \textbf{M}erritt, T. \textbf{A}lexander, S. \textbf{M}ikkola, and C. M. \textbf{W}ill, Phys. Rev. D., 81(6):062002, 2010 \\
\textbf{MBH} & \textbf{M}assive \textbf{B}lack \textbf{H}ole \\
\textbf{WIMP} &  \textbf{W}eakly  \textbf{I}nteracting  \textbf{M}assive  \textbf{P}article

}

\pagestyle{myplain}
\setstretch{1.2} 

\acknowledgements{\addtocontents{toc}{} 

In my PhD program I have had the great opportunity to work with two advisors which I will always treasure the lessons that I learned form both of them. Foremost, I would like to express my sincere gratitude to my advisor Prof. Clifford M. Will for the continuous support of my PhD study and research, his patience, motivation, enthusiasm, and immense knowledge. His guidance helped me in all the time of research and writing of this thesis. I should also mention that his sense of humor was always appreciated. Overally I could not have imagined having a better advisor and mentor for my PhD study and I simply do not have the words to thank him enough. 
 
Also, my sincere gratitude and heartfelt thanks goes to my other advisor Professor Francesc Ferrer for his continuous support, guidance and encouragement. his unflagging enthusiasm and energy impressed me all the time and I have always felt very lucky and fortunate for having him as my advisor. I have benefitted a lot from his knowledge and experience and I am very grateful to him because of his generosity with his time.

I wish to thank the members of my dissertation committee, Prof. Mark Alford, Prof. Ram Cowsik, Prof. Gregory Comer and Prof. Renato Feres for their time, guidance and helpful comments and suggestions.

I am also very grateful to all the faculty, staff and graduate students in the Department of Physics at Washington University for providing a very calm and friendly atmosphere and my special thanks goes to Sai Iyer who has been very kind and patient and always willing to lend his service whenever I approached him. I acknowledge and appreciate him for all of his helps.

I would also like to thank Claud Bernard, Luc Blanchet, Joe Silk, David Merritt, Scott Hughes, K. G. Arun, Ryan Lang and Daniel Hunter for their helpful comments and discussions during this work.

I would never have achieved what I have achieved without the unconditional love and support I have received from my parents, Soraya and Mohammad, and my siblings, Nadia, Nahid, and Shahin. Finally, I am infinitely grateful for the love and support I have gotten from my husband Saeed who has been also a great officemate and colleague for me.

The research presented in this thesis was supported in part by the National Science Foundation, Grant Nos.\ PHY 06--52448, 09-65133, 12--60995 \& 0855580, the U.S. DOE under contract No.\ DE--FG02--91ER40628, the National Aeronautics and Space Administration, Grant No.\ NNG-06GI60G, and the Centre National de la Recherche Scientifique, Programme Internationale de la Coop\'eration Scientifique (CNRS-PICS), Grant No.\ 4396. I also gratefully acknowledge the Institut d'Astrophysique de Paris  and University of Florida for their hospitality during the completion part of my research.
}
\clearpage 

\setstretch{1.3} 

\addtotoc{Abstract} 

\abstract{\addtocontents{toc}{} 

This thesis includes two main projects. In the first part, we assess the feasibility of a recently suggested strong-field general relativity test, in which future observations of a hypothetical class of stars orbiting very close to the supermassive black hole at the center of our galaxy, known as Sgr A$^\star$, could provide tests of the so-called no-hair theorem of general relativity through the measurement of precessions of their orbital planes. By considering how a distribution of stars and stellar mass black holes in the central cluster would perturb the orbits of those hypothetical stars, we show that for stars within about 0.2 milliparsecs (about 6 light-hours) of the black hole, the relativistic precessions dominate, leaving a potential window for tests of no-hair theorems. Our results are in agreement with N-body simulation results.

In the second part, we develop a fully general relativistic phase-space formulation to consider the effects of the Galactic center supermassive black hole Sgr A$^\star$ on the dark-matter density profile and its applications in the indirect detection of dark matter. We find significant differences from the non-relativistic result of Gondolo and Silk (1999), including a higher density for the spike and a larger degree of central concentration. Having the dark matter profile density in the presence of the massive black hole, we calculate its perturbing effect on the orbital motions of stars in the Galactic center, and find that for the stars of interest,  relativistic effects related to the hair on the black hole will dominate the effects of dark matter.
}

\clearpage 

\mainmatter 

\pagestyle{myfancy} 


 
\chapter{Introduction and Overview}
\label{chapter1} 
\thispagestyle{myplain}
\lhead{Chapter 1. \emph{Introduction and Overview}} 
\ClearWallPaper

\section{Black Holes}
The simplest description of black holes says a black hole is a region of spacetime from which gravity prevents anything, including light, from escaping. It is an object created when a massive star collapses to a size smaller than twice its geometrized mass, thereby creating such strong spacetime bending that its interior can no longer communicate with the external universe.
Black holes were first predicted using solutions of the equations of General Relativity (GR); these equations predict specific  properties for their external geometry. If the black hole is non-rotating, then its exterior metric is be that of Schwarzschild, which is the exact, unique, static and spherically symmetric solution of Einstein's equation in vacuum. In Schwarzschild coordinates, the line element for the Schwarzschild metric has the form
\be \label{ch1-1}
{\rm d}s^2=-\left(1-2Gm/r \right) {\rm d}t^2+\frac{{\rm d}r^2}{1-2Gm/r}+r^2\left({\rm d}\theta^2+\sin^2 \theta \ {\rm d}\phi^2 \right) \ ,
\ee
where $G$ is Newton's constant and we use units in which $c=1$. The surface of the black hole, i.e., the horizon, is located at $r=2Gm$. Only the region on and outside the black hole's surface, $r\geq2Gm$, is relevant to external observers. Events inside the horizon can never influence the exterior.

In that region of spacetime, $r\gg 2Gm$, where the geometry is nearly flat, Newton's theory, ${\rm d}{\bm v}/{\rm d}t=\nabla \Phi(r)$, where $\Phi(r)$ is the Newtonian gravitational potential, can be obtained from the approximate line element 
\be \label{ch1-2}
{\rm d}s^2=-\left(1-2Gm/r \right) {\rm d}t^2+{\rm d}r^2+r^2\left({\rm d}\theta^2+\sin^2 \theta \ {\rm d}\phi^2 \right) \ .
\ee
For Schwarzschild metric, in the limit $r\gg 2Gm$, $\Phi(r) = -Gm/r$. Consequently, $m$ is the mass that governs the Keplerian motions of test masses in the distant, Newtonian gravitational field and we can call $m$ in Eq. (\ref{ch1-1}) {\it Keplerian mass} of the black hole.

If the black hole is rotating with angular momentum $J$, its exterior geometry is given by the Kerr metric. The Kerr metric is given in Boyer-Lindquist coordinates, which are a generalization of Schwarzschild coordinates, by
\begin{eqnarray}  \nonumber
{\rm d}s^2 &=& - \left( 1 - \frac{2Gmr}{\Sigma^2} \right ) {\rm d}t^2
 + \frac{\Sigma^2}{\Delta} {\rm d}r^2 + \Sigma^2 {\rm d}\theta^2
 - \frac{4Gmra}{\Sigma^2} \sin^2 \theta {\rm d}t {\rm d}\phi  \\ \label{ch1-3}
 && \quad + \left ( r^2 + a^2 + \frac{2Gmra^2 \sin^2 \theta}{\Sigma^2} \right ) \sin^2 \theta d\phi^2  \,,
\end{eqnarray}
where $a$ is the Kerr parameter, related to the angular momentum $J$ by $a \equiv J/m$; $\Sigma^2 = r^2 + a^2 \cos^2 \theta$, and $\Delta = r^2 + a^2 - 2Gmr$.  We will assume throughout that $a$ is positive.

Just as the electromagnetic potentials $\Phi$ and $A^i$ of a charge and current distribution can be expanded in a sequence of multipole moments (dipole, quadrupole, magnetic dipole, etc), so too can part of the exterior metric of the Kerr black hole. In a coordinate system that is a variant of the Boyer Lindquist coordinates, the $00$ and $0 \phi$ components of the Kerr metric describing the exterior of a rotating black hole can be expanded as
\begin{eqnarray} \nonumber
\Phi&=&\frac{Gm}{r}+\frac{GQ_2P_2(\cos \theta)}{r^3}+\frac{GQ_4P_4(\cos \theta)}{r^5}+\ldots \ ,  \\ \label{ch1-4}
A^\phi&=&\frac{GJ}{r^2}+\frac{GJ_3\tilde P_3(\cos \theta)}{r^4}+\frac{GJ_5\tilde P_5(\cos \theta)}{r^6}+ \ldots \ ,
\end{eqnarray} 

where $\Phi = (1+g_{00})/2$, and $A^\phi = -g_{0\phi}/2\sin^2 \theta$. The quantities $Q_\ell$ and $J_\ell$ are mass and current multipole moments respectively and $P_\ell(\cos \theta)$ and ${\tilde P}_\ell(\cos \theta)$ are suitable angular functions. The zero degree mass moment is equal to the mass of the black hole, $Q_0 = m$, and the degree one current moment is its angular momentum, $J_1 = J$.

\subsection{The Black Hole No-Hair Theorem}
One important property of black holes predicted by GR is commonly known as the {\it no-hair theorem}. The no-hair theorem states that, once a black hole achieves a stable condition after formation, it has only three independent physical properties: mass $m$, angular momentum $J$, and charge $Q$. The exterior geometry of a black hole is completely governed by these three parameters. In fact, any two black holes that share the same values for these parameters are indistinguishable. It is widely agreed that processes involving the matter in which they are embedded will rapidly neutralize astrophysical black holes, and so from now on, we only consider neutral black holes, $Q=0$.

The no-hair theorem establishes the claim that black holes are uniquely characterized by their mass $m$ and spin $J$, i.e., by only the first two multipole moments of their exterior spacetimes \cite{Israel67,Israel68, Carter71,Hawking72,Robinson75}. As a consequence of the no-hair theorem, all higher-order moments are already fully determined and turn out to obey the simple relation \cite{Geroch70,Hansen74}
\be  \label{ch1-5}
Q_{\ell}+i J_{\ell}=m(ia)^\ell \ ,
\ee
where $a\equiv J/m$ is the spin parameter, and the multipole moments are written as a set of mass multipole moments $Q_{\ell}$ which are nonzero for even values of $\ell$ and as a set of current multipole moments $J_\ell$ which are nonzero for odd values of $\ell$. The specific relation that we are going to use in testing the no-hair theorem is, for $\ell=2$:
\be  \label{ch1-6}
Q_2=-ma^2=-\frac{J^2}{m} \ .
\ee

\section{The Massive Black Hole at the Galactic Center}
Observation indicates that most galaxies contain a massive compact dark object in their centers whose mass lies in the range $10^6 M_{\odot} <m<{\rm few} \times 10^9 M_{\odot}$ \cite{Ferrarese05,Alexander05}. It is widely believed that these dark objects are Massive Black Holes (MBHs), and that they exist in the centers of most, if not all galaxies. Their number density and mass scale are broadly consistent with the hypothesis that they are now-dead quasars, which were visible for a relatively short time in their past as extremely luminous Active Galactic Nuclei (AGN), powered by the gravitational energy released by the accretion of gas and stars \cite{Yu02}. It is also possible that low-mass MBHs like the one in the Galactic center (GC) have acquired most of their mass by mergers with other black holes. Some present-day galaxies have AGN, although none as bright as quasars. However, most present-day galactic nuclei are inactive, which implies that accretion has either almost ceased or switched to a non-luminous mode. Their inactivity is not due to the lack of gas supply; most galaxies have more than enough to continue powering an AGN. The ``dimness problem'' is one of the key issues of accretion theory, which deals with the physics of flows into compact objects.  

The MBH in the center of Milky Way is the nearest example of a central galactic MBH. It was first detected as an unusual non-thermal radio source, Sagittarius A$^\star$ (Sgr A$^\star$). Over the following decades, observations across the electromagnetic spectrum, together with theoretical arguments, established with ever-growing confidence that Sgr A$^\star$ is at the dynamical center of the Galaxy and that it is associated with a very massive and compact dark mass concentration. This has ultimately led to the nearly inescapable conclusion that the dark mass is a black hole.   

The Galactic MBH is quite normal. Like most MBHs, it is inactive. With $m\sim(3-4)\times10^6M_{\odot}$, it is one of the least massive MBHs discovered. What makes it special is its proximity. At $\sim 8 \ {\rm kpc}$ ($1 \ {\rm ps}=3.26 \ {\rm light \ years}$) from the Sun, the Galactic black hole is $\sim 100$ times closer than the MBH in Andromeda, the nearest large galaxy, and $\sim 2000$ times closer than galaxies in Virgo, the nearest cluster of galaxies. For this reason it is possible to observe today the stars and gas in the immediate vicinity of the Galactic MBH at a level of details that will not be possible for any other galaxy in the foreseeable future.

In spite of its relative proximity, observations of the GC are challenging due to strong, spatially variable extinction by interstellar dust, which is opaque to optical-UV wavelengths. As a result, observation of the GC must be conducted in the infrared. Using the highest angular resolution obtained at near-infrared wavelength at mid 1990s, a large population of faint stars orbiting the center of the Galaxy was discovered \cite{Genzel97,Eckart97,Ghez98}. The orbital periods of these stars are on the scale of tens of years and since the initial discovery, one of these stars has been observed to make a complete orbit around the center. 

The detection of stars orbiting the dynamical center of the Galaxy has given us quantitative information about the mass, size and position of the dark mass at the center and has confirmed the idea that we have a MBH at the Galactic center. Inside $\sim 0.04 {\rm pc}$, there are no bright giants, and only faint blue stars are observed with orbital periods on the scale of tens of years. This population is known as the ``S-stars'' or ``S-cluster'', after their identifying labels. Deep near-IR photometric and spectroscopic observations of that region were all consistent with the identification of these stars as massive main sequence stars. There is no indication of anything unusual about the S-stars, apart from their location very near the MBH.   

Because of the huge mass ratio between a star and the MBH, stars orbiting near it, are effectively test particles. This is to be contrasted with the gas in that region, which can be subjected to non-gravitational forces due to thermal, magnetic or radiation pressure. These can complicate the interpretation of dynamical data and limit its usefulness. The term ``near'' is taken here to mean close enough to the MBH so that the gravitational potential is completely dominated by it, but far enough so that the stars can survive, i.e. beyond the MBH event horizon, or beyond the radius where stars are torn apart by the black hole's tidal gravitational field. In this range, stars directly probe the gravitational field of the MBH. The event horizon of the MBH in the GC is much smaller than the orbital radius for the stars that have been observed to date, and so effects due to GR lead to deviations from Newtonian motion that are unmeasurable at present. To first order, the stellar orbits can be treated as Keplerian, which substantially simplifies the analysis. However, with accurate enough astrometric observations it may be possible to detect post-Newtonian effects in the orbits and to probe GR. We will discuss this more specifically in the next section in the context of testing the no-hair theorem, and with more details in Chapter 2.   

\section{Testing the Black Hole No-Hair Theorem at the Galactic Center}
There seems to be every expectation that, with improved observing capabilities, a population of stars closer to the MBH than the S-stars, will eventually be discovered, making orbital relativistic effects detectable. This makes it possible to consider doing more than merely detect relativistic effects, but rather to provide the first test of the black hole no-hair theorem, which demands that $Q_2=-J^2/m$, to see if the central dark mass at the GC is truly a GR black hole.

If the black hole were non-rotating ($J=0$), then its exterior would be that of Schwarzschild, and the most important relativistic effect would be the advance of the pericenter. If it is rotating, then two new phenomena occur, the dragging of inertial frames and the effects of the hole's quadrupole moment, leading not only to an additional pericenter precession, but also to a precession of the orbital plane of the star. These precessions are smaller than the Schwarzschild effect in magnitude because they depend on the dimensionless angular momentum parameter $\chi= a/(Gm)=J/(Gm^2)$, which is always less than one, and because they fall off faster with distance from the black hole. However, accumulating evidence suggests that the MBH should be rather rapidly rotating, with $\chi$ larger than $0.5$ and possibly as large as $0.9$, so these effects could be significant.

It has been suggested that if a class of stars were to be found with orbital periods of fractions of a year, and with sufficiently large orbital eccentricities, then the frame-dragging and quadrupole-induced precessions could be as large as $10 \  \mu {\rm arcsecond}$ per year \cite{Will08}.

The precession of the orbital plane is the most important effect in testing the no-hair theorem, because it depends only on $J$ and $Q_2$; the Schwarzschild part of the metric affects only the pericenter advance  because its contributions are spherically symmetric, and thus cannot alter the orbital plane. In order to test the no-hair theorem, one must determine five parameters: the mass of the black hole, the magnitude and two angles of its spin, and the value of the quadrupole moment. The Kepler-measured mass is determined from the orbital periods of stars, but may require data from a number of stars to fix it separately from any extended distribution of mass. Then to measure $\bm J$ and $Q_2$, it is necessary and sufficient to measure precessions in the orbital planes for two stars in non-degenerate orbits.   

Detecting such stars so close to the black hole, and carrying out infrared astrometry to $10 \  \mu {\rm arcsec}$ per year accuracy will be a challenge. However, if this challenge can be met with future improved adaptive optics systems currently under study, such as GRAVITY \cite{Eisenhauer08}, it could lead to a powerful test of the black hole paradigm.

\section{Complications in Testing the Black Hole No-Hair Theorem}
As we discussed, observations of the precessing orbits of a hypothetical class of stars very near the MBH in the GC could provide measurements of the spin and quadrupole moment of the hole and thereby test the no-hair theorem of GR. However, in assessing the feasibility of such strong-field GR tests, one must inevitably address potential complications, notably the perturbing effect of the other stars that may also reside in a cluster close to the black hole and a possible distribution of dark matter (DM) particles in the GC. These perturbing effects will be the focus of this thesis, and will be detailed in Chapters 2 and 3.

\subsection{Perturbing Effects of Stars in the Surrounding Cluster}
$N$-body simulations, have shown that for a range of possible stellar and stellar-mass black hole distributions within the central few milliparsecs (mpc) of the black hole, there could exist stars in eccentric orbits with semi-major axes less than $0.2$ milliparsecs for which the orbital-plane precessions induced by the stars and black holes would {\it not} exceed the relativistic precessions \cite{MAMW09}. These conclusions were gleaned from thousands of simulations of clusters ranging from seven to 180 stars and stellar mass black holes orbiting a $4 \times 10^6 \, M_\odot$ maximally rotating black hole, taking into account the long-term evolution of the system as influenced by close stellar encounters, dynamical relaxation effects, and capture of stars by the black hole.

In Chapter 2, we study the extent to which the conclusions of these complex $N$-body simulations can be understood, at least within an order of magnitude, using analytic orbit perturbation theory. After a brief review of orbit perturbation theory, we calculate the average change in the orientation of the orbital plane of a given ``target'' star orbiting the massive black hole, as determined by its inclination and ascending node angles $i$ and $\Omega$, induced by the Newtonian gravitational attraction of a distant third star (which could be either inside or outside the target star's orbit).

The perturbing accelerations are expanded in terms of multipoles through $\ell = 3$. We then calculate the root-mean-square variation of each orbit element, averaged over all possible orientations of the perturbing star's orbit, and averaged over a distribution of orbits in semi-major axis and eccentricity, arguing that this will give an estimate of the ``noise'' induced by the graininess of the otherwise spherically symmetric perturbing environment. Our analytic estimates of this ``noise'' will turn out to be consistent with the results from the N-body simulations, and will demonstrate that, for a range of possible distributions of stars in the central region, a test of the no-hair theorem will still be possible.

\subsection{Perturbing Effects of Dark Matter}
Another perturbing factor which can cause precessions in stellar motions is DM. To study the effect of DM on stellar motions in the GC, we need to have the DM density in that region. In order to derive an accurate density profile of DM particles in the GC, the effect of the MBH on the DM particles distribution, should be taken into account. Calculations by Gondolo and Silk (\cite{Gondolo99}, GS hereafter) have shown that for a pre-existing cusped DM halo, adiabatic (i.e. slow) growth of the MBH pulls the DM particles into a dense ``spike''. The calculation in GS was based on a Newtonian analysis, with some relativistic effects introduced in an ad hoc fashion, but because of the strong gravitational field near the MBH, a more reliable and realistic prediction for the DM density profile demands a fully general relativistic calculations. In Chapter 3, we report the first, fully relativistic calculation of the density profile of DM particles near a Schwarzschild black hole in the adiabatic growth model. We find significant differences with the conclusions of GS very close to the hole, but we are in complete agreement with them at large distances.


We use these relativistically correct density distributions to calculate the perturbing effect of the DM distribution on stellar motion in the GC for the hypothetical target stars to test the no-hair theorem and also the for S2 star in the S-stars cluster. The perturbing effect of the DM distribution depends on whether or not the dark-matter particles self-annihilate. 

The DM density distribution and therefore its perturbing effect also depends on whether the DM particle can self-annihilate or not. We will show that the perturbing effects of the DM mass distribution are too small to affect the possibility of testing the no-hair theorems using stars very close to the black hole.

\section{Dark Matter Evidence and Distribution}
We observe some ``anomalies'' in astrophysical systems, with sizes ranging from sub-galactic to cosmological scales, that can be explained by assuming the existence of a large amount of unseen, DM. Therefore, DM can be studied in different scales. In the following, we review the evidence for DM at these different scales although we will be primarily interested in the sub-galactic domain.

\subsection{Galaxy Cluster and Galactic Scales}
A galaxy cluster gave the first evidence of DM. In 1933, F. Zwicky \cite{Zwicky33} calculated the gravitational mass of the galaxies within the Coma cluster using the observed velocities of outlying galaxies and obtained a value more than 400 times greater than expected from their luminosity, which his interpretation was that most of the matter controlling the motion of the galaxies must be dark. Today, using the modern value of the Hubble constant and taking into account that there is baryonic gas in the galaxy cluster, bring down the amount of DM to 25 times the baryonic matter which still makes it clear that the great majority of matter appears to be dark.

\begin{figure}
\centering
\includegraphics[scale=0.8]{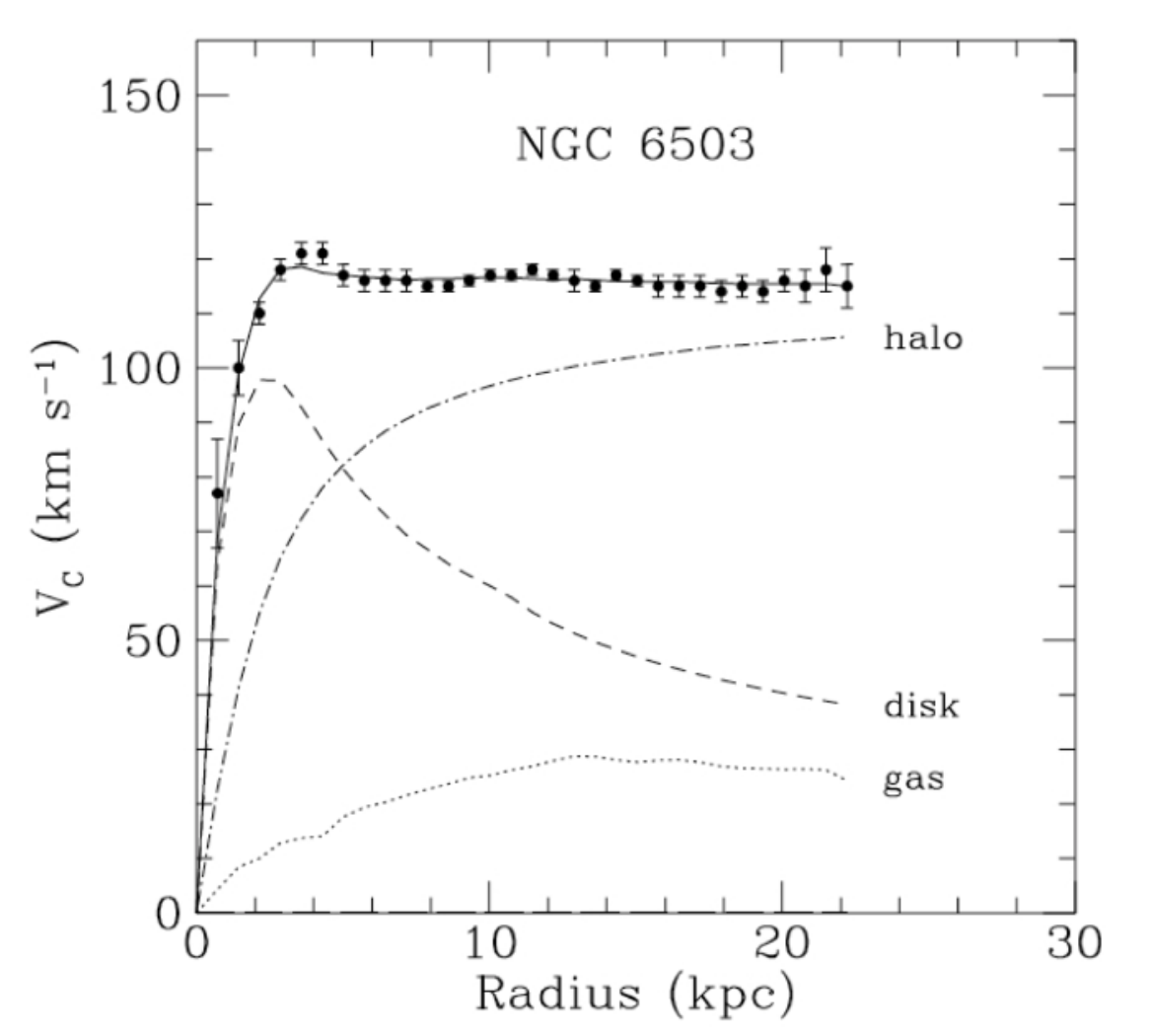}
\caption{Rotation curve of NGC 6503 from \cite{Begeman91}. The dotted, dashed and dash-dotted lines are the contributions of gas, disk and DM respectively.}
\label{rotation_curve}
\end{figure}

The most convincing and direct evidence for the DM existance on galactic scales, comes from the observations of the {\it rotation curves} of galaxies, namely the graph of circular velocities of stars and gas as a function of their distance from the galactic center. Observed rotation curves usually exhibit a characteristic {\it flat} behavior at large distances, i.e. out towards, and even far beyond, the edge of the visible disk. Fig.~\ref{rotation_curve} is a typical example \cite{Begeman91}.

In Newtonian dynamics the circular velocity is expected to be 
\be \label{ch1-7}
v(r)=\sqrt{\frac{Gm(r)}{r}} \ ,
\ee
where as usual, $m(r)=4\pi \int \rho(r)r^2{\rm d}r$, and $\rho(r)$ is the mass density profile. If $\rho$ vanishes outside the visible disk, then $m(r)$ is constant beyond the visible disk, and $v(r)$ should be falling as $1/\sqrt{r}$. The fact that the {\em observed}  $v(r)$ is approximately constant implies the existence of a halo with $m(r)\propto r$ and a mass density profile closely resembling that of an isothermal sphere, i.e., $\rho \propto 1/r^2$ at distances of few kiloparsec. 

Although there is a consensus about the shape of DM halos at intermediate distances, DM distribution is unclear in the innermost regions of galaxies. The observed rotation velocity associated with DM in the inner parts of disk galaxies is found to rise approximately linearly with radius which leads to mass $\propto r^3$ and therefore constant density. This solid-body behavior can be interpreted as indicating the presence of a central core in the DM distribution, spanning a significant fraction of the visible disk \cite{Blok10}. Observations of dwarf spheroidal galaxies also seem to favor a constant density of DM in the inner parts \cite{Walker2011}.

On the other hand, N-body simulations indicate a steep power-law-like behavior for the DM distribution at the center. The results of N-body simulations are based on the ($\Lambda$)CDM paradigm, where the most of the mass-energy of our universe consists of collisionless cold dark matter (CDM) in combination with a cosmological constant $\Lambda$. This $\Lambda$CDM paradigm provides a comprehensive description of the universe at large scales. However, despite its great successes, it should be kept in mind that the cusp and the central DM distribution are not predicted from first principles by $\Lambda$CDM. Rather these properties are derived from analytical fits made to dark-matter-only numerical simulations. While the quality and quantity of these simulations has improved by orders of magnitude over the years, there is as yet no ``cosmological theory'' that explains and predicts the distribution of DM in galaxies from first principles.   

In the early 1990s, the first results of numerical N-body simulations of DM halos based on the collisionless cold dark matter (CDM) prescription became available. ``Cold'' dark matter is dark matter composed of constituents with a free-streaming length much smaller than the ancestor of a galaxy-scale perturbation. These did not show the observed core-like behavior in their inner parts, but were better described by a steep power-law mass density distribution, the so-called {\it cusp}. The presence of a cusp in the center of a CDM halo is one of the earliest results derived from cosmological N-body simulations. The first simulations indicated an inner distribution $\rho \sim r^\alpha$ with $\alpha=-1$ \cite{Dubinski91}. They did not rule out the existence of central cores, but noted that these would have to be smaller than the resolution of their simulations ($\sim 1.4 \ {\rm kpc}$). Subsequent simulations, at higher and higher resolutions, made the presence of cores in simulated CDM halos increasingly unlikely. In addition to finite resolution, the other limitation of N-body simulations is that the role of the baryons at small radius is ignored in their calculations.

A systematic study by Navaro {\it et al.} \cite{Navarro96,Navarro97} of simulated CDM halos, derived assuming many different sets of cosmological parameters, found that the innermost DM density distribution could be well described by a characteristic $\alpha=-1$ slope for all simulated halos, independent of mass and size. A similar general result was found for the outer mass profile, with a steeper slope of $\alpha=-3$:
\be \label{ch1-8}
\rho_{\rm NFW}(r)=\frac{\rho_0}{(r/a)(1+r/a)^2} \ ,
\ee    
where $\rho_0$ is related to the density of the universe at the time of halo collapse and $a$ is the characteristic radius of the halo. This kind of profile is also known as the ``NFW profile''.   

In Chapter 3, we will consider the constant and Hernquist distribution functions as examples of cored and cuspy models, respectively. The advantage of considering the Hernquist density profile which, like the NFW profile, is $\propto 1/r$ for small $r$, is that for the Hernquist model we have a closed analytical distribution function which allows us to study the effect of adiabatic growth of the MBH on the DM distribution using adiabatic invariants. 

\subsection{Cosmological Scales}
As we have seen, on distance scales of the size of galaxies and clusters of galaxies, the evidence of DM appears to be compelling. Despite this, the observations discussed do not allow us to determine the total amount of DM in the Universe.  

The theory of Big Bang nucleosynthesis gives a good estimate of the amount of ordinary (baryonic) matter at around 4 - 5 percent of the critical density (the density required to have a universe with a flat spatial section); while evidence from large-scale structure and other observations indicates that the total matter density is substantially higher than this \cite{Planck2013}. The Cosmic Microwave Background (CMB) fluctuations imply that at present the total energy density is equal to the critical density. This means that the largest fraction of the energy density of the universe is dark and nonbaryonic. It is not yet clear what constitutes this dark component. Combining the data on CMB, large scale structure, gravitational lensing and high-redshift supernovae, it appears that the dark component is a mixture of two types of constituents. More precisely, it is composed of dark matter and dark energy. The cold dark matter has zero pressure and can cluster, contributing to gravitational instability, but it does not emit light, which means that it does not have electromagnetic interactions. Various (supersymmetric) particle theories provide us with natural candidates for the cold dark matter, among which Weakly Interacting Massive Particles (WIMPs) are the most favored at present. The nonbaryonic cold dark matter contributes only about 25 percent of the critical density. The remaining 70 percent of the missing density comes in the form of nonclustered dark energy with negative pressure. It may be either a cosmological constant ($pressure=- \ energy \ density$) or a scalar field (quintessence) with $pressure=\omega \times energy \ density$, where $\omega$ is less than $-1/3$ today \cite{Weinberg2008}.



\section{Dark Matter Candidates}
The evidence for non-baryonic DM is compelling at all observed astrophysical scales. Candidates for nonbaryonic DM are hypothetical particles such as axions, or supersymmetric particles. The most widely discussed models for nonbaryonic DM are based on the cold dark matter hypothesis, and the corresponding particle is most commonly assumed to be for instance a WIMP.

WIMPs interact through a weak-scale force and gravity, and possibly through other interactions no stronger than the weak force. Because of their lack of electromagnetic interaction with normal matter, WIMPs would be dark and invisible through normal electromagnetic observations and because of their large mass, they would be relatively slow moving and therefore cold. Their relatively low velocities would be insufficient to overcome their mutual gravitational attraction, and as a result WIMPs would tend to clump together. 

Although WIMPs are a more popular DM candidate, there are also experiments searching for other particle candidates such as axions. The axion is a hypothetical elementary particle postulated to resolve the strong CP problem in quantum chromodynamics. Observational studies to detect DM axions through the products of their decay are underway, but they are not yet sufficiently sensitive to probe the mass regions where axions would be expected to be found if they are the solution to the DM problem.

\section{Indirect Detection of Dark Matter}
Indirect dark matter searches measure the annihilation and/or decay products of DM from astrophysical systems. Schematically, they measure the rate for {\it DM} {\it DM} $\rightarrow$ {\it SM} {\it SM} or {\it DM} $\rightarrow$ {\it SM} {\it SM}, depending on whether dark matter particles annihilate or decay where {\it DM} represents the dark matter particle and {\it SM} represents any standard model particle. In many instances, the particle represented by {\it SM} is unstable, and decays into other particles (for example, photons or neutrinos) that are observable in detectors. In order to best interpret the results from indirect searches, we must have a good idea as to both how the dark matter is distributed in halos, and what standard model particles the dark matter preferentially annihilates or decays into.

One of the main possibilities for indirect detection of DM particles is to search for high-energy gamma rays, positrons, antiprotons, or neutrinos produced by WIMP pair annihilations in the Galactic halo. In particular, the flux of gamma rays in a given direction is proportional to the square of the DM particle density and since the DM density is expected to be largest towards the Galactic center, the flux of such exotic gamma rays should be highest in that direction. In other words, the innermost region of our galaxy is one of the most promising targets for the indirect detection of DM and it is important that we know the DM density profile in the vicinity of the Galactic center MBH. In Chapter 3, to study the effect of the MBH, we developed a fully general relativistic phase-space formulation, allowed the central black hole to grow adiabatically, holding the general relativistic adiabatic orbital invariants fixed, and incorporated a relativistically correct condition for particle capture by the black hole. The result showed significant differences with the semi-relativistic result of Gondolo and Silk \cite{Gondolo99}, including a bigger spike in the halo density close to the black hole.  Finally having the dark matter profile density in presence of the MBH, we also calculated its perturbing effect on the orbital motions of stars in the Galactic center.


\clearpage 


\chapter{Testing the Black Hole No-Hair Theorem at the Galactic Center} 
\label{chapter2} 
\thispagestyle{myplain}
\lhead{Chapter 2. \emph{Testing the Black Hole No-Hair Theorem at the Galactic Center}} 

In this chapter we start with the well-known Kepler problem to introduce the notation and review the necessary equations which need to be generalized to the non-spherical cases in order to study Keplerian orbits in space. Then we introduce the basic equations of orbit perturbation theory and derive the general relativistic effects of the central massive black hole on the orbits of stars as one of the applications of this theory. This provides the test of the no-hair theorem in the innermost region of the galactic center. Then we study the perturbing effect of a distribution of stars on the orbit of a target star.

\section{General Relativistic Effects in Stellar Motion Around Massive Black Holes}

\subsection{The Kepler Problem}
The simplest Newtonian problem is that of two ``point'' masses in orbit about each
other, frequently called the  ``Kepler problem''. In Kepler's problem, we have a body of mass $m_1$, position $\bm r_1$, velocity $\bm v_1={\rm d} {\bm r}_1/{\rm d}t$, and acceleration ${\bm a}_1={\rm d}{\bm v}_1/{\rm d}t$, and a second body of mass $m_2$, position $\bm r_2$, velocity $\bm v_2={\rm d} {\bm r}_2/{\rm d}t$, and acceleration ${\bm a}_2={\rm d}{\bm v}_2/{\rm d}t$. We place the origin of the coordinate system at the center of mass,  so that $m_1{\bm r}_1+m_2 {\bm r}_2=0$. The position of each body is then given by
\be \label{ch2-0}
{\bm r}_1=\frac{m_2}{m}{\bm r}, \;\;\;\;\;  {\bm r}_2=-\frac{m_1}{m}{\bm r} \ ,
\ee
in which $m\equiv m_1+m_2$ is the total mass and $\bm r \equiv {\bm r}_1-{\bm r}_2 $ the separation between bodies. Similar relations hold between ${\bm v}_1$, ${\bm v}_2$, and the relative velocity  $\bm v \equiv {\bm v}_1-{\bm v}_2={\rm d}{\bm r}/{\rm d}t $. For the relative acceleration $\bm a \equiv {\bm v}_1-{\bm v}_2 ={\rm d}{\bm v}/{\rm d}t$ we have
\be \label{ch2-1}
{\bm a}=-G\frac{m}{r^2} \hat{\bm n} \ ,
\ee
where $r\equiv |{\bm r}|$ is the distance between the bodies, and $\hat{\bm n}\equiv {\bm r}/r$, is a unit vector that points from body 2 to body 1. The total energy and the angular momentum of the system are given by
\begin{eqnarray} \label{ch2-2}
E&=&\frac{1}{2} \mu v^2-G\frac{\mu m}{r} \ , \\ \label {ch2-3}
\bm L &=&\mu \bm r \times \bm v \ ,
\end{eqnarray}
where 
\be \label{ch2-4}
\mu \equiv \frac{m_1m_2}{m_1+m_2} \ ,
\ee
is the {\it reduced mass} of the system. It is simple to verify explicitly using Eq. (\ref {ch2-1})
that ${\rm d}E/{\rm d}t = 0$ and ${\rm d}\bm L/{\rm d}t = 0$. The constancy of $E$ and $\bm L$ are a result of the fact that the potential $Gm/r$ that governs the effective one-body problem of Eq. (\ref {ch2-1}) is static and spherically symmetric. The constancy of $\bm L$ implies that all the motion lies in a plane perpendicular to $\bm L$ and it is fixed. So, we are free to choose our coordinates so that the
$z$-axis is parallel to $\bm L$, and the motion occurs in the $xy$-plane. Converting from Cartesian to polar coordinates in the orbital plane using $x = r \cos \phi$ and $y = r \sin \phi$, we see that
\be  \label {ch2-5}
\bm r \times \bm v= r^2\frac{{\rm d}\phi}{{\rm d}t}\hat{\bm e}_z\equiv h \hat{\bm e}_z \ ,
\ee
where $h$, called the angular momentum per unit reduced mass, is constant. Writing $\bm r=r \hat{\bm n}$, where $\hat{\bm n}=\cos \phi \hat{\bm e}_x+\sin \phi \hat{\bm e} _y$, we see that
\be \label{ch2-6}
\bm v=\frac{{\rm d} \bm r}{{\rm d}t}=\dot r \hat{\bm n}+r \dot \phi \hat{\bm \lambda} \ ,
\ee
where
\be \label{ch2-7}
\hat{\bm \lambda} \equiv {\rm d}\hat {\bm n}/{\rm d} \phi \ ,
\ee
is a vector in the orbital plane orthogonal to $\hat{\bm n}$. From this we see that
\be \label{ch2-8}
v^2=\dot r^2+r^2 \dot \phi^2=\dot r^2+\frac{h^2}{r^2} \ .
\ee

We now take the component of Eq. (\ref{ch2-1}) in the radial direction, and note that
\begin{eqnarray} \nonumber
\hat{\bm n} \cdot \frac{{\rm d}^2 \bm r}{{\rm d}t^2}&=&\frac{{\rm d}^2}{{\rm d}t^2}(\hat{\bm n}  \cdot \bm r)-\frac{{\rm d}}{{\rm d}t}(\bm r \cdot \frac{{\rm d}\hat{\bm n}}{{\rm d}t})-\frac{{\rm d} \hat{\bm n}}{{\rm d}t} \cdot \bm v\\ \nonumber
&=&\frac{{\rm d}^2r}{{\rm d}t^2}-\frac{{\rm d}}{{\rm d}t}(r \hat{\bm n} \cdot \frac{{\rm d}\hat{ \bm n}}{{\rm d}t})-\frac{v^2-\dot r^2}{r} \\  \label{ch2-9}
&=&\ddot r-\frac{h^2}{r^3} \ ,
\end{eqnarray}
where $\dot r\equiv\hat{\bm n} \cdot \bm v$, and we have used the fact that $\hat{\bm n}\cdot {\rm d} \hat{\bm n}/{\rm d}t=0$, and that $h^2=|\bm r \times \bm v|^2=r^2(v^2-\dot r^2)$. The result is a differential equation for the radial motion,
\be \label{ch2-10}
\ddot r-\frac{h^2}{r^3}=-G\frac{m}{r^2} \ .
\ee
Multiplying by $\dot r$ and integrating once, we find the \textquotedblleft first integral'' of the equation,
\be \label{ch2-11}
\frac{1}{2} \left( \dot r^2+\frac{h^2}{r^2}\right)-G\frac{m}{r}=\tilde E \ ,
\ee
where from Eq. (\ref{ch2-2}), we can see that $\tilde E$ is the energy per unit reduced mass.

It is useful to rewrite Eq. (\ref{ch2-11}) in the form
\be \label{ch2-12}
\dot r^2=2\big[\tilde E- V_{\rm eff}(r)\big] \ ,
\ee
where we define the effective radial potential 
\be \label{ch2-13}
V_{\rm eff}(r)\equiv\frac{h^2}{2r^2}-G\frac{m}{r} \ .
\ee
This must be combined with the equation for the angular motion,
\be \label{ch2-14}
\dot \phi =\frac{h}{r^2} \ .
\ee

Now we try to find a parametric solution to the equations, which is a solution of the form $r=r(\lambda)$, $\phi=\phi(\lambda)$, where $\lambda$ is a parameter which will depend on $t$. Consider Eq.\ (\ref{ch2-10}), and insert the fact that ${\rm d}/{\rm d}t=\dot \phi {\rm d}/{\rm d}\phi=(h/r^2){\rm d}/{ \rm d}\phi$, to obtain
\be \label{ch2-15}
\frac{h}{r^2} \frac{{\rm d}}{{\rm d}\phi} \left(\frac{h}{r^2} \frac{{\rm d}r}{{\rm d}\phi} \right)-\frac{h^2}{r^3}+G\frac{m}{r^2}=0 \ .
\ee
Using $1/r$ as the variable, we can recast this equation into the form
\be \label{ch2-16}
\frac{{\rm d}^2}{{\rm d}\phi^2}\left(\frac{1}{r}\right)+\frac{1}{r}=G\frac{m}{h^2} \ .
\ee
The homogenous solution can be written as $A \cos(\phi-B)$, where $A$ and $B$ are arbitrary constants. Combining this with the inhomogeneous solution $m/h^2$, and redefining the constants, we obtain the solution for $1/r$ in terms of the parameter $\phi$, given by 
\be \label{ch2-17}
\frac{1}{r}=\frac{1}{p}[1+e\cos {(\phi-\omega)}] \ ,
\ee
where $e$ and $\omega$ fill in for the two arbitrary constants $A$ and $B$, and
\be \label{ch2-18}
p \equiv \frac{h^2}{Gm} \ .
\ee
Notice that a solution with $e<0$ is equivalent to one with $e>0$, but with $\omega \rightarrow \omega+\pi$; henceforth we will adopt the convention that $e$ is positive. The angle $f\equiv \phi-\omega$ is called the {\it true anomaly}.

The curve described by Eq. (\ref {ch2-17}) can be shown to be a {\it conic section}, an ellipse if the quantity $e<1$, a hyperbola if $e>1$, and a parabola if $e=1$, with the origin $r=0$ at one of the foci of the curve. The parameter $e$ is called the {\it eccentricity} of the orbit. Notice that $r$ is a minimum when $\phi=\omega$; this is the point of closest approach in the orbit, called the {\it pericenter}, and $\omega$ is called the {\it angle of pericenter} and simply fixes the orientation of the orbit in the $xy$-plane.

For the $e<1$ case, the point where $\phi=\omega+\pi$ is the point of greatest separation, called the {\it apocenter}. The pericenter and apocenter distances are thus given by
\be \label{ch2-19} 
r_{\rm peri}=\frac{p}{1+e} \ , \quad r_{\rm apo}=\frac{p}{1-e} \ .
\ee
The sum of these is the major axis of the ellipse, so we define the {\it semi-major axis} $a$ to be
\be \label{ch2-20}
a\equiv \frac{1}{2}(r_{\rm peri}+ r_{\rm apo})=\frac{p}{1-e^2} \ .
\ee 
As a result, we can also write the solution for $1/r$ in the form
\be \label{ch2-21}
\frac{1}{r}=\frac{1+e\cos{(\phi-\omega)}}{a(1-e^2)} \ .
\ee
The quantity $p=a(1-e^2)$ is called the {\it semi-latus rectum}. 

From Eqs.\ (\ref{ch2-14}) and (\ref{ch2-17}), it is straightforward to derive the following useful formulae, valid for arbitrary values of $e$:
\begin {eqnarray} \label{ch2-22}
\dot r& =&\frac{he}{p} \sin{(\phi-\omega)} \ , \\  \label{ch2-23}
v^2&=&G\frac{m}{p}\big[1+2e\cos{(\phi-\omega)}+e^2\big]=m\left(\frac{2}{r}-\frac{1}{a}\right) \ , \\  \label{ch2-24}
E&=&-G\frac{\mu m}{2a} \ , \\ \label{ch2-25}
e^2&=&1+\frac{2h^2E}{\mu (Gm)^2} \ .
\end{eqnarray}
So far we have determined the orbit as a function of $\phi$, with three arbitrary constants, $a$, $e$, and $\omega$, called {\it orbit elements}. To complete the parametric solution we need to determine $\phi$ as a function of time or as a function of some parameter related to time. From Eq. (\ref{ch2-14}), we obtain
\be \label{ch2-26}
t-T=\int_\omega^\phi \frac{r^2{\rm d}\phi'}{h}=\left(\frac{p^3}{Gm}\right)^{1/2}\int_\omega^\phi \frac{{\rm d}\phi'}{[1+e\cos(\phi'-\omega)]^2} \ ,
\ee
where $T$, called the {\it time of pericenter passage}, is the fourth orbit element required to complete our solution in the orbital plane.

For $e<1$, we can integrate over a complete orbit, and obtain the orbital period 
\be \label{ch2-27}
P=2\pi\left(\frac{a^3}{Gm}\right)^{1/2} \ .
\ee
It is common to define the mean angular frequency or {\it mean motion} $n\equiv 2\pi/P$, so that $n^2a^3=Gm$. Now carrying out the integral in Eq. (\ref {ch2-26}) explicitly, we can find that
\be \label{ch2-27A}
n(t-T)=u-e \sin u \ ,
\ee
where the variable $u$ is called the {\it eccentric anomaly}, and is related to $f$ by
\be \label{ch2-27B}
\tan {\frac{f}{2}}=\sqrt{\frac{1+e}{1-e}} \tan{\frac{u}{2}} \ .
\ee
In terms of the eccentric anomaly, the radius of the orbit is given by
\be \label{ch2-27C}
r=a(1-e\cos u) \ .
\ee
This set of equations, called Kepler's solution for the two body problem is a convenient parametric solution for orbit determinations, since for given values of the orbit elements $a$, $e$, $\omega$ and $T$, one chooses $t$, solves Eq.~(\ref{ch2-27A}) for $u$, then substitutes that into Eqs.\ (\ref {ch2-27B}) and (\ref {ch2-27C}) to obtain $f(t)$ and $r(t)$, and thence $x(t)$ and $y(t)$.

Similar parametric solutions can be obtained for hyperbolic orbits, in terms of hyperbolic functions. 

There is one curious feature of our solution for the Kepler problem, and that is that the orientation of the orbit is fixed in the orbital plane, i.e. the angle of pericenter $\omega$ is a constant. It is not related to the spherical symmetry of the potential or to its time independence; these led only to the conservation of $\bm L$ and $E$ and to the integrability of the equations.

The constancy of $\omega$ is the result of a deeper symmetry embedded in the Kepler problem, associated with the $1/r$ nature of the potential. One can define another vector associated with the orbital motion, often called the Runge-Lenz vector, given by
\be \label{ch2-27D}
\bm R \equiv \frac{{\bm v} \times {\bm h}}{Gm}-\frac{\bm r}{r} \ ,
\ee
where ${\bm h}= {\bm r} \times {\bm v}$. Substituting $\bm r=r \bm n$, with $r$ given by Eq.~(\ref{ch2-17}), along with $\hat{\bm n}=\hat{\bm e}_x \cos \phi+\hat{\bm e}_y \sin \phi$ and Eqs.~(\ref{ch2-22}) and (\ref{ch2-23}), it can be shown that 
\be \label{ch2-27E}
\bm R=e[\hat{\bm e}_x \cos \omega+\hat{\bm e}_y \sin \omega] \ ,
\ee
which is a vector of magnitude $e$ pointing toward the pericenter. However, using the equation of motion Eq.~(\ref{ch2-1}), it is easy to show that
\be \label{ch2-27F}
\frac{{\rm d} \bm R}{{\rm d} t}=0 \ ,
\ee
so that $\bm R$ is another constant of the motion. Since $e$ is constant by virtue of Eq.~(\ref{ch2-25}), this implies that $\omega$ is constant. But in this case, the $1/r$ nature of the potential is crucial; had one substituted an equation of motion derived from a potential $1/r^{1+\epsilon}$, or $1/r+\alpha/r^2$, $\bm R$ would no longer be constant, even though $E$ and $\bm L$ would stay constant and the problem would remain completely integrable. 
 
\subsection{Keplerian Orbits in Space}

In order to consider more realistic problems, we are interested in perturbations in our two-body problem which may be caused by gravitational forces exerted by external bodies, by the effects of multipole moments resulting from tidal or rotational perturbations, or by general relativistic contributions. Such effects will not be spherically symmetric in general, and so the orientation of the orbit will be important. So in this section we will review the full Keplerian orbit in space.

The conventional description of the full Keplerian orbit in space goes as follows: we first establish a {\it reference} $XY$ plane and a reference $Z$ direction. For planetary orbits, the reference plane is the plane of the Earth's orbit, called the {\it ecliptic plane}, and the $Z$ direction is perpendicular to the ecliptic plane is in the same sense as the Earth's north pole (ignoring the $23^\circ$ tilt). For Earth orbiting satellites, it is the equatorial plane. For binary star systems, it is the plane of the sky. Within each reference plane, the $X$-direction must be chosen in some conventional manner.

\begin{figure}
\centering
\includegraphics[trim = 40mm 46mm 27mm 20mm, clip, width=17cm]{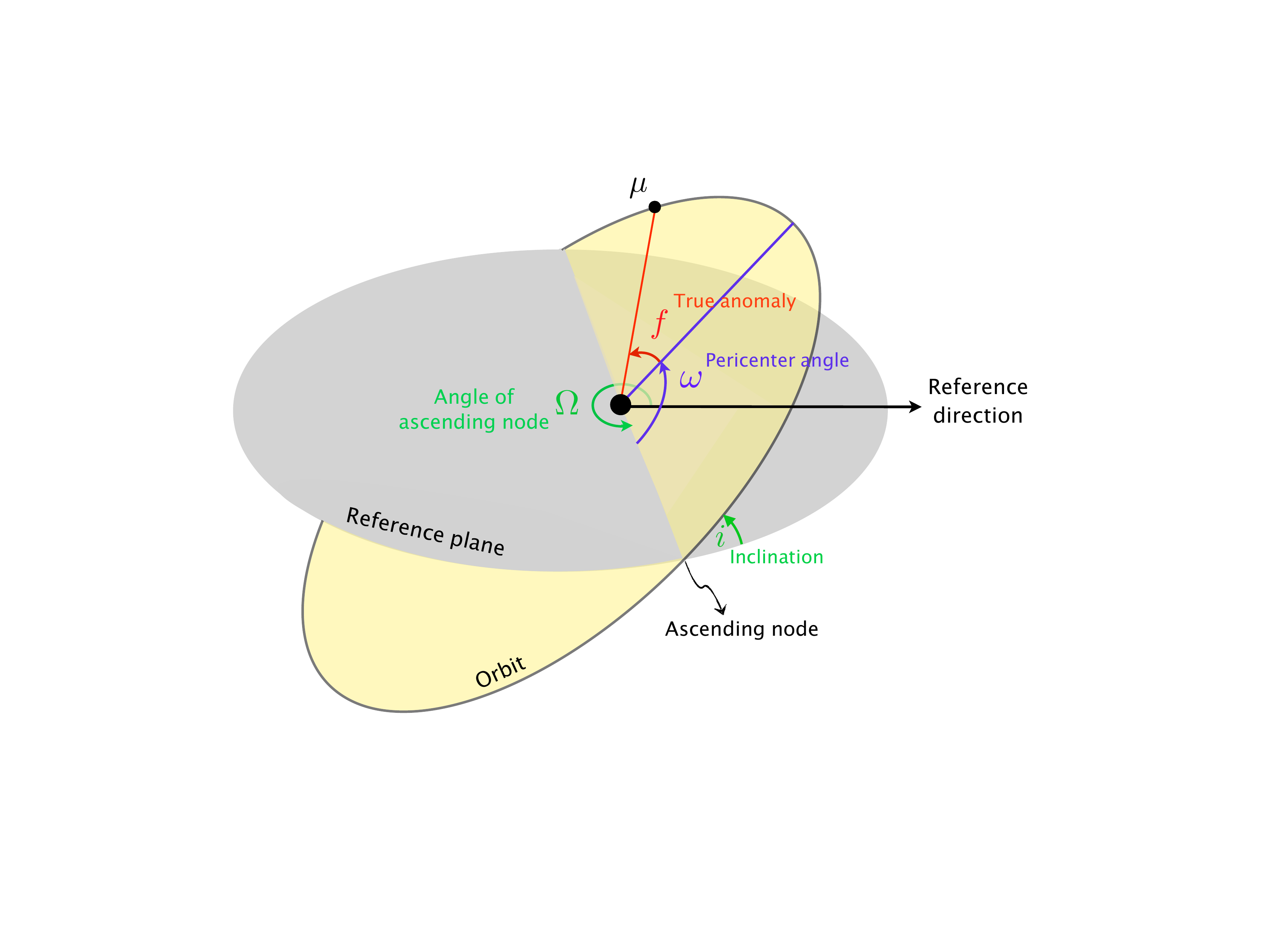}
\caption{The orbit elements which uniquely identify a specific orbit in space. Here, the orbital plane (yellow) intersects a reference plane (gray). }
\label{orbit_elements}
\end{figure}

We now define the {\it inclination} $i$ of the orbital plane to be the angle between the
positive $Z$ direction and a normal to the plane (where the direction of the normal
is defined by the direction of the angular momentum of the orbiting body). This tilted plane then intersects the reference $XY$ plane along a line. We define the {\it angle of the ascending node} or nodal angle $\Omega$ to be the angle between the $X$ axis and the intersection line where the body ``ascends'' from below the reference plane (the negative $Z$ side) to above it. The pericenter angle $\omega$ is the angle measured in the orbital plane from the line of nodes to the pericenter. These three angles
then fix the orientation of the orbit in space. Within the orbital plane, the orbit is determined by the three remaining orbit elements $a$, $e$, and $T$. The true anomaly $f$ is measured in the orbital plane from the pericenter to the location of the body. The orbit elements which uniquely identify a specific orbit are illustrated in Fig.~\ref{orbit_elements}. 

Given a unit vector $\hat{\bm n}$ pointing from the center of mass to the body, it is straightforward to express $\hat{\bm n}$ in terms of the $XYZ$ basis:
\begin{eqnarray} \nonumber
\hat{\bm n}&=&\hat{{\bm e}}_X [\cos{(\omega+f)} \cos{\Omega}-\sin{(\omega+f)\sin \Omega \cos i}]\\ \nonumber
&&+\hat{{\bm e}}_Y [\cos{(\omega+f)} \sin{\Omega}+\sin{(\omega+f)\cos \Omega \cos i}] \\ \label{ch2-28}
&&+\hat{{\bm e}}_Z [\sin{(\omega+f)}\sin i] \ .
\end{eqnarray}

We can relate all the six orbit elements $a$, $e$, $\omega$, $\Omega$, $i$ and $T$ directly to the position $\bm r$ and velocity $\bm v$ of a body in a Keplerian orbit at a given time $t$. The first step is to use $\bm r$ and $\bm v$ to form the vectors
\begin{eqnarray} \nonumber
\bm h &\equiv& \bm r \times \bm v\\ \nonumber
&=&h[\sin i(\hat{{\bm e}}_X \sin{\Omega}-{\hat{\bm e}}_Y \cos {\Omega})+\hat{{\bm e}}_Z \cos {i}] \; , \\ \nonumber
{\bm R} &\equiv& {\bm v} \times {\bm h}/(Gm)-{\bm r}/r \\ \nonumber
&=&e[\hat{{\bm e}}_X(\cos {\omega} \cos {\Omega}- \sin {\omega} \sin{\Omega} \cos {i})\\ \label{ch2-29}
&&+{\hat{\bm e}}_Y(\cos {\omega} \sin{\Omega}+\sin {\omega} \cos{\Omega} \cos {i})+ \hat{{\bm e}}_Z \sin{\omega} \sin {i}] \ ,
\end{eqnarray}

where $\bm R$ is the Runge-Lenz vector. Given $h^2=Gmp=Gma(1-e^2)$, we can identify the orbit elements in terms of quantities constructed from $\bm r$ and $\bm v$ in the XYZ coordinates:
\begin {eqnarray} \label{ch2-30}
e&=& |\bm R| \ , \\  \label{ch2-31}
a&=& \frac{h^2}{Gm(1-e^2)} \ , \\ \label{ch2-32}
\cos {i} &=& \frac{{\bm h}\cdot {\bm e}_Z}{h} \ , \\ \label{ch2-33}
\cos {\Omega} &=& -\frac{{\bm h}\cdot {\bm e}_Y}{h \sin i} \ , \\ \label{ch2-34}
\sin {\omega} &=& \frac{{\bm R} \cdot {\bm e}_Z }{e \sin i} \ .
\end{eqnarray}
Given these elements, and the Keplerian solution Eq. (\ref {ch2-17}), the final orbit element $T$, the {\it time of pericenter passage} is given by the integral
\be \label{ch2-35}
T=t-\int_0^f(r^2/h){\rm d}f \ ,
\ee
where $f=\phi-\omega$. The actual orbit is then given by ${\bm r}(t)=r \hat{\bm n}$, with $r$ given by either  Eq. (\ref{ch2-21}) or Eq. (\ref {ch2-27C}), and with the appropriate relation between the true anomaly $f$ or the eccentric anomaly $u$ and time $t$.

\subsection{Osculating Orbit Elements and the Perturbed Kepler Problem}
\label{Osculating}

Suppose the equation of motion for our effective two-body problem is no longer given by Eq.~(\ref {ch2-1}), but by something else:
\be \label{ch2-36}
{\bm a}=-G\frac{m}{r^2} \hat{\bm n}+{\bm A}({\bm r},{\bm v},t) \ ,
\ee
where $\bm A$ is a perturbing acceleration, which may depend on ${\bm r}$, ${\bm v}$ and time. The solution of this equation is no longer a conic section of the Kepler problem. However, whatever the solution is, at any given time $t_0$, for ${\bm r}(t_0)$, ${\bm v}(t_0)$, there exists a Keplerian orbit with orbit elements $e_0$, $a_0$, $\omega_0$, $\Omega_0$, $i_0$ and $T_0$ that corresponds to those values, as we constructed in the previous section. In other words there is a Keplerian orbit that is {\it tangent} to the orbit in question at the time $t_0$, commonly called the {\it osculating orbit}.

However, because of the perturbing acceleration, at a later time, the orbit will not be the same Keplerian orbit, but will be tangent to a new osculating orbit, with new elements $e'$, $a'$ and so on. The idea then is to study a general orbit with the perturbing acceleration $\bm A$ by finding the sequence of osculating orbits parametrized by $e(t)$, $a(t)$, and so on. If the perturbing acceleration is small in a suitable sense, then since the orbit elements of the original Kepler motion are constants, we might hope that the osculating orbit elements will vary slowly with time and by small amounts.

Mathematically, this approach is identical to the method of variation of parameters in solving differential equations, such as the harmonic oscillator with a slowly varying frequency.

In this case, we replace our Keplerian solution for the motion with the following {\it definitions}:
\begin{eqnarray} \label{ch2-37}
{\bm r} &\equiv& r \hat {\bm n} \ , \\  \label{ch2-38}
r &\equiv& \frac{p}{1+e\cos f} \ , \\ \label{ch2-39}
{\bm v} &\equiv& \frac{he\sin f}{p} \hat {\bm n}+\frac{h}{r} \hat {\bm \lambda} \ , \\ \label{ch2-40}
p &\equiv& a (1-e^2) \ , \\ \label{ch2-41}
h^2 &\equiv&Gmp \ , 
\end{eqnarray}
where the unit vectors $\hat{\bm n}$, $\hat{\bsym \lambda}$, and $\hat{\bm h}$ are given by
\begin{eqnarray} \nonumber
\hat {\bm n}&\equiv&\hat{{\bm e}}_X [\cos{(\omega+f)} \cos{\Omega}-\sin{(\omega+f)\sin \Omega \cos i}]\\ \nonumber
&&+\hat{{\bm e}}_Y [\cos{(\omega+f)} \sin{\Omega}+\sin{(\omega+f)\cos \Omega \cos i}] \\ \label{ch2-42}
&&+\hat{{\bm e}}_Z [\sin{(\omega+f)}\sin i] \ ,\\ \nonumber
\hat{\bm \lambda}&\equiv&-\hat{{\bm e}}_X [\sin{(\omega+f)} \cos{\Omega}+\cos{(\omega+f)\sin \Omega \cos i}]\\ \nonumber
&&-\hat{{\bm e}}_Y [\sin{(\omega+f)} \sin{\Omega}-\cos{(\omega+f)\cos \Omega \cos i}] \\ \label{ch2-43}
&&+\hat{{\bm e}}_Z [\cos{(\omega+f)}\sin i] \ , \\ \label{ch2-43A}
\hat {\bm h}&\equiv&\hat{{\bm e}}_X \sin i \sin \Omega-\hat{{\bm e}}_Y  \sin i \cos \Omega+\hat{{\bm e}}_Z \cos i \ .
\end{eqnarray}
Note that $\hat{\bm n} \times \hat {\bm \lambda}=\hat {\bm h}$.

In the pure Kepler problem, we saw that the orbit elements (apart from $T$) were obtained from the constant vectors $\bm h$ and $\bm R$; now we calculate their time derivatives, using the perturbed equation of motion Eq.~(\ref {ch2-36}), with the result
\begin{eqnarray} \nonumber
\frac{{\rm d}{\bm h}}{{\rm d}t}&=&{\bm r} \times {\bm A} \ , \\  \label{ch2-44}
m\frac{{\rm d} {\bm R}}{{\rm d}t}&=&{\bm A} \times {\bm h}+ {\bm v} \times ({\bm r} \times {\bm A}) \ .
\end{eqnarray}
We now decompose the perturbing acceleration into components along the orthogonal directions $\hat{\bm n}$, $\hat{\bm \lambda}$, and $\hat {\bm h}$ by
\be \label{ch2-45}
{\bm A} \equiv {\cal R} \hat{\bm n}+{\cal S} \hat{\bm \lambda}+ {\cal W} \hat {\bm h} \ ,
\ee  
where $\cal R$, $\cal S$, and $\cal W$ are sometimes referred to as the radial or \textquotedblleft cross-track'', tangential or \textquotedblleft in-track'', and out-of-plane components of the acceleration, respectively. With these definitions we obtain
\begin {eqnarray} \label{ch2-46}
\frac{{\rm d} {\bm h}}{{\rm d}t}&=&-r{\cal W} \hat{\bm \lambda}+r {\cal S} \hat {\bm h} \ , \\ \label{ch2-47}
m \frac{{\rm d} {\bm R}}{{\rm d}t}&=&2h{\cal S} \hat{\bm n}-(h {\cal R}+r \dot r {\cal S}) \hat{\bm \lambda}-r \dot r {\cal W} \hat {\bm h} \ .
\end {eqnarray}
Note that, because ${\bm h} \cdot \dot {\bm h}=h \dot h$, we immediately conclude that
\be \label{ch2-48}
\dot h=r {\cal S} \ .
\ee
We can now systematically develop equations for the variations with time of the osculating orbit elements. For example, since ${\bm h} \cdot \hat{{\bm e}}_Z=h \cos i$, then $\dot {\bm h}\cdot \hat{{\bm e}}_Z=\dot h \cos i-h \sin i ({\rm d}i/{\rm d}t)=r{\cal S} \cos i-r {\cal W} \cos {(\omega+f)}\sin i$, with the result that ${\rm d}i/{\rm d}t=(r{\cal W}/h)\cos {(\omega+f)}$. Similarly, since ${\bm h} \cdot \hat{{\bm e}}_Y=-h \sin i \cos \Omega$, then taking the derivative of both sides and subtracting our previous result for $\dot {\bm h}$, $\dot h$ and ${\rm d}i/{\rm d}t$, we obtain $\sin i \ \dot {\Omega}=(r{\cal W}/h)\sin{(\omega+f)}$. To obtain $\dot e$, we note that $e\dot e={\bm R}\cdot \dot {\bm R}$, and use the fact that ${\bm R}=\hat{\bm n} \cos f-\hat{\bm \lambda} \sin f$. For $\dot a$, we use the definition $h^2=Gma(1-e^2)$, from which $\dot a/a=2 \dot h/h+2e\dot e/(1-e^2)$. For $\dot \omega$, we use the fact that ${\bm R} \cdot \hat{{\bm e}}_Z=e \sin i \sin \omega$, combined with previous results for $\dot e$ and ${\rm d}i/{\rm d}t$. The final equations for the osculating orbit elements are
\begin{eqnarray} \label {ch2-49}
\frac{{\rm d}a}{{\rm d}t}&=&\frac{2a^2}{h}\big({\cal S} \frac{p}{r}+{\cal R} e \sin f\big) \ , \\ \label {ch2-50} 
\frac{{\rm d}e}{{\rm d}t}&=&\frac{1-e^2}{h}\left({\cal R} a \sin f+\frac{\cal S}{er}(ap-r^2)\right), \\ \label {ch2-51} \frac{{\rm d}\omega}{{\rm d}t}&=&-{\cal R} \frac{p}{eh} \cos f+{\cal S} \frac{p+r}{eh} \sin f-{\cal W} \frac{r}{h} \cot i \sin {(\omega+f)} \ , \\ \label {ch2-52}
\sin i\frac{{\rm d}\Omega}{{\rm d}t}&=&{\cal W} \frac{r}{h} \sin{(\omega+f)} \ , \\ \label {ch2-53}
\frac{{\rm d}i}{{\rm d}t}&=&{\cal W} \frac{r}{h} \cos{(\omega+f)} \ .
\end{eqnarray}

Notice that the orbit elements $a$ and $e$ are affected only by components of $\bm A$ in the plane of the orbit, while the elements $\Omega$ and $i$ are affected only by the component out of the plane. The pericenter change has both, but this is because of the combination of intrinsic, in-plane perturbations (the first two terms) with the perturbation of the line of nodes from which $\omega$ is measured (the third term). In fact it is customary to define an angle of pericenter
\be \label{ch2-54}
{\rm d}\varpi\equiv {\rm d}\omega+\cos i {\rm d}\Omega \ ,
\ee
which represent a kind of angle measured from the reference X-direction, rather than from the nodal line. The variation of this angle is given by
 \be \label{ch2-55}
 \frac{{\rm d}\varpi}{{\rm d}t}=-{\cal R} \frac {p}{eh}\cos f+{\cal S} \frac{p+r}{eh} \sin f \ .
 \ee
 
 Although we have discussed this from the point of view of perturbations, Eqs. (\ref {ch2-49})-(\ref {ch2-53}) are {\it exact}; they are merely a reformulation of the three second-order differential equations for ${\bm r}(t)$, Eq.~(\ref {ch2-36}), as a set of six first-order differential equations for the osculating elements (we have not displayed the sixth equation, related to the time orbit element $T$). Given a set of functional forms for $\bm A$ in terms of the orbit elements, an exact solution of these equations is an exact solution of the original equations.

What makes this formulation so useful is that, when $\bm A=0$, the solutions for the orbit elements are constants. If the perturbation represented by $\bm A$ is small in a suitable sense, one expects the changes in the elements to be small. Therefore we can find a first-order perturbation solution by inserting the constant zeroth order values of the elements into the right-hand side, and simply integrating the equations with respect to $t$. In principle, we could go to higher order by inserting this first-order solution back into the right-hand side and integrating again, and so on.

It is sometimes more convenient to integrate the equations with respect to the true anomaly $f$ rather than $t$. To relate the two when dealing with an osculating orbit, we recall that $f=\phi-\omega$, and that $\phi$ is measured from the line of nodes, thus $\phi$ can change both because of the orbital motion, but also by an amount $-\cos i \Delta \Omega$ if $\Omega$ is changing. Hence, since from Eqs.\ (\ref{ch2-37}) - (\ref{ch2-39}) we can write $r^2{\rm d}\phi/{\rm d}t \equiv |{\bm r} \times {\bm v}|=h$, we have
\be \label{ch2-56}
\frac{{\rm d}f}{{\rm d}t}=\frac{h}{r^2}-\left(\frac{{\rm d}\omega}{{\rm d}t}+\cos i \frac{{\rm d}\Omega}{{\rm d}t} \right) \ .
\ee
Of course, if we are integrating the equations only to first order, we can drop the terms involving ${\rm d}\omega/{\rm d}t$ and ${\rm d}\Omega/{\rm d}t$ and use ${\rm d}f/{\rm d}t=h/r^2$.

\section{Testing the No-Hair Theorem Using the Galactic Center Black Hole}
If a class of stars orbits the central black hole in short period ($\sim 0.1$ year), high eccentricity ($\sim 0.9$) orbits, they will experience precessions of their orbital planes induced by both relativistic frame dragging and the quadrupolar gravity of the black hole. Here we are going to apply the orbit perturbation theory that we introduced in the previous sections to study this phenomenon for the galactic center massive black hole. We will see that observation of the precessions of the orbital planes will lead to determination of the spin $\bm J$ and the quadrupole moment $Q_2$ of the black hole. By having $\bm J$ and $Q_2$ we can test the specific relation which the black hole no-hair theorem requires between these parameters and the mass of the black hole i.e. $Q_2=-J^2/m$.

\subsection{Orbit Perturbations in Field of a Rotating Black Hole}

For the purpose of testing the no-hair theorem it suffices to work in the post-Newtonian limit. The post-newtonian limit is the weak-field and slow-motion limit of general relativity in which a quantity of interest is expressed as an expansion in powers of a post-Newtonian parameter $\epsilon \sim v^2 \sim U$ where $U$ is the gravitational potential. The leading term in the expansion is the Newtonian term and it is labeled as $0PN$ term. The term of order $\epsilon$ is the first-post-Newtonian correction, and it is labeled as $1PN$ term and so on.

Consider a two-body system where a body of negligible mass is in the field of a body with mass $m$, angular momentum $\bm J$ and quadrupole moment $Q_2$. The equation of motion of the test body in the first-post-Newtonian limit is given by
\begin{eqnarray} \nonumber
{\bm a} &=& -G\frac{m}{r^2}\hat{\bm n}+ \left(4\frac{Gm}{r}-v^2 \right)\frac{Gm}{r^2}\hat{\bm n}+4\frac{Gm\dot r}{r^2}{\bm v} \\\nonumber
&&-\frac{2GJ}{r^3}\left[2{\bm v}\times{\hat {\bm J}}-3\dot r \hat{\bm n}\times{\hat {\bm J}}-3\hat{\bm n}({\bm h} \cdot \hat{\bm J})/r \right] \\ \label{ch2-57}
&&+\frac{3}{2}\frac{GQ_2}{r^4}\left[ 5\hat{\bm n} (\hat{\bm n}\cdot{\hat {\bm J}})^2-2(\hat{\bm n}\cdot{\hat {\bm J}})\hat{\bm J}-\hat{\bm n}\right],
\end{eqnarray}
where ${\bm r}$ and ${\bm v}$ are the position and velocity of the body, $\hat{\bm n}={\bm r}/r$, $\dot r=\hat{\bm n} \cdot {\bm v}$, ${\bm h}={\bm r} \times {\bm v}$, $\hat {\bm h}={\bm h}/h$, and $\hat {\bm J}={\bm J}/|J|$ (see, e.g. \cite {Will93}). The first line of Eq.~(\ref{ch2-57}) corresponds to the Schwarzschild part of the metric (at post-Newtonian order), the second line is the frame-dragging effect, and the third line is the the effect of the quadrupole moment (formally a Newtonian-order effect). For an axisymmetric black hole, the symmetry axis of its quadrupole moment coincides with its rotation axis, given by the unit vector $\hat {\bm J}$.

\begin{figure}
\centering
\includegraphics[trim = 50mm 50mm 27mm 45mm, clip, width=17cm]{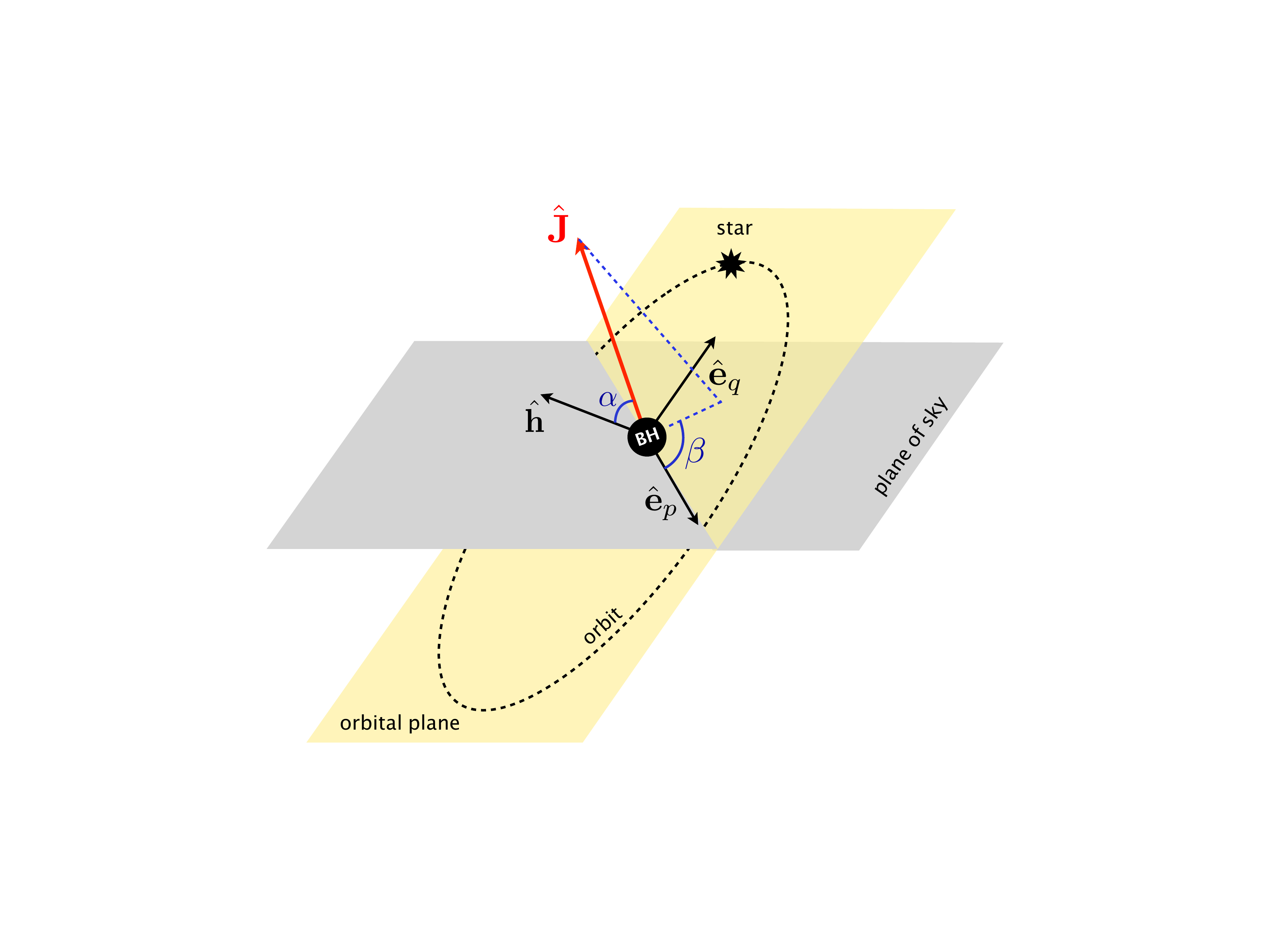}
\caption{The star's orbital plane is defined by the unit vector $\hat{\bm e}_p$ along the line of nodes and the unit vector in the orbital plane $\hat{\bm e}_q$ orthogonal to $\hat{\bm e}_p$ and $\hat{\bm h}$. The polar angels $\alpha$ and $\beta$ define the direction of the black hole's angular momentum $\bm J$ in the $\hat{\bm e}_p$, $\hat{\bm e}_q$, $\hat {\bm h}$ coordinate system.}
\label{SMBHspin}
\end{figure}

As illustrated in Fig.~\ref{SMBHspin}, the star's orbital plane is defined by the unit vector $\hat{\bm e}_p$ along the line of nodes and the unit vector in the orbital plane $\hat{\bm e}_q$ orthogonal to $\hat{\bm e}_p$ and $\hat{\bm h}$ i.e. $\hat{\bm e}_q=\hat {\bm h} \times \hat{\bm e}_p$. With these definitions, then 

\begin{eqnarray} \nonumber
\hat{\bm n}&=&\hat{\bm e}_p \cos(\omega+f)+\hat{\bm e}_q\sin(\omega+f) \ , \\ \label{ch2-58}
\hat{\bm \lambda}&=&-\hat {\bm e}_p \sin(\omega+f)+\hat{\bm e}_q\cos(\omega+f) \ .
\end{eqnarray}
The polar angels $\alpha$ and $\beta$ define the direction of the black hole's angular momentum $\bm J$ in the $\hat{\bm e}_p$, $\hat{\bm e}_q$, $\hat {\bm h}$ coordinate system, so that
\begin{eqnarray} \nonumber
\hat {\bm J} \cdot \hat{\bm e}_p&=&\sin \alpha \cos \beta \ , \\ \nonumber
\hat {\bm J} \cdot \hat{\bm e}_q&=&\sin \alpha \sin \beta \ , \\ \label{ch2-59}
\hat {\bm J} \cdot \hat{\bm h}&=&\cos \alpha \ .
\end{eqnarray}
All the terms in Eq.~(\ref{ch2-57}) except the first term, which is the Newtonian acceleration, are perturbing terms, and by using Eqs.~(\ref{ch2-58}) and  (\ref{ch2-59}) we can find the radial, tangential, and out-of-plane components of the perturbing terms as following
\begin{eqnarray} \nonumber
{\cal R}&=&\frac{Gm}{r^2}\left(\frac{4Gm}{r}-v^2\right)+\frac{4Gm}{r^2}{\dot r}^2+\frac{2GJ h}{r^4}\cos \alpha+\frac{3GQ_2}{2r^4}\left[3\sin^2 \alpha \cos^2{(\beta-\omega-f)}-1 \right]\ , \\ \label{ch2-60} \\
\label{ch2-61}
{\cal S}&=&\frac{4Gm\dot r h}{r^3}-\frac{2GJ\dot r}{r^3} \cos \alpha-\frac{3GQ_2}{2r^4}\sin^2 \alpha \sin{[2(\beta-\omega-f)]} \ , \\ \nonumber
{\cal W}&=& \frac{2GJ}{r^3}\sin \alpha \left[\dot r \sin{(\beta-\omega-f)}+\frac{2h}{r} \cos{(\beta-\omega-f)} \right]-\frac{3GQ_2}{2r^4}\sin{(2\alpha)}\cos{(\beta-\omega-f)} \ . \\  \label{ch2-62}
\end{eqnarray}
By substituting $\cal R$, $\cal S$, and $\cal W$ from Eqs.\ (\ref{ch2-60})-(\ref{ch2-62}) in Eqs.\ (\ref{ch2-49})-(\ref{ch2-53}), we get the rate of change of the each orbit element. To derive the total change of an orbit element over one orbit, we need to integrate over one orbit i.e. integrating over $f$ from $0$ to $2\pi$:
\be \label{ch2-63}
\Delta x=\int_0^{2\pi}{\rm d}f \frac{{\rm d}x}{{\rm d}f}=\int_0^{2\pi} {\rm d}f \frac{{\rm d}t}{{\rm d}f}\frac{{\rm d}x}{{\rm d}t}=\int_0^{2\pi} {\rm d}f \frac{r^2}{h}\frac{{\rm d}x}{{\rm d}t} \ ,
\ee
where $x$ could be any of the orbit elements. We recall the relations $r=p/(1+e\cos f)$, $\dot r=he\sin f/p$, $v^2=(he\sin f/p)^2+(h(1+e\cos f)/p)^2$, and $h=\sqrt{Gmp}$ (see Eqs. (\ref{ch2-37})-(\ref{ch2-41})).

Now to study the precessions of the orbit, we derive the total changes in $i$, $\Omega$, and $\varpi$, which are the three orbit angles defining the orientation of the orbit in space. To first order we get
\begin{eqnarray} \label{ch2-64}
\sin i \Delta \Omega &=&\sin \alpha \sin \beta (A_J-A_{Q_2}\cos \alpha) \ , \\ \label{ch2-65}
\Delta i &=&\sin \alpha \cos \beta (A_J-A_{Q_2}\cos \alpha) \ , \\ \label{ch2-66}
\Delta \varpi &=& A_S-2A_J\cos \alpha-\frac{1}{2}A_{Q_2}(1-3\cos^2 \alpha)\ ,
\end{eqnarray}
where
\begin{eqnarray} \label{ch2-67}
A_S&=&6 \pi \frac{Gm}{(1-e^2)a}\ , \\ \label{ch2-68}
A_J&=&4\pi \chi \left[ \frac{Gm}{(1-e^2)a} \right]^{3/2} \ , \\ \label{ch2-69}
A_{Q_2}&=&3\pi \chi^2 \left[ \frac{Gm}{(1-e^2)a} \right]^2 \ ,
\end{eqnarray}
where $\chi \equiv J/(Gm^2)$ is the dimensionless spin parameter of the black hole which is always less than one for Kerr black hole and $Q_2=-J^2/m$. To get an idea of the astrometric size of these precessions, we define an angular precession rate amplitude $\dot \Theta_i=(a/D)A_i/P$, where $D$ is the distance to the galactic center and $P=2\pi (a^3/Gm)^{1/2}$ is the orbital period. Using $m=4 \times 10^6 \ M_{\odot}$, $D=8 \ {\rm kpc}$, we obtain the rates, in $\mu$arcseconds per year
\begin{eqnarray} \label{ch2-70}
\dot \Theta_S&\approx&92.78P^{-1}(1-e^2)^{-1} \ , \\ \label{ch2-71}
\dot \Theta_J&\approx&0.975\chi P^{-4/3}(1-e^2)^{-3/2} \ , \\ \label{ch2-72}
\dot \Theta_{Q_2}&\approx&1.152 \times 10^{-2} \chi^2 P^{-5/3}(1-e^2)^{-2} \ ,
\end{eqnarray}   
where we have assumed $Q_2=-G^2 m^3 \chi^2$. The observable precessions will be reduced somewhat from these raw rates because the orbit must be projected onto the plane of the sky. For example, the contributions to $\Delta i$ and $\sin i \Delta \Omega$ are reduced by a factor of $\sin i$; for an orbit in the plane of the sky, the plane precessions are unmeasurable.

For the quadrupole precessions to be observable, it is clear that the black hole must have a decent angular momentum ($\chi>0.5$) and that the star must be in a short period high-eccentricity orbit.

\subsection{Testing the No-Hair Theorem}

Although the pericenter advance is the largest relativistic orbital effect, it is {\it not} the most suitable effect for testing the no-hair theorem. The pericenter advance is affected by a number of complicating phenomena including any distribution of mass (such as dark matter or gas) within the orbit. Even if it is spherically symmetric, such a distribution of matter will generally contribute to the pericenter advance because it might induce derivations from the pure Keplerian $1/r$ potential. By contrast, the precessions of the node and inclination are relatively immune from such effects. Any spherically symmetric distribution of mass has no effect on these orbit elements \cite{Will08}.

As a consequence of Eqs.\ (\ref{ch2-64}) and (\ref{ch2-65}) we have the purely geometric relationship,
\be \label{ch2-73}
\frac{\sin i \ {\rm d}\Omega/{\rm d}t}{{\rm d}i/{\rm d}t}=\tan \beta \ ,
\ee
From the measured orbit elements and their drifts for a given star, Eq.~(\ref{ch2-73}) gives the angle $\beta$, independently of any assumption about the no-hair theorem. This measurement then fixes the spin axis of the black hole to lie on a plane perpendicular to the star's orbital plane that makes an angle $\beta$ relative to the line of nodes. The equivalent determination for another stellar orbit fixes another plane; as long as the two planes are not degenerate, their intersection determines the direction of the spin axis, modulo a reflection through the origin.

This information is then sufficient to determine the angles $\alpha$ and $\beta$ for each star. Then, from the magnitude
\be \label{ch2-74}
\left( [ \sin i \frac{{\rm d}\Omega}{{\rm d}t}]^2+[\frac{{\rm d}i}{{\rm d}t}]^2 \right)^{1/2}=\sin \alpha (A_J-A_{Q_2}\cos \alpha) \ ,
\ee
determined for each star, together with the orbit elements, one can solve for $J$ and $Q_2$ to see if the $Q_2=-J^2/m$ relation demanded by the no-hair theorem holds.

So, in principle we see that observations of the precessing orbits of stars very near the massive black hole in the galactic center could provide measurements of the spin and quadrupole moment of the hole and thereby test the no-hair theorems of general relativity. But since the galactic center is likely to be populated by a distribution of stars and small black holes, their gravitational interactions will also perturb the orbit of any given star. In the next sections, we will estimate the effects of such perturbations using analytic orbital perturbation theory to see if the relativistic spin and quadrupole effects of the central massive black hole dominates the effects of stellar cluster perturbation. These estimates will allow us to assess whether the proposed test of the black hole no-hair theorem is going to be feasible.  

\section{Perturbing Effects of a Distribution of Stars in the Surrounding Cluster} 

\subsection{Perturbation by a third body}

In Newtonian theory, the acceleration ${\bm a}_1$ of a target star with mass $m_1$ and the acceleration of the Galactic center black hole with mass $m_2$ in the presence of a perturbing star with mass $m_3$ are given by
\begin{eqnarray} \label{ch2-75}
{\bm a}_1&=&-G\frac{m_2{\bm r}_{12}}{r_{12}^3}-G\frac{m_3{\bm r}_{13}}{r_{13}^3}\ , \\ \label{ch2-76} 
{\bm a}_2&=&-G\frac{m_1{\bm r}_{21}}{r_{21}^3}-G\frac{m_3{\bm r}_{23}}{r_{23}^3} \ ,
\end {eqnarray}
where ${\bm r}_{ab} = {\bm r}_{a}-{\bm r}_{b}$ and $r_{ab} = |{\bm r}_{ab}|$. The equation of motion for the effective two-body problem is
\begin{eqnarray} \nonumber
{\bm a}&\equiv&{\bm a}_1 - {\bm a}_2 \ , \\ \label{ch2-77}
&=&-G\frac{m_2{\bm r}_{12}}{r_{12}^3}-G\frac{m_3{\bm r}_{13}}{r_{13}^3}+G\frac{m_1{\bm r}_{21}}{r_{21}^3}+G\frac{m_3{\bm r}_{23}}{r_{23}^3}\ .
\end {eqnarray}
Since $m_1\ll m_2$, Eq.~(\ref {ch2-77}) is basically the acceleration of the target star. For a perturbing star inside the orbit of the target star (`intenal' star), with $r_{32} \ll r_{12}$, we have
\begin{eqnarray} \nonumber
\frac{1}{r_{13}}=\frac{1}{|{\bm r}_{12}-{\bm r}_{32}|}&=&\frac{1}{r_{12}}-{\bm r}_{32}\cdot \bm \nabla \left( \frac{1}{r_{12}} \right)+\frac{1}{2} \sum_{jk} r_{32}^j r_{32}^k \partial^j \partial^k \left( \frac{1}{r_{12}} \right)- \mathellipsis \\ \label{ch2-78}
&=&\sum_{\ell=0}^{\infty} \frac{(-1)^{\ell}}{\ell!}{r_{32}^L}\partial^{\langle L \rangle} \left( \frac{1}{r_{12}} \right)\ ,
\end{eqnarray}
where the capitalized superscripts denote {\it multi-indices}, so that $r_{32}^L\equiv r_{32}^i r_{32}^j\mathellipsis r_{32}^{k_\ell}$, and similarly for the partial derivatives; $\langle \mathellipsis \rangle$ denotes a symmetric trace-free product (STF). A STF product is symmetric on all indices; furthermore, contracting any pair of indices gives zero. For example applying a gradient successively to $(1/r)$ gives STF products:
\begin{subequations} \label{STF}
\begin {eqnarray} \label{ch2-78A}
\partial_k r^{-1} &=&-n_k r^{-2} \ , \\ \label{ch2-78?B }
\partial_j \partial_k r^{-1}&=&(3n_jn_k-\delta_{jk})r^{-3} \ , \\ \label{ch2-78C}
\partial_i\partial_j \partial_k r^{-1}&=&-[15n_in_jn_k-3(n_i\delta_{jk}+n_j\delta_{ik}+n_k\delta)] \ ,
\end{eqnarray}
\end{subequations} 
where $\partial_k\equiv \partial/\partial x_k$. In Eqs.~(\ref{STF}) the combination of unit vectors in each case is symmetric on all indices, because the partial derivatives commute, and also contracting on any pair of indices automatically gives zero, because, for example for Eq.\ (\ref{ch2-78C}), $\delta^{ij}\partial_{ijk}r^{-1}=\nabla^2 \partial_k r^{-1}=\partial_k \nabla^2 r^{-1}=0$ for $r\neq 0$.

Using Eq.~(\ref{ch2-78}) the $i$-component of $\nabla (1/r_{13})$ is
\begin{eqnarray} \nonumber
\frac{r_{13}^i}{r_{13}^3}&=&-\sum_{\ell=0}^{\infty} \frac{(-1)^{\ell}}{\ell!}{r_{32}^L}\partial^{\langle iL \rangle} \left( \frac{1}{r_{12}} \right), \\ \label{ch2-79}
&=&\frac{r_{12}^i}{r_{12}^3}-\sum_{\ell=1}^{\infty} \frac{(-1)^{\ell}}{\ell!}{r_{32}^L}\partial^{\langle iL \rangle} \left( \frac{1}{r_{12}} \right).
\end{eqnarray}

Substituting Eq.~(\ref {ch2-79}) in Eq.~(\ref{ch2-77}), the equation of motion can be expanded as
\be \label{ch2-80}
a^i=-G\frac{(m_1+m_2+m_3)r^i}{r^3}+G\frac{m_3R^i}{R^3}+Gm_3\sum_{\ell=1}^{\infty}\frac{1}{\ell!} R^L \partial^{\langle iL \rangle} \left (\frac{1}{r} \right ),
\ee
where ${\bm r} \equiv {\bm r}_{12}$ and ${\bm R} \equiv {\bm r}_{23}$.

For a perturbing star outside the orbit of the target star (`external' star), with $r_{12}\ll r_{23}$, the expansion takes the form
\be \label{ch2-81}
a^i=-G\frac{(m_1+m_2)r^i}{r^3}+Gm_3\sum_{\ell=1}^{\infty}\frac{1}{\ell!} r^L \partial^{\langle iL \rangle} \left (\frac{1}{R} \right ).
\ee
Because $m_1\ll m_2$ and $m_3\ll m_2$, and because in what follows we are only concerned with orbital plane effects, we can replace both $m_1+m_2$ and $m_1+m_2+m_3$ with a single $m$, effectively the mass of the massive black hole. 

Establishing a reference XY plane and a reference Z direction, we have the standard ``osculating'' orbital elements including $i$, $\Omega$, $\omega$, $a$, $e$, $f$, and $\varpi$ here. The unit vector $\hat {\bm n}$ pointing from the MBH to the target star, and the orthogonal unit vectors $\hat{\bm \lambda}$ and $\hat {\bm h}$ are given by Eqs.\ (\ref{ch2-42})-(\ref{ch2-43A}) where $\hat {\bm h}$ is normal to the orbital plane. 

We also have the osculating orbit definitions $r \equiv p/(1+e \cos f)$, $h \equiv |{\bf r} \times {\bf v} | \equiv (GMp)^{1/2}$, ${\rm d}\phi/{\rm d}t \equiv h/r^2$, and $p \equiv a(1-e^2)$ for the target star, and $R \equiv p'/(1+e' \cos F)$, $h' \equiv |{\bf R} \times {\bf V} | \equiv (GMp')^{1/2}$, ${\rm d}\phi'/{\rm d}t \equiv h'/R^2$, and $p' \equiv a'(1-{e'}^2)$ for the perturbing star, along with its orbital elements $i'$, $\Omega'$ and $\omega'$.

Here the perturbing acceleration $\bm A$ is everything in Eqs.~(\ref{ch2-80}) and (\ref{ch2-81}) except the leading acceleration $-GM {\bm r}/r^3$. In the internal perturbing star case the first three terms of the expansion are
\begin{eqnarray} \label{ch2-82}
A_{\rm int}^i&=&\frac{Gm_3}{R^2}{\hat{\bm N}}+\underbrace{\frac{3Gm_3}{r^3}\left[R(\hat{\bm n} \cdot \hat{\bm N})n^i-\frac{1}{3}RN^i\right]}_{\ell=1}\\ \label{ch2-83}
&&\underbrace{-\frac{Gm_3}{2r^4}\left[15R^2(\hat{\bm N}\cdot \hat{\bm n})^2n^i-6R^2(\hat{\bm N}\cdot \hat{\bm n})N^i-3R^2n^i\right]}_{\ell=2} \\ \label{ch2-84}
&&\underbrace{+\frac{35Gm_3R^3}{2r^5}\left[ (\hat{\bm N}\cdot \hat{\bm n})^3 n^i-\frac{3}{7} (\hat{\bm N}\cdot \hat{\bm n})^2 N^i-\frac{3}{7}(\hat{\bm N}\cdot \hat{\bm n})n^i+\frac{3}{35}N^i\right]}_{\ell=3},
\end{eqnarray}
where $\hat{\bm n}\equiv{\bm r}/r$ and $\hat{\bm N}\equiv{\bm R}/R$. Similarly, for the external perturbing star we have

\begin{eqnarray} \label{ch2-85}
A_{\rm out}^i&=&\underbrace{\frac{3Gm_3}{R^3}\left[r(\hat{\bm N}\cdot \hat{\bm n})N^i-\frac{1}{3}r n^i \right]}_{\ell=1} \\ \label{ch2-86}
&&\underbrace{-\frac{Gm_3}{2R^4}\left[15r^2(\hat{\bm n}\cdot \hat{\bm N})^2N^i-6r^2(\hat{\bm n}\cdot \hat{\bm N})n^i-3r^2N^i\right]}_{\ell=2} \\ \label{ch2-87}
&&\underbrace{+\frac{35Gm_3r^3}{2R^5}\left[ (\hat{\bm n}\cdot \hat{\bm N})^3 N^i-\frac{3}{7} (\hat{\bm n}\cdot \hat{\bm N})^2 n^i-\frac{3}{7}(\hat{\bm n}\cdot \hat{\bm N})N^i+\frac{3}{35}n^i\right]}_{\ell=3}.
\end{eqnarray}

We use Eqs.\ (\ref{ch2-49})--(\ref{ch2-53}) to calculate the variations with time of the target star's orbit elements, which means we need to derive the components of the perturbing terms along $\hat {\bm n}$, $\hat {\bm \lambda}$, and $\hat{\bm h}$ denoted as $\cal R$, $\cal S$, and $\cal W$ in subsection \ref{Osculating} respectively. We will work in first-order perturbation theory, whereby we express $\cal R$, $\cal S$ and $\cal W$ in terms of osculating orbit variables, set the orbit elements equal to their constant initial values in the right-hand side of Eqs.\ (\ref{ch2-49})--(\ref{ch2-53}), and then integrate with respect to time.

\subsection{Time Averaged Variations in Orbit Elements}

We want to use Eqs.\ (\ref{ch2-49})--(\ref{ch2-53}) to calculate the time averaged rates of change of the orbit elements of the target star, given by $\overline{{\rm d}x/{\rm d}t} \equiv T^{-1} \int_0^T ({\rm d}x/{\rm d}t) {\rm d}t$, where $T$ is the longest relevant timescale, and $x$ is the element in question. 

For an internal perturbing star, $T$ would be the period of the target star $P$.
\be \label{ch2-88}
\overline{\frac{{\rm d}x}{{\rm d}t}} =\frac{1}{P} \int_0^P \frac{{\rm d}x}{{\rm d}t} {\rm d}t=\frac{1}{P} \int_0^{2\pi} \frac{{\rm d}x}{{\rm d}f} {\rm d}f \ ,
\ee
where $f$ is the true anomaly. Assuming that the shorter period $P'$ is much shorter than the longer period $P$, we can split the longer period $P$ to small pieces, each equal to $P'$. Then ${\rm d}x/{\rm d}f$ in Eq. (\ref{ch2-88}) will be the rate of change of $x$ with $f$ while the perturbing star completes one orbit ($\Delta f=2\pi$) and we can write
\begin{eqnarray} \nonumber
\overline{\frac{{\rm d}x}{{\rm d}t}} &=&\frac{1}{P} \int_0^{2\pi} \left( \frac{1}{P'}\int_0^{P'} \frac{{\rm d}x}{{\rm d}f} {\rm d}t'\right) {\rm d}f \ , \\
 \nonumber  \\ \nonumber
&=&\frac{1}{P} \int_0^{2\pi} \left( \frac{1}{P'}\int_0^{2\pi} \frac{{\rm d}x}{{\rm d}f} \frac{r'^2}{h'} {\rm d}F\right) {\rm d}f \ , \\ 
\nonumber \\ \label{ch2-89}
&=&\frac{1}{PP'}\int_0^{2\pi}\int_0^{2\pi}\frac{{\rm d}x}{{\rm d}f} \frac{r'^2}{h'}{\rm d}F {\rm d}f \ ,
\end{eqnarray}  
where ${\rm d}t'$ and $F$ are the time element and the true anomaly of the perturbing star, respectively and ${\rm d}t'=(r'^2/h'){\rm d}F$, valid to first order in perturbation theory. Using the osculating orbit definitions, Eq. (\ref{ch2-89}) can be written as 
\be \label{ch2-90}
\overline{\frac{{\rm d}x}{{\rm d}t}} \equiv \frac{1}{2 \pi P} (1-e'^2)^{\frac{3}{2}}\int_0^{2\pi} \int_0^{2\pi} \frac{{\rm d}x}{{\rm d}f}\frac{1}{(1+e'\cos{F})^2}{\rm d}F {\rm d}f \ .
\ee

For an external perturbing star, $T$ would be the orbital period of the perturbing star $P'$ and by a similar argument, it is straightforward to show that Eq. (\ref{ch2-90}) gives the time-averaged rates of change of the orbital elements of the target star in this case too.

By way of illustration, we show here the time-averaged changes of orbital elements for the $\ell=1$ term induced by an external star (Eq.~(\ref{ch2-82})), for the special case $i'=0$ and $\Omega'=0$:
\begin{eqnarray} \label{a1}
\overline{\frac{{\rm d}a}{{\rm d}t}}&=&0 \ , \\ \label{e1}
\overline{\frac{{\rm d}e}{{\rm d}t}}&=&\frac{15}{4}B_{\rm ext}\frac{e(1-e'^2)^{3/2}}{(1-e^2)^{5/2}}\sin{\omega} \cos{\omega} \sin^2{i} \ , \\ \label{i1}
\overline{\frac{{\rm d}i}{{\rm d}t}}&=& -\frac{15}{4}B_{\rm ext}\frac{(1-e'^2)^{3/2}}{(1-e^2)^{7/2}} e^2 \, \sin{\omega}\cos{\omega}\sin{i}\cos{i} \ , \\ \label{Omega1}
\overline{\frac{{\rm d}\Omega}{{\rm d}t}}&=&- \frac{3}{4}B_{\rm ext}\frac{(1-e'^2)^{3/2}}{(1-e^2)^{7/2}}(1+4e^2-5e^2\cos^2{\omega})\cos{i} \ , \\ \label{varpi1} 
\overline{\frac{{\rm d}\varpi}{{\rm d}t}}&=&\frac{3}{4} B_{\rm ext} \frac{(1-e'^2)^{3/2}}{(1-e^2)^{5/2}} \bigl ( 5\cos^2 \omega - 3 + 5 \cos^2 i \sin^2 \omega -\cos^2 i \bigr ) \ ,
\end{eqnarray}
where $B_{\rm ext}=(2\pi/P)(m_3/m)(p/p')^3$. For arbitrary orientations $i'$ and $\Omega'$ the expressions are much more complicated.  We have also found the analogous expressions for the $\ell = 2$ and $\ell = 3 $ terms. These are smaller than the $\ell = 1$ results by factors of $p/p'$ and $(p/p')^2$, respectively. We used a trick described in Appendix \ref{AppendixA}, which allows us to get analytical forms of the integrations over $f$ and $F$ easily by Maple or Mathematica.

For an internal star, the $\ell=1$ term (Eq.\ (\ref{ch2-85})) contributes no time-averaged variation of any of the elements. The $\ell = 2$ contributions scale as $B_{\rm int} = (2\pi/P)(m_3/m)(p'/p)^2$, while the $\ell =3$ contributions are smaller by a factor of $p'/p$.  Again, the general expressions are long, so we will not display them here. 

Since the orbital energy of the target star is proportional to $1/a$, Eq.  (\ref{a1})  simply reflects the absence of a secular energy exchange mechanism between the target and perturbing stars at first order in the perturbations.  As a side remark, Eqs.\ (\ref{e1}) and (\ref{i1}) together imply that $(1-e^2)^{1/2} \cos{i}$ is a constant, so that a decreasing inclination produces an increasing eccentricity; in planetary dynamics this is known as the Kozai mechanism \cite{Kozai62}. 

\subsection{Average Over Orientations of Perturbing Stellar Orbits}

With the time-averaged changes in the orbital elements due to one perturbing star in hand, we now turn to the changes caused by a distribution of perturbing stars. We will assume a cluster of stars whose orbital orientations ($i'$, $\Omega'$, $\omega'$) are randomly distributed. We will discuss the distributions in $a'$ and $e'$ later. The ``orientation-average'' of a function $F(i',\Omega', \omega')$ will be defined by
\be \label{ch2-91}
\langle F \rangle \equiv \frac{1}{8\pi^2}\int_{0}^\pi  \sin{i'} \,{\rm d}i'\int_{0}^{2\pi} {\rm d}\Omega' \int_{0}^{2\pi} {\rm d}\omega' \, F(i', \Omega', \omega' ) \ .
\ee
We then find that $\langle \overline{{\rm d}x/{\rm d}t} \rangle = 0$ for all four  orbit elements $e$, $i$, $\Omega$ and $\varpi$, for both internal and external stars. The reason is easy to understand: the averaging process is equivalent to smearing the perturbing stars' mass over a concentric set of spherically symmetric shells. The target star will thus be moving in what amounts to a spherically symmetric, $1/r$ potential and its orbit elements will therefore be constant, just as in the pure Kepler problem.  

But for a finite number of stars, the potential will not be perfectly spherically symmetric, even if the orientations are randomly distributed. It is the effect of this discreteness that we wish to estimate. We do this by calculating the root-mean-square (r.m.s.) angular average $[\langle (\overline{{\rm d}x/{\rm d}t})^2 \rangle ]^{1/2}$. This will give an estimate of the ``noise'' induced in the orbital motion of the target star by the surrounding matter. We will then compare this noise with the relativistic effects that we wish to measure.

Here we list the r.m.s. orientation averages for ${\rm d}i/{\rm d}t$ and ${\rm d}\Omega/{\rm d}t$ for internal and external stars, and for all $\ell \le 3$. It turns out that cross terms between different $\ell$ values vanish. We can also see that the contribution of the $(Gm_3/R^2){\hat{\bm N}}$ term in Eq. (\ref{ch2-82}) is zero.

\begin{itemize}
\item
Internal: Lowest order ($\ell =2$)
\begin{eqnarray}
\langle (\overline{\frac{{\rm d}i}{{\rm d}t}})^2 \rangle_{\rm int} &=&
\frac{3}{80}B_{\rm int}^2\frac{1+3e'^2+21e'^4}{(1-e'^2)^4} \ , 
\label{h2int}\\
\langle (\overline{\frac{{\rm d}\Omega}{{\rm d}t}})^2 \rangle_{\rm int} &=&
\frac{3}{80}B_{\rm int}^2\frac{1+3e'^2+21e'^4}{(1-e'^2)^4} \frac{1}{\sin^{2} i} \ ,
\label{om2int}
\end{eqnarray}
\noindent
\item
Internal: First order ($\ell = 3$)
\begin{eqnarray}
\langle (\overline{\frac{{\rm d}i}{{\rm d}t}})^2 \rangle_{\rm int} &=&
\frac{75}{7168} B_{\rm int}^2 \left ( \frac{p'}{p} \right )^2
\frac{e^2 e'^2 (6+9e'^2+34e'^4) }{(1-e'^2)^6}  (5 + 12 \cos^2 \omega )
\,, 
\label{h3int}\\
\langle (\overline{\frac{{\rm d}\Omega}{{\rm d}t}})^2 \rangle_{\rm int} &=&
\frac{75}{7168} B_{\rm int}^2 \left ( \frac{p'}{p} \right )^2
\frac{e^2 e'^2  (6+9e'^2+34e'^4) }{(1-e'^2)^6}   \frac{ (5 + 12 \sin^2 \omega )}{\sin^2 i} \,,
\label{om3int}
\end{eqnarray}
\item
External: Lowest order ($\ell = 1$)
\begin{eqnarray}
\langle (\overline{\frac{{\rm d}i}{{\rm d}t}})^2 \rangle_{\rm ext} &=&
\frac{3}{80} B_{\rm ext}^2  
\frac{(1-e'^2)^3}{(1-e^2)^7} (C_1 + D_1 \cos^2 \omega )
\,, 
\label{h1ext}\\
\langle (\overline{\frac{{\rm d}\Omega}{{\rm d}t}})^2 \rangle_{\rm ext} &=&
\frac{3}{80} B_{\rm ext}^2 
\frac{(1-e'^2)^3}{(1-e^2)^7} \frac{ (C_1 + D_1 \sin^2 \omega )}{\sin^2 i} \,,
\label{om1ext}
\end{eqnarray}
\item
External: First order ($\ell = 2$)
\begin{eqnarray}
\langle (\overline{\frac{{\rm d}i}{{\rm d}t}})^2 \rangle_{\rm ext} &=&
\frac{225}{3584} B_{\rm ext}^2 \left ( \frac{p}{p'} \right )^2 
\frac{e^2 (1-e'^2)^3}{(1-e^2)^9} (C_2 + D_2 \cos^2 \omega )
\,, 
\label{h2ext}\\
\langle (\overline{\frac{{\rm d}\Omega}{{\rm d}t}})^2 \rangle_{\rm ext} &=&
\frac{225}{3584} B_{\rm ext}^2 \left ( \frac{p}{p'} \right )^2 
\frac{e^2 (1-e'^2)^3}{(1-e^2)^9} \frac{ (C_2 + D_2 \sin^2 \omega )}{\sin^2 i} \,,
\label{om2ext}
\end{eqnarray}
\item
External: Second order ($\ell = 3$)
\begin{eqnarray}
\langle (\overline{\frac{{\rm d}i}{{\rm d}t}})^2 \rangle_{\rm ext} &=&
\frac{45}{4096} B_{\rm ext}^2 \left ( \frac{p}{p'} \right )^4 
\frac{(1-e'^2)^3  }{(1-e^2)^{11}}\left (1+3e'^2 +\frac{7}{2}e'^4 \right ) 
\nonumber \\
&& \quad \quad \times
 ( C_3 + D_3 \cos^2 \omega  ) \,, 
\label{h3ext} \\
\langle (\overline{\frac{{\rm d}\Omega}{{\rm d}t}})^2 \rangle_{\rm ext} &=&
\frac{45}{4096} B_{\rm ext}^2 \left ( \frac{p}{p'} \right )^4 
\frac{(1-e'^2)^3 }{(1-e^2)^{11}} \left (1+3e'^2 +\frac{7}{2}e'^4 \right ) 
\sin^{-2} i
\nonumber \\
&& \quad \quad \times
 ( C_3 + D_3 \sin^2 \omega  )
 \,.
 \label{om3ext}
\end{eqnarray}
\end{itemize}
where 
\begin{eqnarray}
C_1 & = (1-e^2)^2\ ,  \quad \quad \quad \quad \quad \quad \quad \;\;\;  &D_1 = 5e^2 (2 + 3e^2) \,,
\nonumber \\
C_2 &= 5(1-e^2)^2  \ ,  \quad \quad \quad \quad \quad \quad  \quad \;\; &D_2  = (4+3e^2)(3+11e^2) \,,
\nonumber \\
C_3 &= (1-e^2)^2 (2+3e^2+44e^4) \ , \quad &D_3 = 21e^2 (2+e^2)(1+5e^2 +8e^4) \,.
\end{eqnarray}

We will focus on the r.m.s change in the direction of $\hat{\bm{h}}$, the normal to the orbital plane which can be expressed in terms of r.m.s changes in $i$ and $\Omega$. Squaring both sides of Eq.~(\ref{ch2-46}) gives
\begin{eqnarray} \nonumber
\left| \dot h \hat{\bm h}+h \frac{{\rm d}\hat{\bm h}}{{\rm d}t} \right|^2&=&\left|-r {\cal W} \hat{\bm \lambda}+r {\cal S} \hat{\bm h}\right|^2, \\  \label{ch2-91A}
{\dot h}^2+2h \dot h \underbrace{ \hat{\bm h} \cdot \frac{{\rm d}\hat{\bm h}}{{\rm d}t}}_0+h^2 \left| \frac{{\rm d} \hat{\bm h}}{{\rm d}t} \right|^2&=&r^2 {\cal W}^2+r^2{\cal S}^2.
\end{eqnarray}
Substituting Eq.~(\ref{ch2-48}), $\dot h=r \cal S$,  in Eq.~(\ref{ch2-91A}), we have 
\be \label{ch2-91B}
\left| \frac{{\rm d} \hat{\bm h}}{{\rm d}t} \right|^2=\left( \frac{r \cal W}{h} \right)^2.
\ee
Then we note that adding the squares of Eqs.~(\ref{ch2-52}) and (\ref{ch2-53}) gives us exactly the right hand side of Eq.~(\ref{ch2-91B}) and therefore we can write
\be \label{ch2-91C}
\left|\frac{{\rm d} \hat{\bm h}}{{\rm d}t}\right|^2=\left( \frac{{\rm d}i}{{\rm d}t} \right)^2+\sin^2 i \left( \frac{{\rm d}\Omega}{{\rm d}t} \right)^2, 
\ee
which leads to 
\be \label{ch2-92}
\langle (\overline {{\rm d}h/{\rm d}t})^2 \rangle \equiv \langle (\overline {{\rm d}i/{\rm d}t})^2 \rangle+\sin^2{i} \langle (\overline {{\rm d}\Omega/{\rm d}t})^2 \rangle \ .
\ee

The leading contributions, corresponding to the $\ell=2$ contribution from internal stars, and to the $\ell=1$ contribution from external stars are given by 
\begin{eqnarray} \label{ch2-93}
\langle (\overline {{\rm d}h/{\rm d}t})^2 \rangle_{\rm int} &=& \frac{3}{40}B_{\rm int}^2\frac{1+3e'^2+21e'^4}{(1-e'^2)^4} \ , \\ 
\nonumber
\\ \label{ch2-94}
\langle (\overline {{\rm d}h/{\rm d}t})^2 \rangle_{\rm ext} &=& \frac{3}{40}B_{\rm ext}^2\frac{(1-e'^2)^3}{(1-e^2)^7} \left (1+3e^2+ \frac{17}{2} e^4 \right ) \ .
\end{eqnarray}  

Note although the perturbing terms are due to randomly distributed stars, the r.m.s changes of the individual elements $i$ and $\Omega$ depend on $\omega$ and $i$ (see Eqs.~(\ref{h2int})-(\ref{om3ext})). The reason is that $i$ and $\Omega$ depend on the choice of reference plane, and $\omega$ is measured from the line of nodes and $i$ is the inclination angle between the reference plane and the orbital plane. So they also depend on the reference plane and if we chose a different reference plane, $\omega$ and $i$ would be different, and so we might expect $\langle (\overline {{\rm d}\Omega/{\rm d}dt})^2 \rangle$ and $\langle (\overline {{\rm d}i/{\rm d}t})^2 \rangle$ to depend on the orientation of the orbital ellipse relative to the nodal line.

On the other hand, from Eqs. (\ref {ch2-93}) and (\ref {ch2-94}) we can see that $\langle (\overline {{\rm d}h/{\rm d}t})^2 \rangle$ is independent of $\omega$. It is because $\hat {\bm h}$ is a vector in space, it knows nothing about the arbitrary choice of reference plane, and hence its variation can't depend on $\omega$. For future use, we define the angular r.m.s. rate of change of the orbital orientation by ${\rm d}\theta/{\rm d}t \equiv \langle (\overline {{\rm d}h/{\rm d}t})^2 \rangle^{1/2}$.

\subsection{Average Over Size and Shape of Perturbing Stellar Orbits}

We now integrate over the semi-major axes $a'$ and eccentricities $e'$ of the perturbing stars. We will use a distribution function of the form
${\cal N} g(a') h(e'^2) {\rm d}a' {\rm d}e'^2$, where $\cal N$ is a normalization factor, set by the condition ${\cal N} = N/{\cal I}$, where $N$ is the total number of stars in the distribution, and
\be \label{ch2-95}
{\cal I} =  \int h(e'^2) {\rm d}e'^2 \int  g(a') {\rm d}a' \ ,
\ee
where the limits of integration will be determined by the limiting orbital elements for those stars. Since at the end we are going to compare our results with N-body simulations by Merritt {\it et al} (\cite{MAMW09}, hereafter referred to as MAMW), we will consider the same range of parametrized models for the dependences $g(a')$ and $h(e'^2)$ as was used in their simulations, and will consider clusters that contain both stars and stellar-mass black holes.

The variables $a'$ and $e'$ will be constrained by a number of considerations. The minimum pericenter distance $r_{\rm min}$ for any body will be given by the tidal-disruption radius for a star, and the capture radius for a black hole. This will therefore give the bound
\be \label{ch2-96}
a' (1-e') > r_{\rm min} \ .
\ee
For $r_{\rm min}$ we will use the estimates
\begin{eqnarray} \nonumber
r_{\rm min}^{\rm star} &\approx& 4 \times 10^{-3} \, \left ( \frac{m_{\rm star}}{m_\odot} \right )^{0.47} \left (\frac{m}{4 \times 10^6 M_\odot} \right )^{1/3} \, {\rm mpc} \ , \\ \label{ch2-97}
r_{\rm min}^{\rm bh} &\approx& 8Gm \approx 1.5 \times 10^{-3} \left (\frac{m}{4 \times 10^6 M_\odot} \right ) \, {\rm mpc} \ .
\end{eqnarray}
These are derived in Appendix \ref{AppendixB}.

However our analytic formulae for the r.m.s. orientation-averaged variations are valid only in the limits $p'/p \ll 1$ or $p/p' \ll 1$ for internal and external stars, respectively.  But since our target star is embedded inside the cluster of stars, there may well be perturbing stars that do not satisfy either constraint. On the other hand, an encounter between the target star and another star that is too close could perturb the orbit so strongly that it will be unsuitable for any kind of relativity test. Because we are looking only for an estimate of the statistical noise induced by the cloud of stars, we will try three approaches in order to capture the range of perturbations induced by the cluster.  

{\bf {\em Integration I.}}
Because Eqs.\ (\ref{ch2-93}) and (\ref{ch2-94}) are valid only in the extreme limits where the perturbing star is always far from the target star (so that the higher-order terms are suitably small), we cut out of the stellar distribution any stars that violate this constraint.  This yields the following conditions  on the allowed orbital elements of the perturbing stars:
(i) for an internal star, we demand that $r'_{\rm max} = a' (1+e')$ of the perturbing star be less than $r_{\rm min} = a(1-e)$ of the target star;  (ii) for an external star, we demand that $r'_{\rm min} = a' (1-e')$ of the perturbing star be greater than $r_{\rm max} = a(1+e)$ of the target star. 

For an internal star, we thus have the two conditions,
\be \label{ch2-98}
a' (1-e') > r_{\rm min} \ ,
\qquad
a' (1+e') < a(1-e) \ .
\ee
The maximum values of $e'$ and $a'$ allowed under these conditions are
\be \label{ch2-99}
e'_{\rm max,int} = \frac{a(1-e)-r_{\rm min}}{a(1-e)+r_{\rm min}} \ ,
\qquad
a'_{\rm max,int} = a\frac{1-e}{1+e'} \ .
\ee
For an external star, we have the two conditions
\be \label{ch2-100}
a' (1-e') > a(1+e)\ ,
\qquad
    a' < a_{\rm max} \ ,
\ee
where $a_{\rm max}$ is the outer boundary of the cluster, chosen to be large enough that the effects of stars beyond this boundary are assumed to be negligible.  Following MAMW, we choose $a_{\rm max} = 4$ mpc.
The maximum $e'$ and minimum $a'$ allowed are thus 
\be \label{ch2-101}
e'_{\rm max,ext} = 1 - \frac{a(1+e)}{a_{\rm max}} \ ,
\qquad
a'_{\rm min,ext} = a \frac{1+e}{1-e'} \ .
\ee
Thus the average of a function ${\cal F}(a', e')$ over this distribution will be given by
\be  \label{ch2-102}
\langle {\cal F} \rangle \equiv {\cal N} (J_1 + J_2) \ ,
\ee
where
\begin{eqnarray}  \nonumber
J_1 ({\cal F}) &=&  \int_0^{{e'}^2_{\rm max,int}} h(e'^2) {\rm d}e'^2 \int_{r_{\rm min}/(1-e')}^{a'_{\rm max,int}}  g(a') {\cal F} (a',e') {\rm d}a' \ ,
 \\  \label{ch2-103}
J_2 ({\cal F}) &=&  \int_0^{{e'}^2_{\rm max,ext}} h(e'^2) {\rm d}e'^2 \int_{a'_{\rm min,ext}}^{a_{\rm max}}  g(a') {\cal F} (a',e'){\rm d}a' \ .
\end{eqnarray}
However, instead of substituting ${\cal N} = N/{\cal I}$, we substitute
\be  \label{ch2-104}
{\cal N} = N/ ({\cal I}_1 + {\cal I}_2) \ ,
\ee
where 
\begin{eqnarray} \nonumber
{\cal I}_1 &=&  \int_0^{{e'}^2_{\rm max,int}} h(e'^2) {\rm d}e'^2 \int_{r_{\rm min}/(1-e')}^{a'_{\rm max,int}}  g(a')  {\rm d}a' \ ,
 \\  \label{ch2-105}
{\cal I}_2 &=&  \int_0^{{e'}^2_{\rm max,ext}} h(e'^2) {\rm d}e'^2 \int_{a'_{\rm min,ext}}^{a_{\rm max}}  g(a')  {\rm d}a' \ .
\end{eqnarray}
This amounts to assuming that all $N$ stars in the cluster happen to have orbit elements that satisfy our constraint. Thus the average of the function ${\cal F}(a', e')$ will be given by
\be  \label{ch2-106}
\langle  {\cal F} \rangle = N \frac{J_1({\cal F}) + J_2 ({\cal F})}{{\cal I}_1 + {\cal I}_2} \ .
\ee
Note that if ${\cal F} = 1$, we get  $\langle { \cal F} \rangle = N$.   

In our simple model, we are treating the stars and black holes as independent distributions, so the mean value of ${\cal F}$ can be written as a sum over the two normalized distributions,
\be \label{ch2-107}
\langle {\cal F} \rangle = \langle {\cal F} \rangle_S + \langle {\cal F}\rangle_B \ ,
\ee
where the only differences between the integrals for the distributions are the perturbing object mass $m_3$, the value of $r_{\rm min}$, which affects only the integrals ${J}_1$ and ${\cal I}_1$, and the number of particles, $N_S$ for stars, and $N_B$ for black holes, with $N = N_B + N_S$; for later use, we define $N_B/N_S \equiv R$. Hence we obtain
\be \label{ch2-108}
\langle {\cal F} \rangle = \frac{N_S}{N} \frac{J_{1S}({\cal F}_S) + J_{2} ({\cal F}_S)}{{\cal I}_{1S} + {\cal I}_{2}} + \frac{N_B}{N} \frac{J_{1B}({\cal F}_B) + J_{2} ({\cal F}_B)}{{\cal I}_{1B} + {\cal I}_{2}}  \ .
\ee
For the r.m.s. variations in ${\rm d}h/{\rm d}t$, we include all the higher-order terms shown in Eqs. (\ref{h2int})-(\ref{om3ext}).
 
{\bf {\em Integration II.}}  Taking the ratio of the higher $\ell$ contributions to the orbit element variations  to the leading $\ell$ contribution (see Eqs. (\ref{h2int})-(\ref{om3ext})) reveals that the parameter controlling the relative size of the higher-order terms is the ratio $a'/[a(1-e^2)]$ for internal stars, and $a/[a' (1-e'^2)]$ for external stars.   Requiring each of these ratios in turn to be less than one, we repeat the integrals, but with new limits of integration given by
\begin{eqnarray} \nonumber
e'_{\rm max,int} &=& 1-r_{\rm min}/a(1-e^2) \ ,
\qquad a'_{\rm max,int} = a(1-e^2) \ , \\  \label{ch2-109}
e'_{\rm max,ext} &=& (1 - a/a_{\rm max} )^{1/2}\ ,
\qquad a'_{\rm min,ext} = a/(1-e'^2) \ .
\end{eqnarray}
This condition permits closer encounters than the condition imposed in {\em Integration I}.  Here as well, we include all higher-order contributions to the r.m.s. variations.

{\bf {\em Integration III}.}  In an attempt to include even closer encounters between the target star and cluster stars, we adopt a fitting formula for the r.m.s. perturbations of the orbital plane that interpolates between the two limits of very distant internal and very distant external stars. A simple formula that achieves this is given by
\be \label{ch2-110}
{\dot h}^2_{\rm fit}  = \frac{1}{\langle (\overline {{\rm d}h/{\rm d}t})^2 \rangle_{\rm int}^{-1}  + \langle (\overline {{\rm d}h/{\rm d}t})^2 \rangle_{\rm ext}^{-1} } \ ,
\ee
where we use only the lowest-order contributions to the r.m.s. variations, given by Eqs.\ (\ref{ch2-93}) and (\ref{ch2-94}). In this case the average over the distributions becomes
\be \label{ch2-111}
\langle {\cal F} \rangle = \frac{N_S}{N} \frac{J_{S}({\cal F}_S) }{{\cal I}_{S} } + \frac{N_B}{N} \frac{J_{B}({\cal F}_B)}{{\cal I}_{B} }  \ ,
\ee
where the integrals now take the form
\be \label{ch2-112}
J({\cal F}) =  \int_0^{(1-r_{\rm min}/a)^2} h(e'^2) {\rm d}e'^2 \int_{r_{\rm min}/(1-e')}^{a_{\rm max}}  g(a') {\cal F} (a',e') {\rm d}a',
\ee
with ${\cal I} = {J}(1)$, thereby including the full distribution of stars.

\subsection{Numerical Results}

In order to compare our analytic estimates with the results of the N-body simulations of MAMW, we will adopt as far as possible the same model assumptions.  We parametrize the distribution functions $g(a')$ and $h(e'^2)$ according to $g(a')=a'^{2-\gamma}$, and $h(e'^2) = (1-e'^2)^{-\beta}$, where $\gamma$ ranges from 0 to 2, and $\beta$ ranges from -1 to 0.5.  The values $(\gamma, \, \beta) = (2,0)$ correspond to a mass segregated distribution with isotropic velocity dispersion.  We will choose $a'_{\rm max} = 4$ mpc, arguing that the perturbing effect of the cluster outside this radius is negligible by virtue of the increasing distance from the target star and the more effective ``spherical symmetry'' of the mass distribution.  We will assume that the cluster contains stars each of mass $1M_\odot$ and black holes each of mass $10M_\odot$, and will consider values of the ratio of the number of black holes to the number of stars to be $R=0$ and $R=1$ (MAMW also consider the ratio $R=0.1$). The main difference between stars and black holes in our integrals is the factor $m_3^2$, so there will simply be a relative factor of 100 between the black hole contribution and the stellar contribution, apart from the small effect of the difference in $r_{\rm min}$ between stars and black holes.    

\begin{table}[t]
\begin{center}
\begin{tabular}{|crrcrrr|crrcrr|}
\hline
Model&$\gamma$ & $\beta$ & $M_\star (M_\odot)$ & $\cal R$ & $N$ 
&&Model&$\gamma$ & $\beta$ & $M_\star (M_\odot)$ & $\cal R$ & $N$ \\
\hline

1&0 & -1  & 10 & 0   & 159  &&9& 2 & 0   & 10  & 1    & 7    \\
2&0 & -1  & 10 & 1    & 29    &&10& 2 & 0   & 30  & 0    & 119   \\

3&1 & -1  & 10 & 0    & 119  &&11& 2 & 0   & 30  & 1    & 21    \\
4&1 & -1  & 10 & 1  & 21  &&12& 2 & 0   & 100 & 0   & 400   \\

5&1 & 0 & 30 & 0 & 209 &&13&2 & 0   & 100 & 1    & 72  \\
6& 1 & 0 & 30 & 1 & 43 &&14&2 & 0.5 & 100 & 0   & 400\\

7&2 & -1  & 30 & 0    & 119   &&15&2 & 0.5 & 100 & 1   & 72    \\
8&2 & -1  & 30 & 1    & 21 && & & & & & \\
\hline
\end{tabular}
\caption{ Parameters of the distributions}
 \label{tbl-1}
\end{center}
\end{table}

Of the 22 stellar distribution models listed in Table I of MAMW, we consider only the 15 models with either $R=0$ or $R=1$; these are listed in Table~\ref{tbl-1}.  While $N$ denotes the total number of objects within $4$ mpc, the parameter $M_\star$, chosen to parallel the notation of MAMW, denotes the approximate total mass within one mpc of the black hole, and gives an idea of the perturbing environment around a close-in target star.  

\begin{figure}
\includegraphics[width=6in]{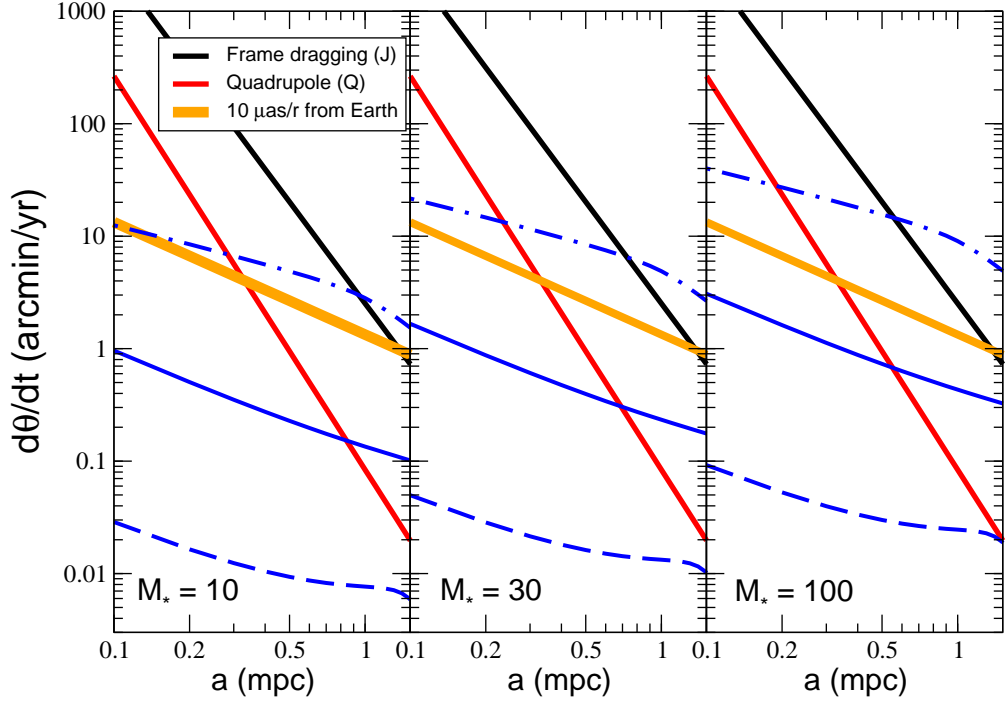}
\caption{R.m.s.\ precession ${\rm d}\theta/{\rm d}t =  (\langle \dot{i}^2 \rangle + \sin^2 i \langle \dot{\Omega}^2 \rangle)^{1/2}$ for a target star with $e=0.95$ plotted against semi-major axis, for three models with $\gamma =2$, $\beta = 0$, $R = 1$.  $M_\star$ denotes the total mass within one mpc, in solar mass units. Shown (blue in color version) are results from {\it Integration I} (dashed curves), {\it Integration II} (solid curves) and {\it Integration III} (dot-dash curves).  Also shown are the amplitudes of frame-dragging (black in color version) and quadrupole (red in color version) relativistic precessions for the corresponding star, assuming a maximally rotating black hole.  Wide line (orange in color version) denotes the precession corresponding to an observed astrometric displacement of 10 $\mu$arcsec/yr. }
\label{models}
\end{figure}

Figure~\ref{models} shows the results for the three stellar distribution models 9, 11 and 12 in Table~\ref{tbl-1}; In these models $\gamma =2$ and $\beta = 0$ and they have an equal number of $1M_\odot$ stars and $10M_\odot$ black holes. The three cases correspond to a total number of perturbing bodies within a radius of four mpc of 7, 21 and 72, respectively. The target star has eccentricity $e = 0.95$, and its semi-major axis $a$ ranges from $0.1$ to $2$ mpc. Plotted is the rate of precession of the vector perpendicular to the orbital plane, ${\rm d}\theta/{\rm d}t \equiv \langle (\overline {{\rm d}h/{\rm d}t})^2 \rangle^{1/2}$, observed at the source, in arcminutes per year, calculated using three ways of carrying out the integrals over the stellar distribution. The dashed line denotes {\it Integration I}, in which all perturbing stars are assumed to be sufficiently far from the target star at all times that their pericenters are outside its apocenter or that their apocenters are inside its pericenter. The solid line denotes {\it Integration II}, in which closer encounters are permitted, limited by demanding that all perturbing stars be on orbits such that the higher $\ell$ contributions to ${\rm d}\theta/{\rm d}t$ be at worst comparable to the contribution at lowest order in $\ell$. The dot-dashed line denotes {\it Integration III}, which uses a fitting formula that interpolates between the extreme limits of a perturbing star well outside the target star, and a perturbing star well inside the target star; in this case the integration is over the entire stellar distribution. The orange band in each panel denotes the value of ${\rm d}\theta/{\rm d}t$ corresponding to an astrometric precession rate ${\rm d}\Theta/{\rm d}t$ of $10 \, \mu$arcsecond per year as seen from Earth, given by 
\be \label{ch2-113}
\frac{({\rm d}\theta/{\rm d}t)_{\rm source}}{({\rm arcmin/yr})}
\approx  \frac{1.3}{\tilde a} \frac{({\rm d}\Theta/{\rm d}t)_{\rm Earth}}{(10 \, \mu {\rm as/yr})} \ ,
\ee
where ${\tilde a}$ is the semi-major axis in units of mpc; we use 8 kiloparsecs as the distance to the galactic center.

Also plotted are the rates of precessions due to the frame-dragging and quadrupolar effects of a Kerr black hole, given by \cite{Will08}
\begin{eqnarray} \nonumber
{\dot A}_J&\equiv&\frac{A_J}{P} = \frac{4\pi}{P} \chi\left[\frac{Gm}{a(1-e^2)}\right]^{3/2} \\ \label{eq:defaj}
& \approx& 0.0768 (1-e^2)^{-3/2}\chi {\tilde a}^{-3} 
{\rm arcmin/yr},
\\ \nonumber
{\dot A}_{Q_2} &\equiv&\frac{A_{Q_2}}{P}= \frac{3\pi}{P} \chi^2 \left[ \frac{Gm}{a(1-e^2)}\right]^{2} 
\\ \label{eq:defaq}
& \approx& 7.97 \times 10^{-4} (1-e^2)^{-2}\chi^2 {\tilde a}^{-7/2}
{\rm arcmin/yr}, 
\end{eqnarray} 
where $P=2\pi (a^3/Gm)^{1/2}$ is the orbital period, $A_J$ and $A_{Q_2}$ are the amplitude of precessions given in Eqs. (\ref{ch2-68}) and (\ref{ch2-69}), and where $\chi = J/Gm^2$ is the dimensionless Kerr spin parameter, set equal to its maximum value of unity in Fig.~\ref{models}.

Because {\it Integration I} keeps the stars far from the target star, the precessions are small. By contrast, the fitting formula of {\it Integration III} is large for very close encounters, so not surprisingly, the precessions from that method are large. {\it Integration II} gives results intermediate between the two.  Interestingly, the spread between these methods is roughly consistent with the spread between individual precessions obtained in the $N$-body simulations of MAMW. This can be seen in the top panel of MAMW, Fig.~7, which corresponds to the middle panel of Fig.~\ref{models} (to properly compare the two figures, one must translate between ${\rm d}\theta/{\rm d}t$ and ${\rm d}\Theta/{\rm d}t$). It can also been in the bottom panel of MAMW Fig.~5, where the points labelled by $\times$ indicate the mean precessions in the absence of black hole spin, for the same three stellar distributions as are shown in Fig.~\ref{models}. Thus we regard our three integration methods as giving a reasonable estimate of the range of stellar perturbations.

Comparing the three stellar distributions shown in Fig.~\ref{models}, we see that the effects vary roughly as $N^{1/2} \propto M_\star^{1/2}$, as expected, from the nature of our r.m.s. calculation. 

\begin{figure}
\centering
\includegraphics[width=6in]{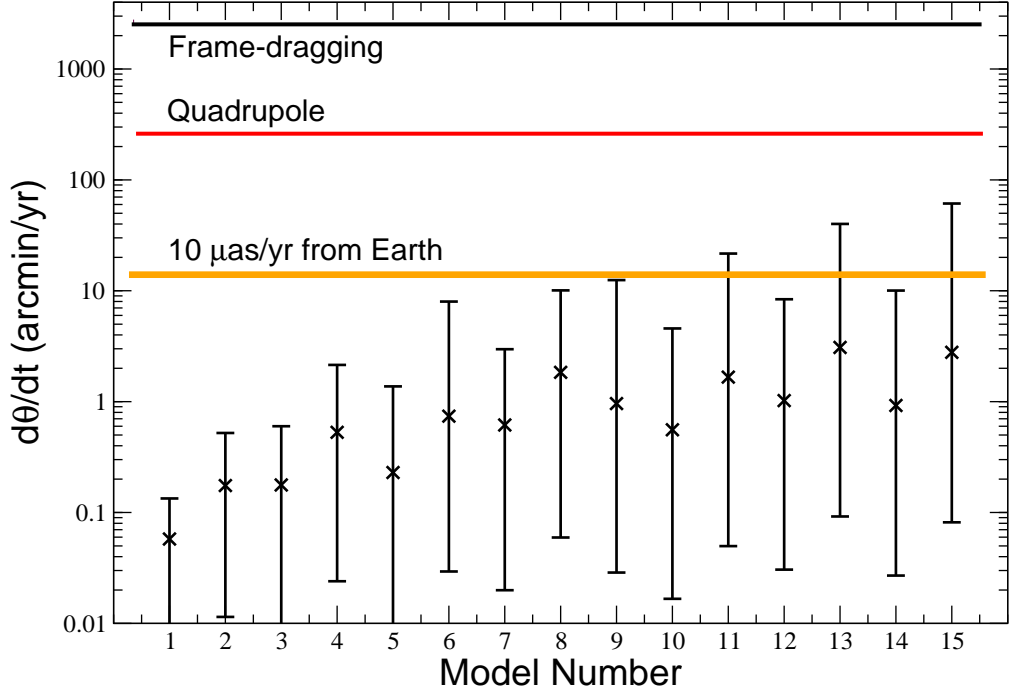}
\caption{R.m.s.\ precession ${\rm d}\theta/{\rm d}t =  (\langle \dot{h}^2 \rangle)^{1/2}$  for target star with $e=0.95$ and $a = 0.1$ mpc  for 15 stellar distribution models.  Symbol $\times$ denotes estimates from {\it Integration II}, and error bars indicate the range of estimates from {\it Integrations I} and {\it III}.  Rates of precessions due to the frame-dragging and quadrupolar effects and astrometric displacement of 10 $\mu$arcsec/yr are shown as in Fig.~\ref{models}.}
\label{errorbars}
\end{figure}

We consider eight different stellar distribution models, and for seven of them, consider models with equal numbers of stars and black holes, and models with only stars, totaling 15 models.  In all but one case, the precessions are generally smaller than the ones shown in Fig.~\ref{models}, and that case is a centrally condensed model with a non-isotropic velocity dispersion leading to a preponderance of highly eccentric orbits. We conclude that, for a target star in a very eccentric orbit with $a < 0.2$ mpc, there is a reasonable possibility of seeing relativistic frame-dragging and quadrupole effects above the level of $10 \, \mu$arcsec/yr without undue interference from stellar perturbations. We also show in Appendix \ref{AppendixC} that the effects of tidal deformations on the orbital planes of stellar orbits are negligible.

To illustrate the differences between different models of the stellar distribution, Fig.~\ref{errorbars} shows the predicted precessions for a target star at $0.1$ mpc with $e = 0.95$, for all 15 model distributions.  The crosses and the error bars indicate the range of results from the three integration models.   Models with $\gamma = 0$ or  $1$ generally give smaller precessions than those with $\gamma =2$. The latter models are more centrally condensed, and lead to larger perturbations of a close-in target star.  For the same value of $(\gamma,\, \beta, \, M_\star)$, models with equal numbers of stars and black holes $(R=1)$ lead to larger perturbations than those with pure stars ($R=0$); the former models are more ``grainy'' (smaller $N$), and so the effects are larger by roughly $N_{R=0}^{1/2}/N_{R=1}^{1/2}$. Models 14 and 15 $(\beta = 0.5)$ have an excess of stars in highly eccentric orbits, thus leading to larger precessions.

\subsection{Conclusions}

We have used analytic orbital perturbation theory to investigate the rate of precession of the orbital plane of a target star orbiting the galactic center black hole Sgr A$^\star$ induced by perturbations due to other stars in the central cluster. We found that, although the results have a wide spread, they compare well with the distribution of precessions obtained using $N$-body simulations. One feature not included in our analysis is the fact that orbital planes in a real cluster are not randomly distributed, but become somewhat correlated over the long-term evolution of the cluster. Whether these correlations are large enough to have a significant effect on our estimates is an open question. Within our assumptions, however, we find a range of possible models for the cluster of objects within the central 4 mpc of the black hole in which it may still be possible to detect relativistic precessions of the orbital planes at the $10 \, \mu$arcsec/yr level.

\clearpage 


\chapter{Dark Matter Distributions Around Massive Black Holes: A General Relativistic Analysis} 
\label{chapter3} 
\thispagestyle{myplain}
\lhead{Chapter 3. \emph{Dark Matter Distributions Around MBHs: A General Relativistic Analysis}} 

In this chapter we start with reviewing the non-relativistic phase-space formulation to study the effects of the adiabatic i.e. slow growth of the massive black hole on the dark-matter density profile. Then we develop a fully general relativistic phase-space formulation to consider these effects and we find the dark matter distribution in vicinity of the Galactic center supermassive black hole Sgr A$^\star$ which has significant differences with the non-relativistic results. Having the dark matter profile density in the presence of the massive black hole, we calculate its perturbing effect on the orbital motions of stars in the Galactic center, and find that for the stars of interest,  relativistic effects related to the hair on the black hole will dominate the effects of dark matter.

\section{Growing a Black Hole in a Dark Matter Cluster: Newtonian Analysis}

In this section we begin with a purely Newtonian analysis of the process of growing a black hole slowly within a pre-existing DM halo. This is an example of a process in which a system responds adiabatically to a slowly varying potential. In such a situation, the use of action-angle variables enables us to predict how a distribution of particles will respond to changes in the gravitational field that confine it.

As discussed below, when the process of growing a black hole within a pre-existing DM halo is adiabatic, the gravitational potential changes slowly enough so that the constants of the motion of the DM particles vary smoothly while keeping the action variables invariant. A brief consideration of the physical conditions close to the GC will convince us that the requirements for adiabatic evolution are likely to be met.

The central MBH will dynamically dominate a region of radius $r_h = G m_{\rm BH}/\sigma^2$, where $m_{\rm BH}$ is the mass of the black hole and $\sigma$ is the velocity dispersion of the DM particles outside the radius of influence. The dynamical timescale inside $r_h$ can be estimated as $t_{\rm dyn} = r_h/\sigma$, which for the Milky Way turns out to be about $10^4$ yr, taking $m_{\rm BH} \sim 4 \times 10^6 M_\odot$ and estimating from the velocity dispersion of the stars $\sigma \approx 66$ km/s. On the other hand we can estimate the shortest timescale for growth of the black hole as the Salpeter timescale $t_S = m_{\rm BH}/\dot{m}_{\rm Edd} \approx 5 \times 10^7 \ yr$, where $\dot{m}_{\rm Edd}$ is the usual Eddington accretion timescale. Hence, the dynamical timescale inside $r_h$ is much shorter than the typical timescale for black hole growth. In addition, since the DM is assumed to be collisionless, the relaxation timescale will always be longer than the evolutionary timescale (This is not necessarily the case for the stellar population close to the central cusp) \cite{Sigurdsson2004, BT08}.

We generally follow the approach used by Binney and Tremaine \cite{BT08} and Quinlan {\it et al.} \cite{Quinlan95}. In addition to reproducing the non-relativist results in \cite{Quinlan95}, which extended the study of the isothermal sphere carried out in \cite{Young80}, this will set the stage for our fully general relativistic analysis. We will use $c=1$ throughout this chapter.

\subsection{Basic Equations}
Given a distribution function $f(E,L)$, which is normalized to give the total mass $M$ of the halo upon integration over phase-space, the physical mass density is given by:
\be \label{ch3-1}
\rho(r) =\int f(E,L){\rm d}^3 {\bm v} \ ,
\ee
where the energy and angular momentum per unit mass $E$ and $L\equiv |{\bm L}|$ are functions of velocity and position, defined by
\begin{eqnarray} \nonumber
{\bm L}&=&{\bm x}\times{\bm v} \ , \\ \label{ch3-2}
E&=&\frac{v^2}{2}+\Phi(r) \ ,
\end{eqnarray}
where $\Phi(r)$ is the Newtonian gravitational potential. We now change integration variables from $\bm v$ to $E$, $L$, and the z-component of angular-momentum $L_z$, using the relation
\be \label{ch3-3}
{\rm d}^3v=J^{-1}{\rm d}E\ {\rm d}L\ {\rm d}L_z \ ,
\ee
where the Jacobian is given by the determinant of the matrix
\begin{eqnarray} \nonumber
J\equiv 
\left| \frac{\partial(E,\ L,\ L_z)}{\partial(v^x,\ v^y,\ v^z)} \right|
&=&\frac{r}{L}\ {\begin{vmatrix}
v^x& v^y& v^z \\
(rv^x-x\dot r)& (rv^y-y\dot r)& (rv^z-z\dot r)\\
-y& x& 0
\end{vmatrix}} \ , \\ \label{ch3-4}
&=&\frac{r^2\dot r}{L}(z\dot r-v^zr) \ ,
\end{eqnarray}
where $\dot r=v_r={\bm r}\cdot {\bm v} /r$. For the $\partial L/\partial v^i$ components, we used the relation $L^2=r^2(v^2-\dot r^2)$; e.g. for $v^x$ we have
\begin{eqnarray} \label{ch3-5}
L^2&=&r^2(v^2-\dot r^2) \ , \\ \nonumber
\Rightarrow \;\;\; 2L\frac{\partial L}{\partial v^x}&=&r^2(2v^x-2\dot r \frac{\partial \dot r}{\partial v^x}) \ , \\ \nonumber
&=&r^2(2v^x-2\dot r \frac{x}{r}) \ , \\ \label{ch3-6}
 \Rightarrow \;\;\;  \frac{\partial L}{\partial v^x}&=&\frac{r}{L}(r v^x-\dot r x) \ .
\end{eqnarray} 

To express the Jacobian in terms of the components of $\bm v$ we use
\begin{eqnarray} \nonumber
v^\theta=\frac{1}{r^2} {\bm v}\cdot{\hat {\bm e}_\theta}&=&\frac{1}{r^2}\left( v^i \frac{\partial x^i}{\partial \theta} \right) \ , \\ \nonumber
&=&\frac{1}{r^2}\left(v^x \frac{\partial x}{\partial \theta}+v^y \frac{\partial y}{\partial \theta}+v^z \frac{\partial z}{\partial \theta} \right) \ , \\ \label{ch3-7}
&=&\frac{1}{r^2}\left( v^xr\cos \theta \cos \phi+v^yr\cos \theta \sin \phi-v^zr\sin \theta \right) \ ,
\end{eqnarray}
also we have
\begin{eqnarray} \nonumber
\dot r=\frac{{\bm r}\cdot{\bm v}}{r}&=&\frac{1}{r}(xv^x+yv^y+zv^z) \ , \\ \nonumber
&=&\sin \theta(v^x\cos \phi+v^y \sin \phi)+v^z\cos \theta \ , \\ \label{ch3-8}
\Rightarrow \;\;\; v^x \cos \phi+v^y \sin \phi &=& \frac{1}{\sin \theta}\left(\dot r \frac{v^z z}{r} \right) \ ,
\end{eqnarray}
therefore
\begin{eqnarray} \nonumber
v^\theta&=&\frac{1}{r^2}\left[\frac{z}{\sin \theta}\left( \dot r -\frac{v^z z}{r} \right)-v^zr\sin \theta \right] \ , \\ \nonumber
&=&\frac{1}{r^2}\frac{z\dot r r-v^z z^2-v^z r^2 \overbrace{\sin^2 \theta}^{1-z^2/r^2}}{r\sin \theta} \ , \\ \label{ch3-9}
&=&\frac{1}{r^2 \sin \theta}(z\dot r-v^z r) \ ,
\end{eqnarray}
So, Eq. (\ref{ch3-4}) can be written as
\be \label{ch3-10}
J=\frac{r^4}{L}v_rv^\theta \sin \theta \ .
\ee

To perform the integrations in Eq. (\ref{ch3-1}) using Eq. (\ref{ch3-3}), we need to write the Jacobian in terms of $L$ and $L_z$: 
\begin{eqnarray} \nonumber
z\dot r-v_z r&=&\frac{z}{r}(xv_x+yv_y+zv_z)-v_zr \ , \\ \nonumber
&=&\frac{1}{r}[x(zv_x-xv_z)+y(zv_y-yv_z)] \ , \\ \nonumber
&=&\frac{1}{r}(xL_y-yL_x) \ , \\ \nonumber
\Rightarrow \;\;\; (z\dot r -v_z r)^2 &=&\frac{1}{r^2}[x^2L_y^2+y^2L_x^2-\underbrace{(xL_x+yL_y)^2-x^2L_x^2+y^2L_y^2}_{2xyL_yL_x}]\ , \\ \nonumber
&=&\frac{1}{r^2}[(x^2+y^2)(L_x^2+L_y^2)-z^2L_z^2] \ , \\ \nonumber
&=&\frac{1}{r^2}[(r^2-z^2)(L^2-L_z^2)-z^2L_z^2] \ , \\ \nonumber
&=&L^2\underbrace{(1-z^2/r^2)}_{\sin^2 \theta}-L_z^2 \ , \\ \label{ch3-11}
\Rightarrow \;\;\; z\dot r-v_zr&=&\sin \theta [L^2-\frac{L_z^2}{\sin^2 \theta}]^{1/2}\ ,
\end{eqnarray}
where we used $\bm r \cdot \bm L=0$.  

An alternative derivation of the Jacobian in terms of $L$ and $L_z$ uses the metric components in spherical coordinates, $g_{rr}=1$, $g_{\theta \theta}=r^2$, and $g_{\phi \phi}=r^2 \sin^2 \theta$, in Eq. (\ref{ch3-5}) which leads to
\begin{eqnarray} \nonumber
L^2&=&r^2v^2-({\bm r}\cdot {\bm v})^2 \ , \\ \nonumber
&=&r^2\left( g_{rr}{v^r}^2+g_{\theta\theta}{v^\theta}^2+g_{\phi\phi}{v^\phi}^2\right )-\left (g_{rr}rv^r \right)^2 \ ,\\  \nonumber
&=&r^4 \left({v^\theta}^2+\sin^2 {\theta}\ {v^\phi}^2 \right), \\ \label{ch3-12}
&=&r^4 \left( {v^\theta}^2+\frac{L_z^2}{\sin ^2 \theta} \right) \ ,
\end{eqnarray}
where we used $L_z\equiv v_\phi=g_{\phi \phi} v^\phi$. Solving Eq. (\ref{ch3-12}) for $v^\theta$ gives
\be \label{ch3-13}
v^\theta=\frac{1}{r^2} \left( L^2-\frac{L_z^2}{\sin^2\theta} \right)^{1/2} \ .
\ee
Therefore, we can write
\be \label {ch3-14}
v^\theta=\frac{1}{r^2}\bm v \cdot \bm e_{\theta}=\frac{z\dot r-rv^z}{r^2\sin \theta}=\frac{1}{r^2} \left( L^2-\frac{L_z^2}{\sin^2\theta} \right)^{1/2} \ ,
\ee
Combining Eqs. (\ref{ch3-14}) and (\ref{ch3-4}) we again find Eq. (\ref{ch3-10}).

\be \label{ch3-15}
J=\frac{r^4}{L}v_rv^\theta \sin \theta \ .
\ee
Including a factor of $4$ to take into account the $\pm$ signs of $v^\theta$ and $v_r$ available for each value of $E$ and $L$, we obtain 
\be \label{ch3-16}
{\rm d}^3v=\frac{4L}{r^4|v_r||v^\theta|\sin \theta} \ {\rm d}E\ {\rm d}L\ {\rm d}L_z \ ,
\ee
and thus the physical density
\be \label{ch3-17}
\rho(r)=4 \int {\rm d} E \int L{\rm d} L \int {\rm d} L_z \frac{f(E,L)}{r^4|v_r||v^\theta|\sin \theta} \ .
\ee
The limits on $L_z$ are derived by demanding that $v^\theta$ should be real in Eq. (\ref{ch3-14}). We will also assume throughout that the distribution function is independent of $L_z$; as a result we can integrate over $L_z$ between the limits $\pm L \sin \theta$, to obtain Eq. (1) in \cite{Quinlan95}:
\be \label{ch3-18}
\rho(r)=4 \pi \int {\rm d}E \int L{\rm d}L \frac{f(E,L)}{r^2|v_r|} \ .
\ee
The limits of integration are set in part by the fact that $|v_r|$ must be real. Solving the energy per unit mass of each particle, $E=\Phi(r)+(1/2)(v_r^2+L^2/r^2)$, for $v_r$ we have
\be \label{ch3-19}
|v_r|=\left( 2E-2\Phi(r)-\frac{L^2}{r^2} \right)^{1/2} \ ,
\ee
and thus $L$ ranges from $0$ to $[2r^2(E-\Phi(r))]^{1/2}$, while $E$ ranges from $\Phi(r)$ to $E_{\rm max}$, the maximum energy that a bound particle could have. We thus have
\be \label{ch3-20}
\rho(r)=\frac{4\pi}{r^2}\int_{\Phi(r)}^{E_{\rm max}} {\rm d}E \int_0 ^{L_{\rm max}} L{\rm d}L\frac{f(E,L)}{\sqrt{2E-2\Phi(r)-L^2/r^2}} \ .
\ee

Hence, given a distribution function, $f(E,L)$, we can use Eq. (\ref{ch3-20}) to find the density, $\rho(r)$, which acts as the source of the gravitational potential $\Phi(r)$. We will also encounter the situation where $\rho(r)$ is known, e.g. from fits to numerical simulations, and we would like to find the distribution function. In the next subsection we review Eddington's method, which allows us to construct the distribution function from the density density.

\subsection{Eddington's Method}
Following the terminology in Binney and Tremaine (\cite{BT08}, BT hereafter), we define a new gravitational potential and a new energy. If $\Phi_0$ is some constant, then let the {\it relative potential} $\Psi({\bm r})$ and the {\it relative energy} $\mathbb E$ of a particle be defined by
\begin{eqnarray} \nonumber
\Psi({\bm r}) &\equiv& -\Phi({\bm r})+\Phi_0 \ , \\ \label{ch3-21}
{\mathbb E}&\equiv&-H({\bm r}, {\bm v})+\Phi_0=\Psi({\bm r})-\frac{1}{2}v^2 \ .
\end{eqnarray}
where $H$ is the Hamiltonian of the system. In practice, $\Phi_0$ is chosen to be such that $f>0$ for ${\mathbb E} > 0$ and $f=0$ for ${\mathbb E} \leq 0$. If an isolated system extends to infinity, $\Phi_0=0$ and the relative energy is equal to the binding energy. The relative potential of an isolated system satisfies Poisson's equation in the form
\be \label{ch3-22}
\nabla \Psi({\bm r})=-4\pi G \rho({\bm r}) \ ,
\ee
subject to the boundary condition $\Psi({\bm r})\rightarrow \Phi_0$ as $|\bm x|\rightarrow \infty$.

Suppose we observe a spherical system that is confined by a known spherical potential $\Phi(r)$. Then it is possible to derive for the system a unique distribution function that depends on the phase-space coordinates only through the Hamiltonian $H({\bm r},{\bm v})$. Here we express this distribution function as a function of the relative energy $f(\mathbb E)$. Using Eq. (\ref{ch3-1}), since $f$ depends on the magnitude $v$ of $\bm v$ and not its direction, we can immediately integrate over angular coordinates in velocity space. We then have
\begin{eqnarray} \nonumber
\rho(r)&=&4\pi \int {\rm d}v \ v^2 f \left(\Psi(r)-v^2/2\right) \ , \\ \label{ch3-23}
\displaystyle&=&4 \pi \int_0^{\Psi} {\rm d}{\mathbb E} f({\mathbb E}) \sqrt{2(\Psi (r)-{\mathbb E})} \ ,
\end{eqnarray}
where we have used Eq. (\ref{ch3-21}) and assumed that the constant $\Phi_0$ in the definition of $\mathbb E$ has been chosen such that $f=0$ for ${\mathbb E} \leq 0$. It can be shown that $\Psi$ is a monotonic function of $r$ in any spherical system, therefore we can regard $f$ as a function of $\Psi$ instead of $r$. Thus
\be \label{ch3-24}
\frac{1}{\sqrt8 \pi}f(\Psi)=2\int_0^\Psi {\rm d} {\mathbb E} \ f({\mathbb E}) \sqrt {\Psi-\mathbb E} \ .
\ee

Differentiating both sides of Eq. (\ref{ch3-24}) with respect to $\Psi$, we obtain
\be \label{ch3-25}
\frac{1}{\sqrt8 \pi}\frac{{\rm d}f}{{\rm d} \Psi}=\int_0^\Psi {\rm d}{\mathbb E} \frac{f({\mathbb E})}{\sqrt{\Psi- {\mathbb E}}} \ .
\ee
Equation (\ref{ch3-25}) is an Abel integral equation having the solution 
\begin{subequations} \label{Abel}
\be \label{ch3-26}
f({\mathbb E})=\frac{1}{\sqrt 8 \pi^2} \frac{{\rm d}}{{\rm d} {\mathbb E}}\int_0^{\mathbb E} \frac{{\rm d}\Psi}{\sqrt{{\mathbb E}-\Psi}} \frac{{\rm d}f}{{\rm d}\Psi} \ .
\ee
An equivalent formula is
\be \label{ch3-27}
f({\mathbb E})=\frac{1}{\sqrt 8 \pi^2} \left[ \int_0^{\mathbb E} \frac{{\rm d}\Psi}{\sqrt{{\mathbb E}-\Psi}} \frac{{\rm d}^2 f}{{\rm d} \Psi^2}+\frac{1}{\sqrt {\mathbb E}} \left(\frac{{\rm d} f}{{\rm d}\Psi} \right)_{\Psi=0} \right] \ .
\ee
\end{subequations}
This result is due to Eddington \cite{Eddington1916}, and it is called {\it Eddington's formula}. It implies that, given a spherical density distribution, we can recover a distribution function depending only on the Hamiltonian that generates a model with the given density. In general, there might be multiple distribution functions that generate a given density, and Eddington's formula gives us the one which is isotropic in the velocity space. However, there is no guarantee that the solution $f(\mathbb E)$ to Eqs. (\ref{Abel}) will satisfy the physical requirement that it be nowhere negative. Indeed, we may conclude from Eq. (\ref{ch3-26}) that a spherical density distribution $f(r)$ in the potential $\Phi(r)$ can arise from a distribution function depending only on the Hamiltonian if and only if
\be \nonumber
\int_0^{\mathbb E} \frac{{\rm d}\Psi}{\sqrt{{\mathbb E}-\Psi}} \frac{{\rm d}f}{{\rm d}\Psi} \ ,
\ee
is an increasing function of $\mathbb E$.

\subsection{Adiabatic Invariants}
We next imagine a point mass growing slowly at the center of a pre-existing distribution of particles. Systems like this where potential variations are slow compared to a typical orbital frequency are called {\it adiabatic}. It can be shown using the action-angle formalism (\cite{BT08}, Section 3.6.) that the actions of particles, $\oint p dq$, for each independent coordinate and conjugate momentum are constant during such adiabatic changes of potential. For this reason such action integrals are often called {\it adiabatic invariants}.  

So, as the gravitational potential near the point mass changes because of the growth of the point mass, each particle responds to the change by altering its energy $E$ and angular momentum $L$ and $L_z$, holding the adiabatic invariants $I_r$, $I_\theta$, and $I_\phi$ fixed, where
\begin{eqnarray} \nonumber
I_r(E,L) &\equiv& \oint v_r {\rm d}r= \oint {\rm d}r \sqrt{2E-2\Phi(r)-L^2/r^2}\ , \\ \nonumber
I_\theta(L,L_z) &\equiv& \oint v_\theta {\rm d}\theta= \oint {\rm d}\theta \sqrt{L^2-L_z\sin ^{-2} \theta}=2\pi(L-L_z) \ , \\ \label{ch3-28}
I_\phi(L_z) &\equiv& \oint v_\phi {\rm d}\phi=\oint L_z {\rm d}\phi=2\pi L_z \ .
\end{eqnarray}

The constancy of $I_\theta$ and $I_\phi$ implies that $L$ and $L_z$ remain constants, no surprise considering the assumed spherical symmetry. But when the potential evolves from the initial potential $\Phi'$ to a new potential $\Phi$ that includes the point mass, $E'$ evolves to $E$ such that
\be \label{ch3-29}
I_r(E,L)=I'_r(E',L) \ .
\ee
In \cite{Young80}, it has been shown that for an adiabatic growth of a point mass inside a cluster, the conservation of the adiabatic invariants of each particle leads to the invariance of the distribution function $f(E,L)=f'(E',L')$. In Appendix  \ref{AppendixD} we review this argument of \cite{Young80} and also generalize it to the relativistic analysis. 

So, by equating radial actions in Eq. (\ref{ch3-29}) and solving to obtain the relation $E'=E'(E,\ L)$, the new distribution function is then assumed to be given by the original distribution function $f'$, where $E'$ is expressed in terms of $E$ and $L$.
\be \label{ch3-30}
f(E,\ L)=f'(E'(E, \ L), \ L) \ .
\ee
Note that, in a Newtonian analysis for a potential dominated by a point mass, $\Phi(r)=-Gm/r$, and
\be \label{ch3-31}
I_r(E,\ L)=2\pi \left(-L+\frac{Gm}{\sqrt{-2E}} \right) \ .
\ee

Considering what we reviewed here, the density in the presence of the point mass may then be expressed as
\be \label{ch3-32}
\rho(r)=\frac{4 \pi}{r^2}\int_{-Gm/r}^{E_{\rm max}}{\rm d} E \int_0^{L_{\rm max}} L{\rm d} L \frac{f'(E'(E, \ L), \ L)}{\sqrt{2E+2Gm/r-L^2/r^2}} \ .
\ee

\section{Growing a Black Hole in a Dark Matter Cluster: Relativistic Analysis}
Given a system of particles characterized by a distribution function $f^{(4)}(p)$, there is a standard prescription for writing down the mass current four-vector \cite{Fackerell68}:
\be \label{ch3-33}
J^\mu(x)\equiv \int f^{(4)}(p)\frac{p^\mu}{\mu}\sqrt{-g}\ {\rm d}^4p \ ,
\ee
where $\mu$ is the particle's rest mass, $p$ and $p^\mu$ represent the four-momentum, $g$ is the determinant of the metric, and $d^4p$ is the four-momentum volume element; the distribution function is again normalized so that the total mass of the halo is $M$. 

As in the Newtonian case, we wish to change variables from $p^\mu$ to variables that are related to suitable constants of the motion. In the absence of a black hole, and for a spherically symmetric cluster, the constants would be the relativistic energy $\cal E$, the angular momentum and its $z$-component $(L, \ L_z)$, together with the conserved rest-mass $\mu = (-p_\mu p^\mu)^{1/2}$. A black hole that forms at the center will generically be a Kerr black hole, whose constants of motion are $\cal E$, $L_z$, $\mu$, plus the so-called Carter constant $C$. In the limit of spherical symmetry, such as for the case of no black hole or for a central Schwarzschild black hole, $C \to L^2$.

We will therefore begin by changing coordinates in the phase-space integral from $p^\mu$ to $\cal E$, $C$, $L_z$, and $\mu$ assuming that the background geometry is the Kerr spacetime. We will find that the loss of spherical symmetry and the dragging of inertial frames that go together with the Kerr geometry make the problem considerably more complex. Further study of this case will be deferred to future work. Taking the limit of a Schwarzschild black hole simplifies the analysis, and allows us to formulate the adiabatic growth of a non-rotating black hole in a fully relativistic manner. 

\subsection{Kerr Black Hole Background}
The Kerr metric is given in Boyer-Lindquist coordinates by
\begin{eqnarray}  \nonumber
{\rm d}s^2 &=& - \left( 1 - \frac{2Gmr}{\Sigma^2} \right ) {\rm d}t^2
 + \frac{\Sigma^2}{\Delta} {\rm d}r^2 + \Sigma^2 {\rm d}\theta^2
 - \frac{4Gmra}{\Sigma^2} \sin^2 \theta {\rm d}t {\rm d}\phi  \\ \label{ch3-34}
 && \quad + \left ( r^2 + a^2 + \frac{2Gmra^2 \sin^2 \theta}{\Sigma^2} \right ) \sin^2 \theta d\phi^2  \,,
\end{eqnarray}
where $G$ is Newton's constant, $m$ is the mass, $a$ is the Kerr parameter, related to the angular momentum $J$ by $a \equiv J/m$; $\Sigma^2 = r^2 + a^2 \cos^2 \theta$, and $\Delta = r^2 + a^2 - 2Gmr$.  We will assume throughout that $a$ is positive, and use units in which $c=1$.

Timelike geodesics in this geometry admit four conserved quantities: energy of the particle per unit mass, ${\cal E}$, angular momentum per unit mass,  $L_z$, Carter constant per unit (mass)$^2$, $C$, and the norm of the four momentum, 
\begin{subequations} \label{constants}
\begin{eqnarray} \label{energy}
{\cal E} &\equiv& - u_0 = -g_{00} u^0 - g_{0\phi} u^\phi \ , \\ \label{Lz}
L_z &\equiv&  u_\phi = g_{0\phi} u^0 + g_{\phi \phi} u^\phi \ , \\ \label{carter}
C &\equiv& \Sigma^4 \bigl ( u^\theta \bigr )^2 + \sin^{-2} \theta L_z^2 + 
a^2 \cos^2 \theta (1 - {\cal E}^2 ) \ , \\ \label{norm}
g_{\mu\nu} p^\mu p^\nu &=& -\mu^2 \,.
\end{eqnarray}
\end{subequations}
The version of the Carter constant used here has the property that, in the Schwarzschild limit ($a \to 0$), $C \to L^2$, where $L$ is the total conserved angular momentum per unit mass.

We want to convert from the phase space volume element ${\rm d}^4p$ to the volume element ${\rm d}{\cal E} {\rm d}C {\rm d}L_z {\rm d}\mu$, using the relation
\be \label{ch3-35}
{\rm d}^4 p = |J|^{-1} {\rm d}{\cal E} {\rm d}C {\rm d}L_z {\rm d}\mu \,,
\ee
where the Jacobian is given by the determinant of the matrix
\begin{eqnarray} \nonumber
J \equiv \left |\frac{\partial ({\cal E},\,C,\,L_z,\,\mu)}{\partial (p^0,p^r,p^\theta,p^\phi)} \right |
&=&\mu^{-3} \left|
\begin{array}{cccc} 
   -g_{00}&0&0&-g_{0\phi}\\
   \partial C/\partial u^0&0&2 \Sigma^4 u^\theta&\partial C/\partial u^\phi \\
   g_{0\phi}&0&0&g_{\phi\phi}\\
   {\cal E}&-u_r&-u_\theta&-L_z\\
\end{array}
\right | \ , \\ 
&=& -2 \mu^{-3} \Sigma^4 u_r u^\theta (g_{0\phi}^2-g_{00}g_{\phi \phi}) \ , \\ \label{ch3-36}
&=& -2 \mu^{-3} \Delta \Sigma^4 u_r u^\theta \sin^2 \theta \,.
\end{eqnarray}
where we used the fact, which follows from the Kerr metric, that $g_{0\phi}^2-g_{00}g_{\phi \phi}=\Delta \sin^2 \theta$. Again including a factor of $4$ to take into account the $\pm$ signs of $p^\theta$ and $p^r$ in contrast to the quadratic nature of $C$ and the norm of $p^\mu$, and using the fact that $\sqrt{-g} = \Sigma^2 \sin \theta$, we obtain
\be \label{ch3-37}
\sqrt{-g} \, {\rm d}^4p = \frac{2 \mu^3 }{\Sigma^2 \Delta |u_r| |u^\theta| \sin \theta} {\rm d}{\cal E} {\rm d}C {\rm d}L_z {\rm d}\mu  \,.
\ee
If the particles described by the distribution have the same rest mass, and if we again assume that the three-dimensional distribution function is normalized as before, then $f^{(4)}(p) \equiv \mu^{-3} f({\cal E}, C) \delta (\mu - \mu_0)$, and thus we can integrate over $\mu$, to obtain 
\be \label{ch3-38}
J^\mu = 2 \int {\rm d} {\cal E} \int {\rm d} C  \int {\rm d} L_z   \frac{u^\mu f({\cal E},C) }{\Sigma^2 \Delta  |u_r| |u^\theta| \sin \theta} \,.
\ee
We again assume that $f$ is independent of $L_z$. This may be compared with Eq.(\ref{ch3-17}); $J^0$ is related to the density $\rho$, the relativistic energy $\cal E$ replaces $E$, $C$ plays the role of $L^2$, $\Sigma^2 \Delta$ replaces $r^4$, and four-velocities $u_r$ and $u^\theta$ replace ordinary velocities $v_r$ and $v^\theta$.  

By definition, $J^\mu \equiv \rho u^\mu$, where $\rho$ is the mass density as measured in a local freely falling frame, and $u^\mu$ is the four-velocity of an element of the matter, which can be expressed in the form $u^\mu \equiv \gamma (1, v^j)$, where $v^j \equiv u^j/u^0 = J^j /J^0$, and using $u_\mu u^\mu=-1$ leads to $\gamma = (-g_{00} - 2 g_{0j} v^j - g_{ij} v^i v^j )^{-1/2}$.  Thus, once the components of $J^\mu$ are known, then the $v^j$ components and therefore, $u^0 = \gamma$ can be determined, and from that $\rho = J^0/u^0$ can be found.  Alternatively, because the norm of $u^\mu$ is $-1$, $\rho = (-J_\mu J^\mu)^{1/2}$.   In particular, if $J^\mu$ has no spatial components, then $u^0 = (-g_{00})^{-1/2}$ and $\rho = \sqrt{-g_{00}} J^0$.  

The four-velocity components $u_r$ and $u^\theta$ can be expressed in terms of the constants of the motion by suitably manipulating Eqs.\ (\ref{carter}) and (\ref{norm}), leading to
\begin{eqnarray} \nonumber 
u^\theta &=& \pm \Sigma^{-2} \left [ C -  L_z^2 \sin^{-2} \theta - a^2 \cos^2 \theta (1 - {\cal E}^2)  \right ]^{1/2} \ , \\ \label{ch3-39}
u_r &=& \pm \frac{r^2}{\Delta} V(r)^{1/2} \ ,
\end{eqnarray}
where
\be \label{Vr}
V(r) = \left ( 1+ \frac{a^2}{r^2} +\frac{2Gma^2}{r^3}  \right ){\cal E}^2 -\frac{\Delta}{r^2} \left (1 + \frac{C}{r^2} \right ) + \frac{a^2 L_z^2}{r^4} - \frac{4Gma{\cal E}L_z}{r^3} \ .
\ee

From Eq. (\ref{ch3-38}), it is clear that, since $u^r$ and $u^\theta$ are equally likely to be positive as negative for a given set of values for $\cal E$, $C$ and $L_z$, the components $J^r$ and $J^\theta$ of the current must vanish.
Furthermore, since $u_0 = -{\cal E}$ and $u_\phi = L_z$, we have that 
\begin{eqnarray} \label{ch3-40}
J_0 &=& -2 \int {\cal E} {\rm d} {\cal E} \int
{\rm d} C  \int {\rm d} L_z   \frac{f({\cal E},C) }{\Sigma^2 \Delta  |u_r| |u^\theta| \sin \theta} \ , \\ \label{ch3-41}
J_\phi &=& 2 \int  {\rm d}{\cal E} \int
{\rm d} C  \int  L_z {\rm d}L_z   \frac{f({\cal E},C) }{\Sigma^2 \Delta  |u_r| |u^\theta| \sin \theta} \ ,
\end{eqnarray}

Even if we assume that $f$ is independent of $L_z$, the presence of the term in $V(r)$ [Eq.~(\ref{Vr})] that is {\em linear} in $L_z$ implies that $J_\phi$ will {\em not} vanish in general, and thus the distribution of matter will have a flux in the azimuthal direction.  This, of course, is the dragging of inertial frames induced by the rotation of the black hole, an effect that will be proportional to the Kerr parameter $a$. In this case the density may be obtained from
\begin{eqnarray} \nonumber
\rho &=& (- g^{00} J_0^2 - 2 g^{0\phi} J_0 J_\phi - g^{\phi \phi} J_\phi^2 )^{1/2} \ ,  \\ \label{ch3-42}
&=& - J_0 \left ( \frac{g_{\phi \phi} + 2 g_{0\phi} \Omega + g_{00} \Omega^2 }{\Delta} \right )^{1/2} \ ,
\end{eqnarray}
where $\Omega \equiv J_\phi/J_0$.    If $a=0$, then $J_\phi = 0$, and $\rho = -J_0 (g_{\phi\phi}/\Delta)^{1/2}
= -J_0 (-g^{00})^{1/2} = \sqrt{-g_{00}} J^0$.     

The three-dimensional region of integration over $\cal E$, $C$ and $L_z$ is complicated. The energy ${\cal E}$ is bounded above by unity if unbound particles are to be excluded from consideration. The variables are bounded by the two-dimensional surfaces defined by $u^\theta =0$ and $u_r =0$, the latter depending on the value of $r$. A final bound is provided by the condition that if a given particle has an orbit taking it close enough to the black hole to be captured, it will disappear from the distribution. For a given $\cal E$ and $L_z$ there is a critical value of $C$, below which a particle will be captured. No analytic form for this condition has been found to date, although for non-relativistic particles for which ${\cal E} =1$ is a good approximation, Will \cite{Will12} found an approximate analytic expression for the critical value of $C$.

\subsection{Schwarzschild Black Hole Background}

We now restrict our attention to the Schwarzschild limit, $a = 0$, in which $\Sigma^2 = r^2$, $C= L^2$, $u^\theta = (L^2 - L_z^2 \sin^{-2} \theta)^{1/2}$ and 
\be \label{ch3-43}
V(r) = {\cal E}^2 - \biggl ( 1 - \frac{2Gm}{r} \biggr ) \biggl ( 1 + \frac{L^2}{r^2} \biggr ) \ .
\ee
The metric components are $g_{00} = -g_{rr}^{-1} = -1 + 2Gm/r$, and $g_{0\phi} =0$. Substituting these relations, along with the fact that $u^0 = -g^{00} {\cal E}$, we write $J_0$ in the form
\be \label{ch3-44}
J_0 = -\frac{2}{r^2} \int {\cal E} {\rm d} {\cal E} \int {\rm d} L^2  \int {\rm d}L_z   \frac{f({\cal E},\ L) }{V(r)^{1/2} (L^2 \sin^2 \theta - L_z^2)^{1/2} } \ ,
\ee
and we observe that $J_\phi = 0$. We then integrate over $L_z$ between the limits $\pm L\sin \theta$ explicitly to obtain
\be \label{J0finalS}
J_0 = -\frac{4\pi }{r^2} \int {\cal E} {\rm d} {\cal E} \int L {\rm d} L \frac{f({\cal E},\ L) }{\sqrt{{\cal E}^2 - (1-2Gm/r)(1 + L^2/r^2)}}  \ . 
\ee
We again assume that $\cal E$ is bounded above by unity;  ${\cal E}$ and $L$ are also bounded by the vanishing of $V(r)$ and by the black hole capture condition.

Unlike the Kerr case, the capture condition in Schwarzschild can be derived analytically.  We wish to find the critical value of $L$ such that an orbit of a given energy ${\cal E}$, and $L$ will not be ``reflected'' back to large distances, but instead will continue immediately to smaller values of $r$ and be captured by the black hole. The turning points of the orbit are given by the values of $r$ where  $V(r)=0$.  The critical values of  ${\cal E}$,  $L$ are those for which the potential has an extremum at that same point, that is where ${\rm d}V(r)/{\rm d}r = 0$.  The chosen sign for $V(r)$ also dictates that this point should be a minimum of $V(r)$, that is that ${\rm d}^2 V(r)/{\rm d}r^2 > 0$, corresponding to an unstable extremum.  
We obtain from the condition ${\rm d}V(r)/{\rm d}r =0$ the standard solution for the radius of the unstable circular orbit in Schwarzschild $r = 6Gm/ \{1 + [1- 12(Gm/L)^2]^{1/2}\}$.  Substituting this into the condition $V(r) =0$ and solving for $L$, we obtain the critical value
\be \label{Lcrit}
{L}_c^2 = \frac{32(Gm)^2}{36{\cal E}^2 - 27{\cal E}^4 -8 + {\cal E}(9{\cal E}^2-8)^{3/2}} \,.
\ee
Notice that, for ${\cal E} =1$, $L_c = 4Gm$, corresponding to the unstable marginally bound orbit in Schwarzschild at $r=4Gm$, while for ${\cal E} = \sqrt{8/9}$, $L_c = 2\sqrt{3}Gm$, corresponding to the innermost stable circular orbit at $r=6Gm$.  

\begin{figure}
\begin{center}
\includegraphics[trim = 10mm 20mm 20mm 25mm, clip, width=16cm]{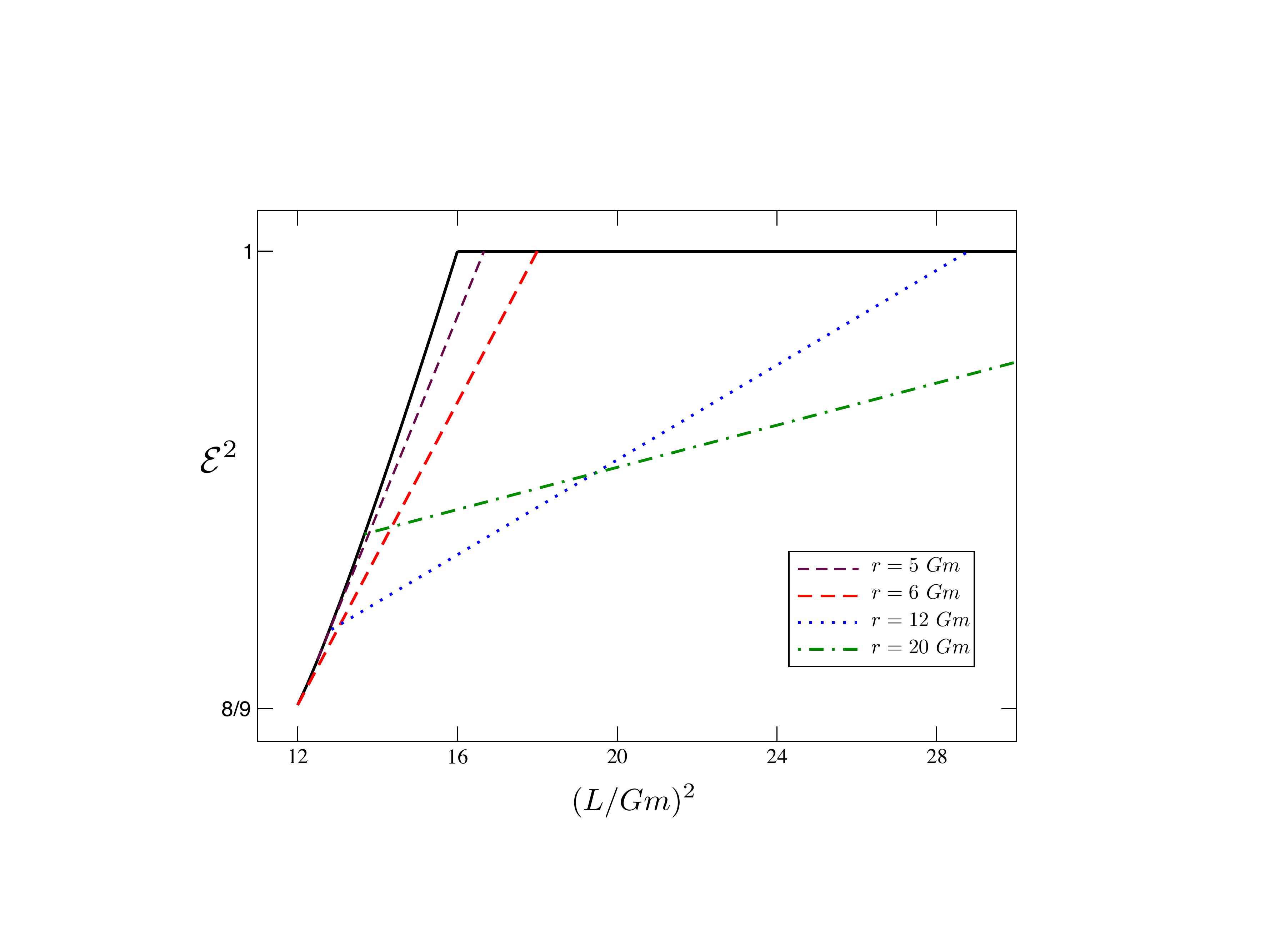}
\caption{\label{fig:phasespace} Integrating over ${\cal E}$-$L$ space for the Schwarzschild geometry.  For a given $r$, the region of integration lies between the solid lines and the various dashed and dotted lines. As $r \to 4m$, the integration area vanishes. }
\end{center}
\end{figure}

The range of integration of the variables is therefore as follows: $L$ is integrated from  $L_{\rm min} = L_c$ to the value given by $V(r) = 0$, namely
\be \label{ch3-46}
L_{\rm max} = r \left ( \frac{{\cal E}^2}{1-2Gm/r} -1 \right )^{1/2} \,.
\ee
In fact, using Eq. (\ref{ch3-39}), $L_{\rm max}$ is the value such that for $L\leq L_{\rm max}$, $u_r$ is real.

The energy ${\cal E}$ is then integrated between its minimum value and unity. That minimum value is found by solving $V(r) =0$ with $L = L_c$, and is given by
\begin{eqnarray}  \label{Elimits}
{\cal E}_{\rm min} &=& \left \{ \begin{array}{ll}
   (1+ 2Gm/r)/(1+6Gm/r)^{1/2}  & : r \ge 6 Gm \\
   (1- 2Gm/r)/(1-3Gm/r)^{1/2} &: 4Gm \le r \le 6Gm \,.\\
    \end{array}
    \right .
\end{eqnarray}

The regions of integration for various values of $r$ are shown in Fig.~\ref{fig:phasespace}.  For a given $r$, the region is a triangle bounded by the critical capture angular momentum on the left, the maximum energy ${\cal E} =1$ at the top, and the condition $V(r)=0$ on the triangle's lower edge.   For $r=6Gm$, the lower edge of the region is the long dashed line shown (red in color version).  As $r$ increases above $6Gm$ the lower edge of the triangle moves upward and the right-hand vertex moves rightward, as shown by the dotted and dot-dashed lines in Fig.~\ref{fig:phasespace} (blue and green in color version).  For values of $r$ decreasing below $6Gm$, the lower edge of the triangle moves upward and leftward as shown by the short dashed line in  Fig.~\ref{fig:phasespace} (violet in color version).   At $r= 4Gm$, ${\cal E}_{\rm min} =  {\cal E}_{\rm max}  =1$ and 
$L_{\rm min} =  L_{\rm max}  =4$, and the volume of phase space vanishes.  This implies that, irrespective of the nature of the distribution function, the density of particles must vanish at $r=4Gm$; this makes physical sense, since any bound particle that is capable of reaching $r=4Gm$ is necessarily captured by the black hole and leaves the distribution.   This is a rather different conclusion from the one reached by Gondolo and Silk (~\cite{Gondolo99}, GS hereafter), who argued that the density would generically vanish at $r=8Gm$.  
The specific shape of this phase space region for small $r$ will play a central role in determining the density distribution near the black hole.

In the Schwarzschild limit, the four-velocity components are given by $u_\phi = L_z$, $u_\theta = (L^2 - L_z^2 \sin^{-2} \theta )^{1/2}$, and $u_r = [{\cal E}^2 - (1-2Gm/r)(1+L^2/r^2)]^{1/2}$, so that the adiabatic invariants are
\begin{subequations}
\begin{eqnarray} \label{rel:invar1}
I_r ({\cal E},L) &=& \oint {\rm d}r \sqrt{{\cal E}^2 - (1-2Gm/r)(1+L^2/r^2)}  \ ,  \\  \label{rel:invar2}
I_\theta (L, L_z)&=&  2\pi (L-L_z) \ , \\  \label{rel:invar3}
I_\phi (L_z) &=& 2\pi L_z \ .
\end{eqnarray}
\end{subequations}
\subsection{Example: Constant Distribution Function}

\begin{figure}
\begin{center}
\includegraphics[trim = 0mm 0mm 0mm 10mm,width=5in]{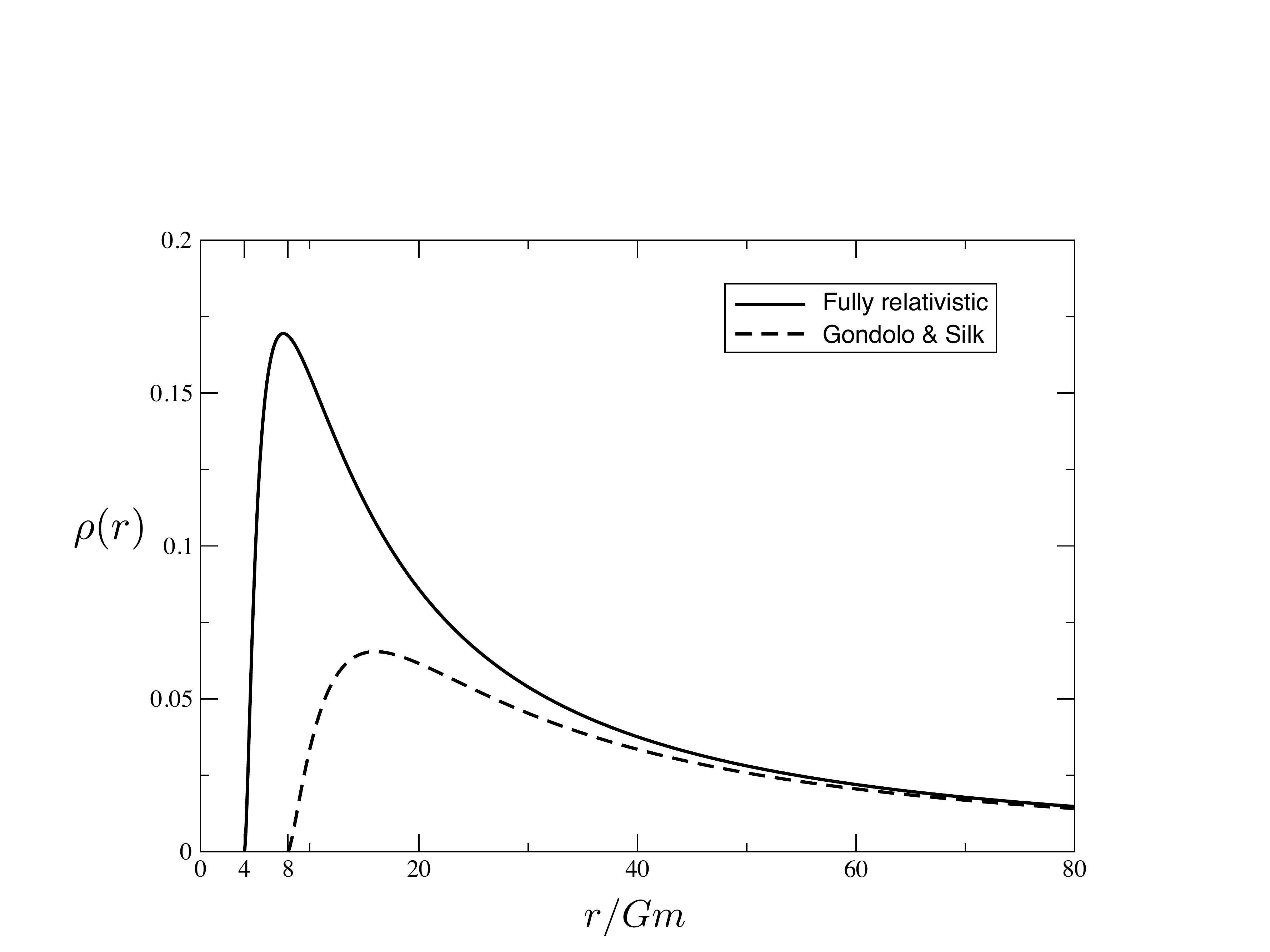}
\caption{Number density around a Schwarzschild black hole for a distribution function $f(p)= f_0 =$ constant.  Shown are the fully relativistic and the GS results. }
\label{fig:plotf0}
\end{center}
\end{figure}

To illustrate the application of these results, we consider the special, albeit unrealistic case of a constant distribution function $f({\cal E},\ L) = f_0$. Then $f$ is still constant after applying the adiabatic condition.  
Since $f$ is independent of $L$, we can do the $L$ integration explicitly to obtain
\be
J_0 = -4\pi  f_0  \int \frac{\sqrt{{\cal E}^2 - (1-2Gm/r)(1 + L_c^2/r^2)}}{1-2Gm/r} {\cal E} {\rm d} {\cal E}  \ , 
\ee
from which we obtain the density
\be \label{nfinal}
\rho(r) = \frac{4\pi  f_0}{(1-2Gm/r)^{3/2}} \int \sqrt{{\cal E}^2 - (1-2Gm/r)(1 + L_c^2/r^2)} \,
{\cal E} {\rm d} {\cal E}  \ .
\ee
Substituting Eq.\ (\ref{Lcrit}) and integrating over ${\cal E}$ numerically between the limits shown in Eq.\ (\ref{Elimits}), we obtain the number density plotted in Fig.~\ref{fig:plotf0}.  

GS ~\cite{Gondolo99} attempted to incorporate the relativistic effects of the black hole within a Newtonian context as follows.  First they approximated the energy $\cal E$ by ${\cal E} = 1 + E$, so that, to Newtonian order, the denominator in Eq. (\ref{J0finalS}) is $\approx [2(E + Gm/r) - L^2/r^2]^{1/2}$, and ${\cal E} {\rm d}{\cal E} \approx {\rm d}E$.  For the critical capture angular momentum they adopted the approximation $L_c = 4Gm$, the value corresponding to ${\cal E} =1$, while for the minimum energy, they adopted the value of $E$ for which the denominator vanishes for {\em that} critical angular momentum. For the constant distribution function the integrals can be done analytically, with the result [GS, Eq.\ (6)]
\be \label{rhoGS}
\rho(r) = \frac{4\pi f_0}{3} \left ( \frac{2Gm}{r} \right )^{3/2} \left ( 1 - \frac{8Gm}{r} \right )^{3/2} \ .
\ee
In Fig.~\ref{fig:plotf0} we plot Eq.\ (\ref{rhoGS}) for comparison with the relativistic result. The two distributions agree completely at large distances, as expected. The GS distribution vanishes at $r = 8Gm$, and is a factor of three smaller at its peak than the fully relativistic distribution.

\section{Application: the Hernquist Model}

The luminosity density of many elliptical galaxies can be approximated as a power law in radius at both the largest and smallest observable radii, with a smooth transition between these power laws at intermediate radii \cite{BM98}. Numerical simulations of the clustering of dark matter (DM) particles suggest that the mass density within a dark halo has a similar structure \cite{BT08}. For these reasons much attention has been devoted to models in which the density is given by
\be \label{dmdensity}
\rho(r)=\frac{\rho_0}{(r/a)^\alpha(1+r/a)^{\beta-\alpha}} \ ,
\ee
where $\rho_0$ and $a$ are the two parameters of the system. With $\beta=4$ these models have particularly simple analytic properties and are known as {\it Dehnen models} \cite{BM98, Dehnen93, Tremaine94}. The model with $\alpha=1$ and $\beta=4$ is called a {\it Hernquist model} \cite{Hernquist90}, while that with $\alpha=2$ and $\beta=4$ is called {\it Jaffe model} \cite{Jaffe83}. Another dark halo model is given by Eq. (\ref{dmdensity}) with $\alpha=1$ and $\beta=3$; this is called the {\it NFW model} after Navarro, Frenk, and White \cite{NFW95}. Note that the Hernquist and NFW models have the same behavior for small $r$. However, the Hernquist model has the advantage that we can find its distribution function as a closed analytical function using Eddington's formula \cite{BT08}. Therefore, in this section we choose the Hernquist model as the initial distribution of DM particles before the growth of the black hole; then, we derive how the growth of a Schwarzschild black hole will redistribute the DM distribution.

\subsection{Newtonian Analysis}
\label{HernquistNewtonian} 
The Hernquist model is a spherically symmetric matter distribution whose density is given by
\be  \label{ch3-47}
\rho(r)=\frac{\rho_0}{(r/a)(1+r/a)^3} \ ,
\ee
where $\rho_0$ and $a$ are the two scale factors. The corresponding Newtonian gravitational potential of this model is 
\be  \label{ch3-47A}
\Phi(r) = - \frac{GM}{a+r} \ ,
\ee
where $M$ is the total mass of the cluster with $M = 2\pi \rho_0 a^3$. The distribution function that is consistent with this potential is given by the (properly normalized) Hernquist form 
\be \label{hernquist1}
f_H \left( \te \right) = \frac{M}{\sqrt{2} (2 \pi)^3 (G M a)^{3/2}} \tilde{f}_H \left(\te \right) \ ,
\ee
where 
\be \label{hernquist2}
\tilde{f}_H \left(\te \right) = \frac{\sqrt{\te}}{\left(1-\te\right)^2}
\left[ \left(1 - 2 \te \right) \left(8 \te^2 - 8 \te - 3 \right)+
\frac{3 \sin^{-1} \sqrt{\te}}{\sqrt{\te \left(1-\te \right)}} \right] \ ,
\ee
where we adopt the following dimensionless quantities:
\begin{subequations} \label{eq:dimensions}
\begin{eqnarray} \label{eq:dimensions1}
	{\te} &\equiv& - \frac{a}{G M}E \ ,  \\ \label{eq:dimensions2}
	\tilde{L} &\equiv& \frac{L}{\sqrt{a G M}}  \ ,  \\ \label{eq:dimensions3}
	x &\equiv& r/a  \ ,  \\ \label{eq:dimensions4}
	\tp &\equiv& - \frac{a}{G M} \Phi(r) = \frac{1}{1+x}  \ ,  \\ \label{eq:dimensions5}
	\tilde{m} &\equiv& m /M \ ,
\end{eqnarray}
\end{subequations}
where $m$ is the mass of the black hole.

With these definitions, the density Eq.~(\ref{ch3-20}) becomes:
\begin{eqnarray}
	\rho(r) &=& 4 \pi \left( \frac{GM}{a}\right)^{3/2} 
	\int_0^{\te_{\rm max} (x)} { {\rm d} \te} \int_{\tl_{\rm min}}^{\tl_{\rm max}}{ \tl {\rm d}f \tl
		\frac{f_H(\te)}{x^2 \sqrt{2 \left(\tp -\te 
	\right) - \tls/x^2}}} \ , \nonumber \\
	& = &\frac{1}{ \sqrt{2} (2\pi)^2 x} \left (\frac{M}{a^3} \right )
	\int_0^{\te_{\rm max} (x)} { {\rm d} \te} \int_{\tl_{\rm min}^2}^{\tl_{\rm max}^2}{ {\rm d} \tl^2
		\frac{\tilde{f}_H(\te)}{\sqrt{\tls_{\rm max}- \tls}}} \,,
	\label{eq:densityx}
\end{eqnarray}
where $\tl_{\rm max}^2 = 2x^2(\tp - \te)$ and $\tilde{f}_H(\te)$ is given by Eq. (\ref{hernquist2}). Normally we would have $\tl_{\rm min} =0$, and $\te_{\rm max} (x) = \tp (x)$. But we will allow the more general limits in order to include for comparison the GS ansatz for incorporating black-hole capture effects, namely $\tl_{\rm min} = 4\tm (GM/a)^{1/2}$ and $\te_{\rm max} (x) = \tp (x) (1-8\tm M/xa)$.   

When we now grow a point mass adiabatially within the Hernquist model, the argument $\te'$ of the initial distribution (\ref{hernquist2}) becomes a function of $\te$ and $L$ by equating the radial actions:
\be
I_r^H \left(\te{}', \tl \right) = I_r^{\rm bh} \left( \te, \tl \right) \ ,
\label{eq:adiabatic}
\ee
and using the fact that $\tl'=\tl$ from the angular action. Hence the density around the point mass in a Hernquist profile takes the form:
\be
	\rho(r) = \frac{1}{ \sqrt{2} (2\pi)^2 x} \left (\frac{M}{a^3} \right )
	\int_0^{\tm/x} {{\rm d} \te} \int_0^{\tl_{\rm max}}{{\rm d} \tls
		\frac{\tilde{f}_H \left(\te{}'(\te,\tl)\right)}
			{\sqrt{\tls_{\rm max} - \tls}}} \,,
	\label{eq:densityh}
\ee
where $\tls_{\rm max} = 2x^2 (\tm/x - \te)$.

From Eq.~(\ref{ch3-31}), the radial adiabatic invariant for a point mass potential in dimensionless variables is
\be
I_r^{\rm bh}= 2\pi \sqrt{GMa} \left( \frac{\tm}{\sqrt{2 \te}} - \tl \right) \,.
\ee
We see that it diverges for $\epsilon \rightarrow 0$, corresponding to the least bound particle. We will have to be careful when matching the radial actions in this limit.

For the Hernquist potential, with $\tp = 1/(1+x)$ an analytic formula cannot be found for the radial invariant
\be
I_r^H = 2\sqrt{GMa}  \int_{x_-}^{x_+} \left ( \frac{2}{1+x} - 2\te - \frac{\tls}
{x^2} \right )^{1/2} \, {\rm d} x \,,
\ee
and thus it will have to be evaluated numerically. To this end, it is convenient to transform the integration in the following way. First, combine the three terms inside the square root to get
\be
\frac{2}{1+x} - 2\te - \frac{\tls}{x^2} =
\frac{-2 \te x^3 + 2(1- \te) x^2 - \tls x -\tls}{x^2 (1+x)} \ .
\ee
We solve for the three roots of the numerator, of which the two positive roots give the turning points $x_+$ and $x_-$, while the third root $x_{\rm neg}$ is always negative.
We then rewrite the function in the square root as:
\be
2 \epsilon \ \frac{(x_+-x)(x-x_-)(x-x_{\rm neg})}{x^2 (x+1)} \ ,
\ee
which is positive in the region $x_- \le x \le x_+$. We now make a change of variables $x = t \left(x_+ - x_-\right) + x_-$, which brings the integral into the domain $[0,1]$:
\be \label{Ir_H}
I_r^H =2\sqrt{GMa} \sqrt{2 \te} \left(x_+ - x_-\right)^2
\int_0^1 {\sqrt{\frac{(1-t) t \left( (x_+ - x_-) t + x_- -x_{\rm neg} \right)}
{(x_+ - x_-) t + x_-}} \frac{{\rm d} t}{(x_+ - x_-) t + x_-+1}} \ .
\ee
This makes it much easier to control the integration numerically, since we can make sure that the roots have the right signs and ordering, and no numerical round-off errors will change that within the domain.

For $\tls=0$, the radial invariant can be integrated analytically, with the turning points $x_-=0$ and $x_+=1/\epsilon -1$, 
\begin{eqnarray} \nonumber
I_r^H&=&2\sqrt{GMa}\int_0^{1/\te-1} \sqrt{\frac{2}{1+x}-2\te} \ {\rm d}x \ , \\
&=&2\sqrt{2GMa} \left[ \frac{\arccos{\sqrt{\te}}}{\sqrt{\te}}-\sqrt {1-\te} \right] \ ,
\end{eqnarray}
and we use this fact in the code. The radial invariant  is again divergent for $\epsilon \rightarrow 0$. Since we are only interested in finding a solution in the domain $(0,1]$, we simply define the value there to be a very large number, and use a bracketing algorithm.

For numerical work, it is also convenient to remap the integral (\ref{eq:densityh}) for $\rho(r)$ into a square domain. This is a particular case of a set of transformations discovered by Duffy~\cite{Duffy82}. We make a change of variables, $( \te, \tls) \rightarrow (u,z)$, that maps the domain of integration in Eq.~(\ref{eq:densityx}) onto the square $[0,1]\times [0,1]$:
\begin{align}
\te \equiv & u \te_{\rm max} \,, \nn 
\tls \equiv & z \tls_{\rm max}(u) + (1-z) \tls_{\rm min} \,,
\label{eq:uz}
\end{align}
where we emphasize that $\tls_{\rm max}$ depends on $u$.

The jacobian is:
\begin{eqnarray}
	\frac{\left(\partial \te, \partial \tls\right)}{(\partial u, \partial
	z)} &=&\left| \begin{array}{cc} \te_{\rm max}  & 0 \\ \ldots & 
		\tls_{\rm max}(u) -
	\tls_{\rm min} \end{array} \right| \ , \nn
	&=&\te_{\rm max} \left(\tls_{\rm max}(u) - \tls_{\rm min} \right) \ .
\end{eqnarray}

With this change, the integral in Eq.~(\ref{eq:densityx}) reads:
\be
\rho (r) = \frac{1}{ \sqrt{2} (2\pi)^2 x} \left (\frac{M}{a^3} \right ) \te_{\rm max} \int_0^1{{\rm d} u \int_0^1{{\rm d}z 
		\sqrt{\frac{\tls_{\rm max}(u) - \tls_{\rm min}}{1-z}}
\tilde{f}_H\left( \tilde{\epsilon}'(u,z) \right)}} \ ,
\label{eq:rhobox}
\ee
where the arguments of the distribution function are given in Eq.~(\ref{eq:uz}).   This will have the effect of making our codes faster and more stable. One of the advantages is that the integrable singularity that was originally in a corner ($\te = \tp$, $ \tls=0$) of the integration domain has now been transferred to a line, depending only on the variable $z$. 

Using the GS conditions for $\tl_{\rm min}$ and $\te_{\rm max}$ and carrying out the numerical integrations, we obtain the curve labeled ``Non-relativistic'' in Fig.~\ref{fig:hernquist}.

\subsection{Relativistic Analysis}

We now apply these considerations to the relativistic formalism. Here we define $\te$ in terms of the relativistic energy $\cal E$ per unit particle mass using
\be
\te \equiv \frac{a}{GM}(1-{\cal E}) \,;
\ee
the other definitions in Eqs.~(\ref{eq:dimensions}) will be the same. Using these definitions, and the relation $\rho = -J_0(-g^{00})^{1/2}$ along with Eq.~(\ref{J0finalS}), we find 
\begin{eqnarray} \nonumber
\rho(r)&=&\sqrt{-g_{00}} \ J^0 \ , \\  \nonumber 
&=&\frac{4\pi}{x^2}\frac{(GM/a)^{3/2}}{\sqrt{1-(2GM/a)(\tm/x)}}  
\int_0^{\te_{\rm max}} [1-(GM/a)\te] \ {\rm d} \te \times 
\\ \nonumber
&& \qquad
\int_{\tl_{\rm min}}^{\tl_{\rm max}} \tl {\rm d} \tl \frac{f_H(\te)}{\sqrt{2(\tm/x-\te)-\tl^2/x^2+(GM/a)\te^2+(2GM/a)(\tm/x)(\tl^2/x^2)}} \ , \nonumber \\  \nonumber
&=&\frac{1}{ \sqrt{2} (2\pi)^2 x} \left (\frac{M}{a^3} \right )\frac{1}{1-(2GM/a)(\tm/x)} \int_0^{\te_{\rm max}} [1-(GM/a)\te] \ {\rm d} \te \times
\\ \label{ch3-48}
&& \qquad
 \int_{\tls_{\rm min}}^{\tls_{\rm max}} {\rm d} \tls \  \frac{\tilde{f}_H(\te)}{\sqrt{\tls_{\rm max}-\tls}} \ ,
\end{eqnarray}
where $\tilde{f}(\te)$ is again given by Eq. (\ref{hernquist2}), and where we used ${\cal E}=1$ for the maximum energy of the bound particles which leads to $\te_{\rm min}=0$.  Compare the last equation of (\ref{ch3-48}) to Eq. (\ref{eq:densityx}).  

To consider the growth of the central black hole and its capture effects, we use Eqs. (\ref{Lcrit})-(\ref{Elimits}) as the limits of the integrals of Eq. (\ref{ch3-48}), which in terms of the dimensionless parameters have the form
\begin{eqnarray} \nonumber
\tls_{\rm min}&=&\frac{GM}{a} \frac{32 \tm^2}{36(1-\te \ GM/a)^2-27(1-\te \ GM/a)^4-8+(1-\te \ GM/a)[9(1-\te \ GM/a)^2-8]^{3/2}} \ , \\ \nonumber \\ \nonumber
\tls_{\rm max}&=&\frac{ax^2}{GM} \left[ \frac{ \left( 1-\te \ GM/a \right)^2}{1-2 (\tm/x) (GM/a)}-1  \right] \ , \\ \nonumber \\ \nonumber
\te_{\rm max} &=& \frac{a}{GM} \left \{ \begin{array}{ll}
1-[1+ 2(\tm/x)(GM/a)]/\sqrt{1+6(\tm/x)(GM/a)}  & : x \ge 6 \tm \ GM/a \\
1-[1- 2(\tm/x)(GM/a)]/\sqrt{1-3(\tm/x)(GM/a)} &: 4 \tm \ GM/a \le x \le 6 \tm \ GM/a \ .
\end{array}
\right . \\ \label{ch3-49}
\end{eqnarray}
As in the non-relativistic case, in order to grow a point mass adiabatically within the Hernquist model, the argument $\te'$  of the initial distribution function becomes a function of $\te$ and $\tl$ by equating the radial actions and using the fact that $\tl'=\tl$ from the angular action. Hence, the density around a relativistic point mass in a Hernquist profile takes the form:
\be \label{ch3-51}
\rho(r)=\frac{1}{ \sqrt{2} (2\pi)^2 x} \left (\frac{M}{a^3} \right )\frac{1}{1-(2GM/a)(\tm/x)} \int_0^{\te_{\rm max}} [1-(GM/a)\te] \ {\rm d} \te \int_{\tls_{\rm min}}^{\tls_{\rm max}} {\rm d} \tls \  \frac{\tilde{f}_H\left (\te' (\te,\tl ) \right)}{\sqrt{\tls_{\rm max}-\tls}} \ ,
\ee
The difference here is that in equating the radial actions in Eq. (\ref{eq:adiabatic}), we use the relativistic expression for the point-like mass radial action i.e. Eq. (\ref{rel:invar1}) which in terms of dimensionless variables can be written as 
\be \label{ch3-52}
I_{\rm {r, \ rel}}^{\rm bh}=2\sqrt{GMa}\int_{x_-}^{x_+}  \left [2(\tm/x-\te)-\tls/x^2+\te^2 \ GM/a+(2GM/a) (\tm/x) (\tls/x^2) \right ]^{1/2} \, {\rm d}x \ , 
\ee
where $x_+$ and $x_-$ are the two turning points. The integration in Eq.~ (\ref{ch3-52}) will have to be evaluated numerically. Now we take the same steps as we used to get Eq. (\ref{Ir_H}): first we combine the terms inside the square root to get
\begin{eqnarray} \nonumber
2(\tm/x-\te)-\tls/x^2+\te^2 \ GM/a+(2GM/a) (\tm/x) (\tls/x^2) \\ \label{ch3-53}
=\frac{-2\te(1-\te \ GM/2a)x^3+2\tm x^2-\tls x+2\tm \tls \ GM/a}{x^3} \ .
\end{eqnarray}
We solve for the three roots of the numerator, of which the two positive roots give the turning points $x_+$ and $x_-$, while the third $x_{\rm neg}$ is always negative. We then rewrite the function in the square root as:
\be \label{ch3-54}
2\te(1-\te \ GM/2a)\frac{(x_+-x)(x-x_-)(x-x_{\rm neg})}{x^3}
\ee
which is positive in the region $x_- \le x \le x_+$. We now make a change of variables $x = t \left(x_+ - x_-\right) + x_-$, which brings the integral into the domain $[0,1]$:
\be \label{ch3-55}
I_{\rm {r, \ rel}}^{\rm bh}=2\sqrt{GMa} \sqrt{2\te(1-\te \ GM/2a)} (x_+-x_-)^2 \int_0^1{\rm d}t\sqrt{\frac{(x_+-x)(x-x_-)(x-x_{\rm neg})}{x^3}}
\ee
As before, this leads to easier numerical control.

For $\tls=0$, the radial invariant can be integrated analytically, with the turning points $x_-=0$ and $x_+=\tm/(\te(1-\te \ GM/2a))$:
\begin{eqnarray}  \nonumber 
I_{\rm {r, \ rel}}^{\rm bh}&=&2 \sqrt{GMa}\int_0^{\tm/(\te(1-\te \ GM/2a))} {\rm d} x \sqrt{2\left(\frac{\tm}{x}-\te \right)+\te^2 \frac{GM}{a}} \ ,  \\  \label{ch3-56}
&=&2 \pi \sqrt{GMa}\frac{\tm}{\sqrt{2 \te}\sqrt{1- \te \ GM/2a}} \ ,
\end{eqnarray}
and we use this fact in the code. The radial invariant  is again divergent for $\epsilon \rightarrow 0$ but we are only interested in finding a solution in the domain $(0,1]$. For the Hernquist potential we use the same equations as the non-relativistic calculations.

Again we remap the integral in Eq. (\ref{ch3-51}) into a square domain using the Duffy transformations. The only difference here is that $\tl^2_{\rm min}$ also depends on $u$. With these changes, the integral in Eq. (\ref{ch3-51}) reads:
\begin{eqnarray} \nonumber
\rho(r)&=&\frac{1}{ \sqrt{2} (2\pi)^2 x}  \left (\frac{M}{a^3} \right )\frac{ \te_{\rm max}}{1-2(\tm/x)(GM/a)}
\times \\  \label{ch3-50} 
&& \int_0^1 {\rm d}u \int_0^1 {\rm d} z \ [1-(GM/a)\te_{\rm max} u]  \sqrt{\frac{\tls_{\rm max}(u)-\tls_{\rm min}(u)}{1-z}} \tilde{f}_H \left (\te'(u,z)\right) \ ,
\end{eqnarray}
where the arguments of the distribution function are given in Eq. (\ref{eq:uz}). The numerical integrations yield the curve labeled ``Relativistic'' in Fig.~\ref{fig:hernquist}.

\subsection{Profile Modification due to Self-annihilation} 

Our calculations so far give the DM distribution as it reacts to the gravitational field of the growing black hole. In addition, the DM density will decrease if the particles self-annihilate. In fact, if we take into account the annihilation of DM particles, the density cannot grow to arbitrary high values, the maximal density being fixed by the value is \cite{Bertone05}:
\be \label{eq:rhocore}
\rho_{\rm{core}}=\frac{m_\chi}{\sigma v \: t_{\rm{bh}}},
\ee
where $\sigma v$ is the annihilation flux (cross-section times velocity), $m_\chi$ is the mass of the DM particle, and $t_{\rm{bh}}$ is the time over which the annihilation process has been acting, which we take it to be $\approx 10^{10}$yr \cite{Gondolo99}.

The probability for DM self-annihilation is proportional to the square of the density,
\be
\dot{\rho}=-\sigma v \frac{\rho^2}{m_\chi} = 
- \frac{\rho^2}{\rho_{\mathrm{core}}{t_\mathrm{bh}}}.
\label{eq:rhodotcore}
\ee
This expression can be derived by noting that the annihilation rate per particle is $\Gamma = n \sigma v$, therefore $\dot n=-n \Gamma=-n^2\sigma v$ and $\rho = n m_\chi$.

If we call the output of our code neglecting annihilations $\rho'(r)$ 
and the final profile reprocessed by this process $\rho_\mathrm{sp}(r)$,
we can integrate Eq.~(\ref{eq:rhodotcore}) as follows:
\be
\int_{\rho'(r)}^{\rho_\mathrm{sp}(r)}{\frac{\rho_\mathrm{core} \: {\rm d}
\rho}{\rho^2}} = - \int_0^{t_\mathrm{bh}}{\frac{{\rm d} t}{t_\mathrm{bh}}},
\ee
which gives:
\be
\rho_{\mathrm{sp}} (r) = \frac{\rho_\mathrm{core} \rho'(r)}
{\rho_\mathrm{core}+ \rho'(r)}.
\ee

\begin{figure}
\begin{center}
\includegraphics{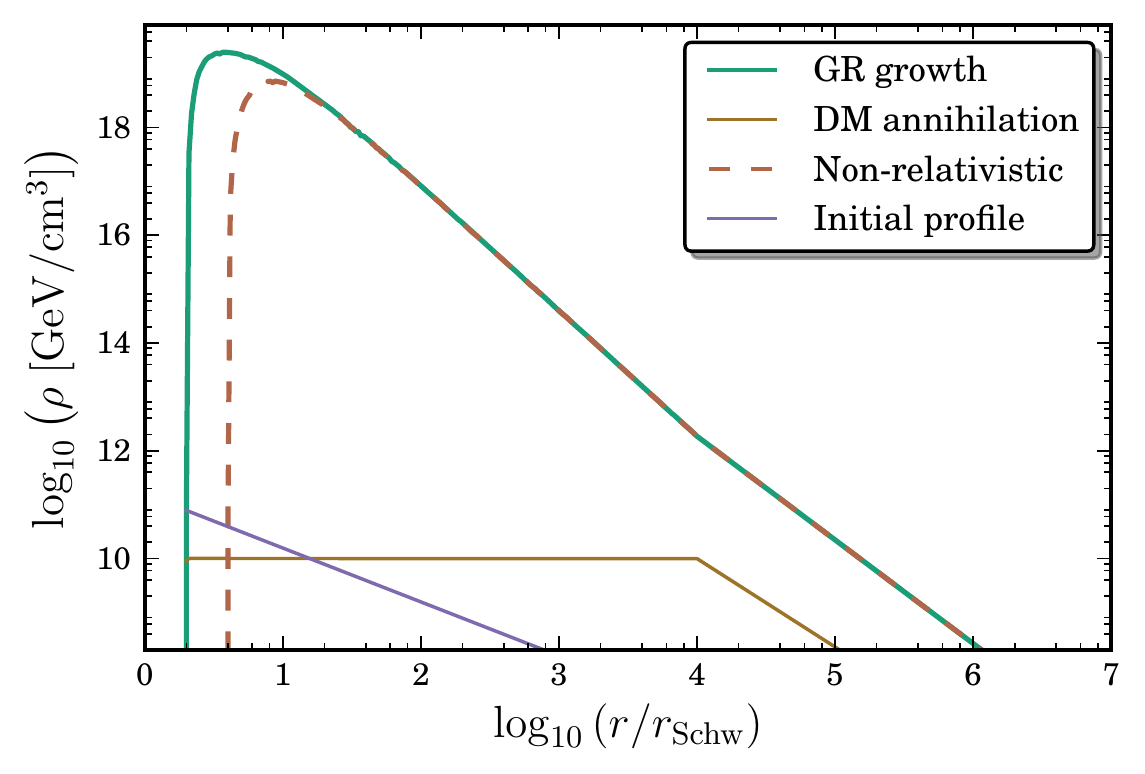}
\caption{Effect of the adiabatic growth of the super-massive black hole at the center of the galaxy on a Hernquist DM profile. Shown are the results of the full relativistic calculation, and the effects of DM annihilations. The dashed line shows the non-relativistic approximation.}
\label{fig:hernquist}
\end{center}
\end{figure}

Our calculations do not include the effect of the gravitational field of the halo in the final configuration. This is a good approximation close to the black hole, but far away from the center the effect of the black hole is negligible and the DM density will be described by the halo only.  We take care of this fact by simply adding the initial Hernquist profile, given in Eq. (\ref{ch3-47}) to the calculated spike. We expect this approximation to be good, except possibly in the transition region. The result is the curve labeled ``DM annihilation'' in Fig.~\ref{fig:hernquist}.

We show in Fig.~\ref{fig:hernquist} the results of our numerical calculations. In the non-relativistic limit, they are a good match to the calculation in GS.

\subsection{Periastron Precession with a Dark Matter Spike}

\begin{figure}
\begin{center}
\includegraphics[trim = 30mm 40mm 20mm 82mm, clip, width=12.0cm]{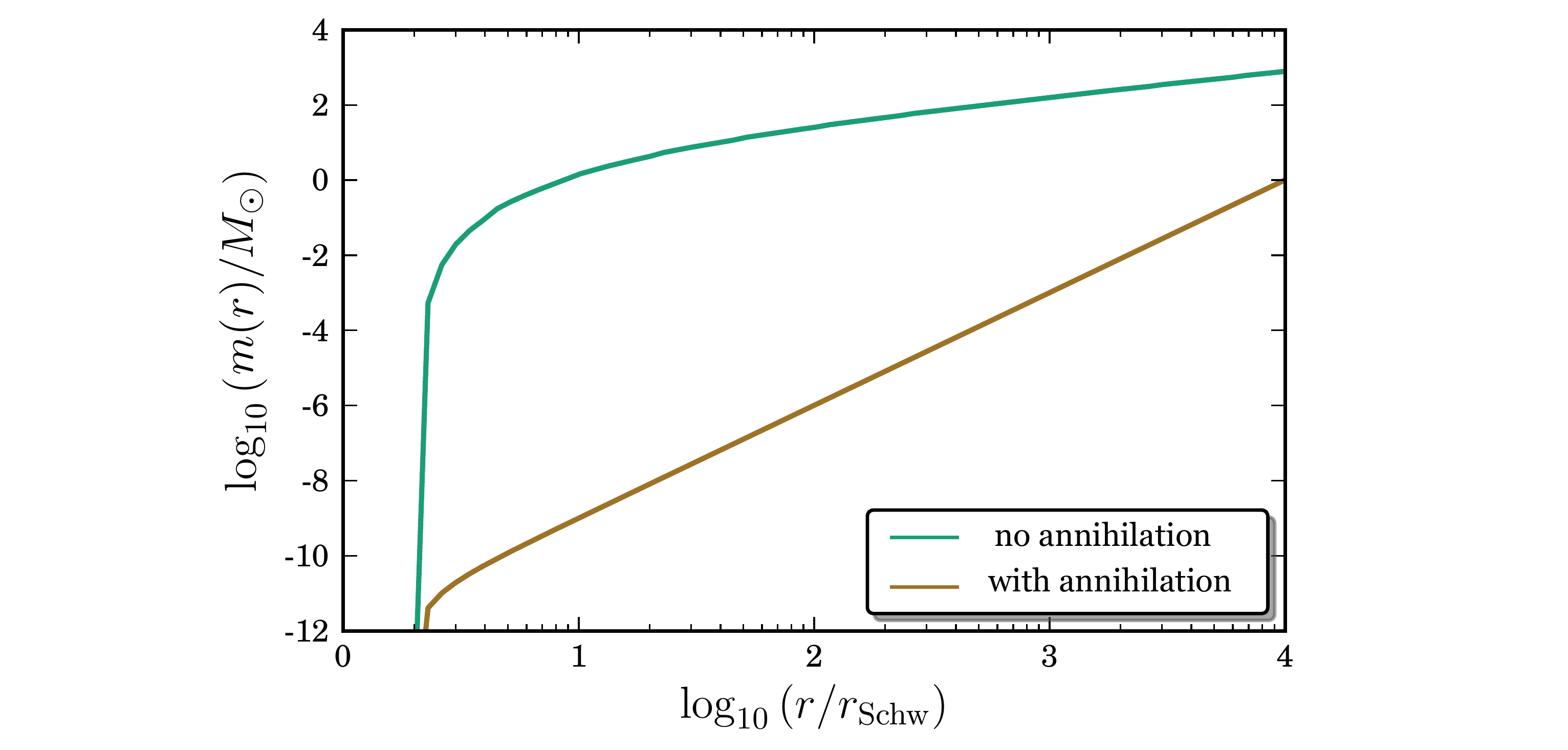}
\caption{Dark matter total mass including the spike as a function of distance for annihilating (brown) and non-annihilating (green) models of dark matter.}
\label{mr}
\end{center}
\end{figure}

As we mentioned in Chapter 2, the presence of the DM density at the GC can perturb the orbits of stars in that region. For related articles see \cite{Iorio2013,Zakharov2007}. A spherically symmetric distribution of dark matter will cause pericenter precessions in orbital motions, but will not change the orientation of the orbital planes. But to get an upper bound on the possible effect of a non-spherical distribution of dark matter on the orbits of potential no-hair-theorem target stars, it is useful to determine the pericenter precession. For this we need the dark matter mass including the spike inside a given radius $r$, which we obtain by integrating our density profile, m(r)= $4\pi \int r^2\rho(r){\rm d}r$. The result for both the self-annihilating and non-self-annihilating cases, is shown in Fig.~\ref{mr}.



As can be seen from Fig.~\ref{mr}, we can approximate the total mass of the DM in the region between $10$ and $10^4$ Schwarzschild radii by a power-law function:
\be \label{ch3-58}
m(r)=m_0(\frac{r}{r_0})^q \ ,
\ee
which leads to the following additional acceleration term in the equation of motion of a star orbiting the black hole:
\be \label{ch3-59}
\bm A=G\frac{m(r)}{r^2} \ \hat{\bm n}=-\frac{Gm_0}{r^2}(\frac{r}{r_0})^q \ \hat{\bm n}\ ,
\ee 
where $\hat{\bm n}\equiv {\bm r}/r$. Since the perturbing term in Eq. (\ref{ch3-59}) has only the radial component $\cal R$, using Eq. (\ref{ch2-51}) for the rate of change with angle of the pericenter of an orbit, ${\rm d}\omega/{\rm d}f$, we have
\be \label{ch3-60}
\frac{{\rm d} \omega}{{\rm d}f}=\frac{r^2}{h} \frac{{\rm d} \omega}{{\rm d}t}=-\frac{r^2 p}{e h^2} {\cal R} \cos f \ ,
\ee
where we used Eq. (\ref{ch2-56}), which for calculations of the first order perturbation, reduces to ${\rm d}f/{\rm d}t=h/r^2$. Substituting Eq. (\ref{ch3-59}) in Eq. (\ref{ch3-60}) and using $r=p/(1+e\cos f)$ and $h^2=Gmp$, we get
\be \label{ch3-61}
\frac{{\rm d} \omega_{\rm DM}}{{\rm d}f}=\frac{1}{e}\left( \frac{m_0}{m} \right) \left( \frac{p}{r_0} \right)^q \frac{\cos f}{(1+e \cos f)^q} \ .
\ee
To get the changes of $\omega$ over one orbit, we integrate Eq. (\ref{ch3-61}) over the true anomaly $f$ from $0$ to $2\pi$ to obtain
\begin{eqnarray} \nonumber
\Delta \omega_{\rm DM}&=&\frac{1}{e}\left( \frac{m_0}{m} \right) \left( \frac{p}{r_0} \right)^q \int_0^{2\pi} \frac{\cos f}{(1+e \cos f)^q} \ , \\  \label{ch3-62}
&=&- \pi q \left( \frac{m_0}{m} \right) \left( \frac{p}{r_0} \right)^q (1-e^2)^{1/2} f_{\rm q}(e) \ ,
\end{eqnarray}
where, using the change of variable of integration described in Appendix {\ref{AppendixA}}, for various values of $q$, we get the forms for $f_{\rm q}(r)$ shown in Table~\ref{fq}.
\begin{table}
\begin{center}
\begin{tabular}{c  c  c}
{\rm q}    & $f_{\rm q}(e)$  & Range of $f_{\rm q}(e)$  \\
\hline
1      & $2/(1+\sqrt{1-e^2})$    & $1<f_1(e)<2$      \\
2      & 1     & 1      \\
3      & 1     & 1      \\
4      & $1+e^2/4$   & $1<f_4(e)<5/4$    \\
\hline
\end{tabular}
\caption{The function $f_{\rm q}(e)$}
\label{fq}
\end{center}
\end{table}

Now from Fig.~\ref{mr}, we can see that the power $q$ in Eq. (\ref{ch3-58}) can be chosen to be $1$ or $3$ depending on whether the DM particles self-annihilate or not, respectively. Using $r_0=r_{\rm Sch}\times 10^4= (2Gm) \times 10^4\approx 4.6 \ {\rm mpc}$, assuming a black hole mass $m=4\times 10^6 \ M_{\odot}$, we can read off the values of $m_0$:
\be  \label{ch3-63}
m_0=\left\{ \begin{array}{l}
    10^3 \ M_\odot \quad,\quad q=1 \quad \text{no self-annihilation}\\
    1 \ M_\odot  \quad,\quad q=3 \quad \text{self-annihilation (constant density core)} \ ,
     \end{array} \right.
\ee

To get an estimation of the pericenter precession effect of stars at the GC as seen from Earth caused by the DM distribution including the spike, we use our previous definition for the angular precession rate amplitude as seen from the Earth in Chapter 2, which is $\dot \Theta_{\rm DM}=(a/D)\Delta \omega/P$, where $D$ is the distance to the GC and $P=2\pi (a^3/m)^{1/2}$ is the orbital period. Using $m=4 \times 10^6 M_\odot$ and $D=8 \ {\rm kpc}$, we obtain the rates for the non-self-annihilating ($q=1$) and self-annihilating ($q=3$) DM particles distributions in microarcseconds per year:
\begin {eqnarray}
\dot \Theta_{\rm DM, \ no-ann.}&=&6.26 \ P^{1/3} \frac{\sqrt{1-e^2}}{1+\sqrt{1-e^2}}  \;\;\; \mu {\rm arcsec/yr} \ , \\
\dot \Theta_{\rm DM, \ ann.}&=&3.81 \times 10^{-4} \ P^{5/3} \sqrt{1-e^2}  \;\;\; \mu {\rm arcsec/yr} \ ,
\end{eqnarray}
where we used Eq. (\ref{ch3-62}) and the numbers in Eq. (\ref{ch3-63}).

To compare the rate of precession of periastron of a star rotating the MBH induced by the DM particles distributions with the relativistic effects of the MBH at the center, in Table~\ref{omega-DM}, we provide numerical results for the S2 star and for a hypothetical target star which is closer to the center and could be used for the test of the no-hair theorem. Shown are the periastron precessions rates as seen from Earth from the Schwarzschild part of the metric and from the two dark matter distributions ($\dot \Theta_{\rm S}$, $\dot \Theta_{{\rm DM}, \ {\rm ann.}}$, and $\dot \Theta_{{\rm DM}, \ {\rm non-ann.}}$, respectively) and the orbital plane precessions from the frame dragging and quadrupole effects ($\dot \Theta_{\rm J}$ and $ \dot \Theta_{{\rm Q}_2}$, respectively).
\begin{table}
\begin{center}
\begin{tabular}{c|c|c|}
\cline{2-3}
& S2 Star  & No-Hair Target Star  \\
& ($a=4.78 \ {\rm mpc}, \ e=0.88, \  P=15.5 \ {\rm yr}$) & ($a=0.2 \ {\rm mpc}, \ e=0.95, \ P=0.13 \ {\rm yr}$) \\ \cline{1-3}
\multicolumn{1}{|c|}{$\dot \Theta_{\rm S}$} & 26.533    & 7319.92      \\
\multicolumn{1}{|c|}{$\dot \Theta_{\rm J}$} & 0.235     & 486.303      \\
\multicolumn{1}{|c|}{$\dot \Theta_{{\rm Q}_2}$} & 0.002     & 36.325      \\
\multicolumn{1}{|c|}{$\dot \Theta_{{\rm DM}, \ {\rm non-ann.}}$} & 5.026   & 0.755    \\
\multicolumn{1}{|c|}{$\dot \Theta_{{\rm DM}, \ {\rm ann.}}$} & 0.017   & $3.969 \times 10^{-6}$  \\
\hline
\end{tabular}
\caption{Astrometric precession rates as seen from the Earth in units of $\mu$arcsec/yr; $\dot \Theta_{\rm J}$ and $\dot \Theta_{\rm Q_2}$ denote orbital plane precessions, while the others denote pericenter precessions.} 
\label{omega-DM}
\end{center}
\end{table}

\begin{figure}
\begin{center}
\includegraphics[trim = 30mm 40mm 14mm 45mm, clip, width=11.0cm]{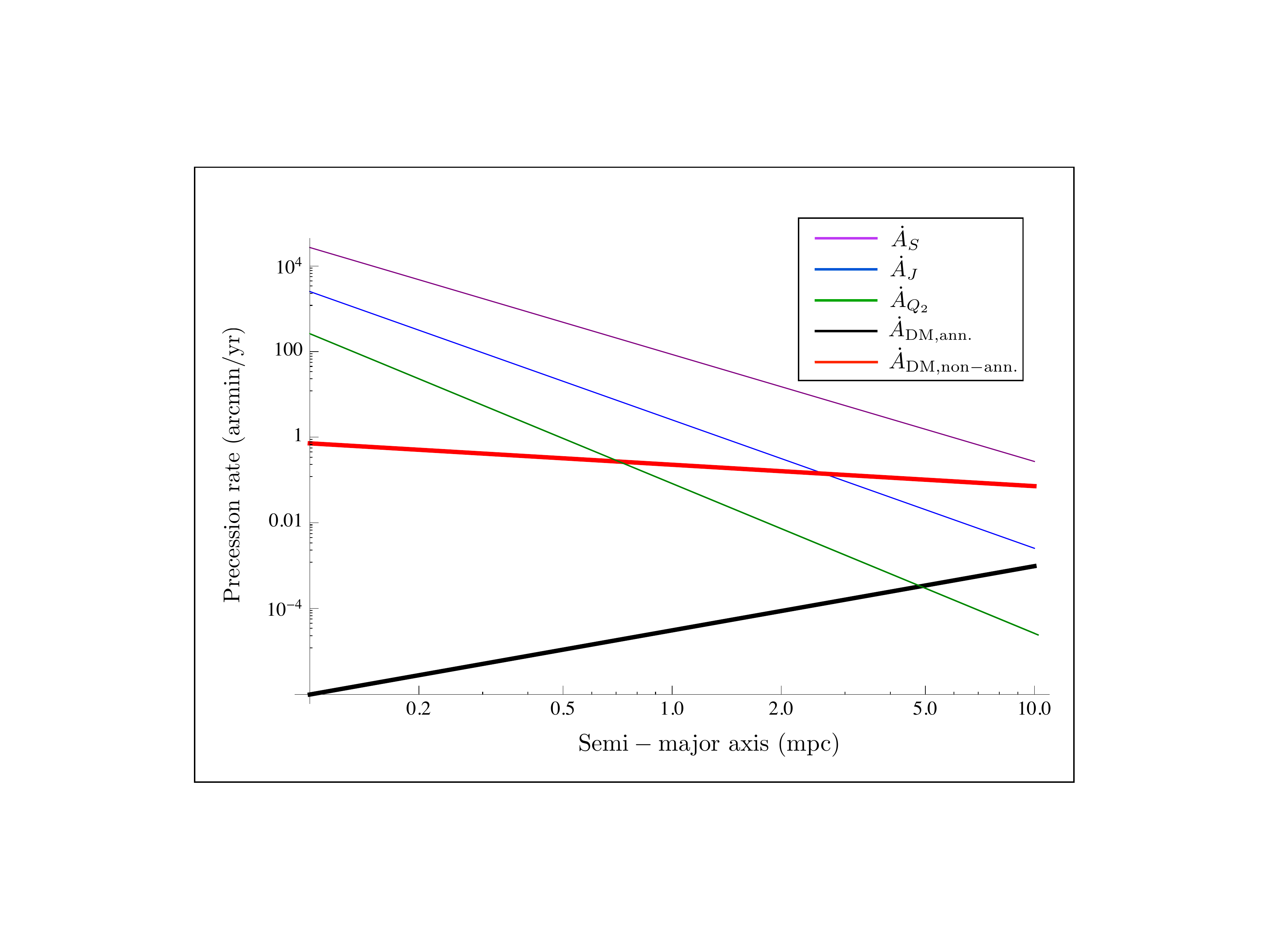}
\caption{Precession rates  at the source for a target star with $e=0.95$ induced by Shwarzschild-part effects of the MBH and by non-self-annihilating and self-annihilating DM particles distribution. Shown are the periastron precession rates from relativistic (purple) and DM (red, black) effects, and the orbit plane precession rates from relativistic frame dragging (blue) and quadrupole (green) effects.}
\label{precession_rates}
\end{center}
\end{figure}

In Fig. \ref{precession_rates}, using Eqs. (\ref{ch2-67})-(\ref{ch2-69}) and Eq. (\ref{ch3-62}), we plot the periastron precessions at the source given in the following equations, for a maximum rotating MBH ($\chi=1$) and a high-eccentricity target star with $e=0.95$ :
\begin{eqnarray} \nonumber
{\dot A}_S&\equiv&\frac{A_S}{P}=\frac{6\pi}{P} \frac{Gm}{a(1-e^2)} \ , \\ 
& \approx & 8.335 \ \tilde a^{-5/2} (1-e^2)^{-1} \ {\rm arcmin/yr} \ , \\ \nonumber
\dot{A}_J&\equiv&\frac{A_J}{P}= \frac{4\pi}{P} \chi \left[ \frac{Gm}{a(1-e^2)}\right]^{3/2} \ , \\ 
& \approx &0.0768 \ \chi {\tilde a}^{-3} (1-e^2)^{-3/2} \ {\rm arcmin/yr} \ , \\ \nonumber
\dot{A}_{Q_2}&\equiv&\frac{A_{Q_2}}{P}= \frac{3 \pi}{P} \chi^2  \left[ \frac{Gm}{a(1-e^2)}\right]^2 \ , \\ 
& \approx & 7.9 \times 10^{-4} \chi^2 {\tilde a}^{-7/2} (1-e^2)^{-2} \ {\rm arcmin/yr} \ , \\ \nonumber
{\dot A}_{\rm DM, \ no-ann.} &\equiv&\frac{\Delta \omega_{\rm DM,\ no-ann.}}{P}= \frac{-2\pi}{P} \left(\frac{m_0}{m} \right)\left( \frac{a}{r_0} \right)\frac{(1-e^2)^{1/2}}{1+(1-e^2)^{1/2}} \ , \\ 
& \approx &0.953 \ {\tilde a}^{-1/2}(1-e^2)^{1/2}[1+(1-e^2)]^{-1/2} \ {\rm arcmin/yr}   \ , \\ \nonumber
{\dot A}_{\rm DM,\ ann.}&\equiv&\frac{\Delta \omega_{\rm DM,\ ann.}}{P}= \frac{-3 \pi}{P} \left(\frac{m_0}{m} \right)\left( \frac{a}{r_0} \right)^3(1-e^2)^{1/2} \ , \\  
& \approx & 9.8 \times 10^{-5} \ {\tilde a}^{3/2} (1-e^2)^{1/2} \ {\rm arcmin/yr}  \ . 
\end{eqnarray} 

As can be seen from Table~\ref{omega-DM} and Fig.~\ref{precession_rates}, for hypothetical target stars in eccentric orbits with semi-major axes less than $0.2$ milliparsec, which could be used to test the no-hair theorem, the periastron precessions induced by the DM distribution at the center do not exceed the relativistic precessions. Because the pericenter advance due the dark matter distribution is so small , we argue that it is reasonable to consider this as a good estimate for the upper limit on the precession of orbital planes that might be induced by a non-spherical component of the DM distribution that would be generated by a rotating central black hole. That non-spherical part is likely to be a small perturbation of the basic DM distribution because the effects of frame dragging and the quadrupole moment are relativistic effects that  fall off faster with distance than the basic Newtonian gravity of the hole. As a result, we can conclude that a dark matter distribution near the black hole will not significantly interfere with a test of the black hole no-hair theorem. Furthermore, if the dark matter particles are self-annihilating, their effects will be utterly negligible.

On the other hand, for S2-type stars, if future capabilities of observational precision reach the level of 10 $\mu$arcsec per year, the perturbing effect of the DM distribution on stellar motion at the GC could be marginally detectable if the DM particles are not self-annihilating, as would be the case if they were axions, for example. If they are self-annihilating, the effects of a DM distribution on the outer cluster of stars will be unobservable.

\clearpage 


\makeatletter
\let\orig@chapter\@chapter
\def\@chapter[#1]#2{\ifnum \c@secnumdepth >\m@ne
                       \if@mainmatter
                         \refstepcounter{chapter}%
                         \typeout{\@chapapp\space\thechapter.}%
                         \addcontentsline{toc}{chapter}%
                                   {Appendix~\protect\numberline{\thechapter.} #1}%
                       \else
                         \addcontentsline{toc}{chapter}{#1}%
                       \fi
                    \else
                      \addcontentsline{toc}{chapter}{#1}%
                    \fi
                    \chaptermark{#1}%
                    \addtocontents{lof}{\protect\addvspace{10\p@}}%
                    \addtocontents{lot}{\protect\addvspace{10\p@}}%
                    \if@twocolumn
                      \@topnewpage[\@makechapterhead{#2}]%
                    \else
                      \@makechapterhead{#2}%
                      \@afterheading
                    \fi}
\makeatother

\def\mypart#1#2#3{\par 
\pagestyle{plain}
\newpage\clearpage

\vspace*{5cm} 
\refstepcounter{part}
{\centering}%
\vspace{1cm}
{\centering \linespread{2}\selectfont \textbf{\Huge #1}\par \linespread{1}\selectfont}%

\thispagestyle{empty}
\vspace{5cm}

\null
\vfill
#3
#2
\addcontentsline{toc}{part}{\textbf{#1}}}

\addtocontents{toc}{\vspace{2em}} 

\mypart{APPENDICES}{}{}
 \ClearWallPaper

\appendix 



\chapter{A Useful Change of Variables} 
\ClearWallPaper
\thispagestyle{myplain}

\label{AppendixA} 

\lhead{Appendix A. \emph{A Useful Change of Variables}} 

In calculating the time averaged rates of change of the orbit elements of the target star given by Eq.\ (\ref{ch2-88}), we encounter integrals such as
\be \label{App.A-1}
P_{n,m}\equiv \int_0^{2\pi} \frac{\cos^n f}{(1+e \cos f)^m} {\rm d}f
\ee
which can not be done analytically by Maple or Mathematica. To find the analytical result for these kind of integrals we rewrite Eq. (\ref{App.A-1}) as
\begin{eqnarray} \nonumber
P_{n,m}&=&2 \int_0^\pi \frac{\cos^n f}{(1+e \cos f)^m} {\rm d}f, \\  \nonumber
&=&2 \int_0^{\pi/2} \frac{\cos^n f}{(1+e \cos f)^m}{\rm d}f+ 2 \int_{\pi/2}^{\pi} \frac{\cos^n f}{(1+e \cos f)^m}{\rm d}f \ , \\  \label{App.A-2}
&=&2 \int_0^{\pi/2} \frac{\cos^n f}{(1+e \cos f)^m}{\rm d}f+ 2 (-1)^n \int_{0}^{\pi/2} \frac{\cos^n f}{(1-e \cos f)^m}{\rm d}f \ .
\end{eqnarray}
where the second term comes from letting $f\rightarrow\pi-f$. Depending on the value of $n$, this gives a sum of integrals of the form
\be \label{App.A-3}
Q_{n,m}\equiv \int_0^{\pi/2} \frac{\cos^n f}{(1-e^2 \cos^2 f)^m } {\rm d}f
\ee
which can be evaluated analytically easily by Maple. For example
\begin{eqnarray} \nonumber
P_{2,4}&=&\int_0^{2\pi} \frac{\cos^2 f}{(1+e \cos f)^4}{\rm d}f \ , \\ \label{App.A-4}
&=&4\ Q_{2,4}+24\ e^2\ Q_{4,4}+4\ e^4\ Q_{6,4} \ ,
\end {eqnarray}
where
\begin{eqnarray} \nonumber
Q_{2,4}&=&\int_0^{\pi/2} \frac{\cos^2 f}{(1-e^2 \cos^2 f)^4}{\rm d}f=\frac{\pi}{32} \frac{(8-4e^2+e^4)}{(1-e^2)^{7/2}} \ , \\ \nonumber
Q_{4,4}&=&\int_0^{\pi/2} \frac{\cos^4 f}{(1-e^2 \cos^2 f)^4}{\rm d}f=\frac{\pi}{32} \frac{(6-e^2)}{(1-e^2)^{7/2}} \ , \\ \label{App.A-5}
Q_{6,4}&=&\int_0^{\pi/2} \frac{\cos^6 f}{(1-e^2 \cos^2 f)^4}{\rm d}f=\frac{5 \pi}{32} \frac{1}{(1-e^2)^{7/2}} \ .
\end{eqnarray}
Writing every $P_{n,m}$ integral as a sum of $Q_{n,m}$ integrals simplify the calculations, and minimally, it allows us to give analytical expressions for many steps.

\clearpage 


\chapter{Minimum Distance for a Stellar or Black Hole Orbit} 

\label{AppendixB} 

\thispagestyle{myplain}

\lhead{Appendix B. \emph{Minimum Distance for a Stellar or Black Hole Orbit}} 

A star that approaches too close to the black hole will be tidally disrupted and be removed from the stellar distribution.  An estimate of this distance is given by the ``Roche radius'', $r_{\rm Roche} \approx R (2M/m)^{1/3}$, where $R$ is the radius of the star, and $M$ and $m$ are the black-hole and stellar masses, respectively.  For a solar-type star, the radius $R$ may be estimated using the empirical formula $R \approx R_{\odot} (m_{\rm {star}}/m_{\odot} )^{0.8}$.  Thus we obtain $r_{\rm min}^{\rm star} \approx R_{\odot} (m_{\rm {star}}/m_{\odot} )^{0.47} (2m/m_\odot)^{1/3}$.  Putting in numbers gives the first of Eqs.\ (\ref{ch2-97}).  

A stellar-mass black hole will not be tidally disrupted, but can be captured directly if its energy and angular momentum are such that there will be no turning point in its radial motion. For equatorial orbits in the Kerr geometry (in Boyer-Lindquist coordinates), the equation of radial motion has the form $({\rm d}r/{\rm d}\tau)^2 = \tilde{E}^2 - V(r)$, where $\tau$ is proper time, $\tilde{E}$ is the relativistic energy per unit $m_{\rm bh}$ of the orbiting black hole where $m_{\rm bh}$ is the mass of the orbiting stellar mass black hole , and
\begin{equation}
V(r) = 1- \frac{2\tilde{m}}{r} + \frac{a^2}{r^2} + \frac{\beta}{r^2}
- \frac{2\tilde{m} \alpha^2}{r^3} \,,
\end{equation}
where $\tilde{m} = Gm$, $a = J/m$, $\beta = \tilde{L}_z^2 - a^2 \tilde{E}^2$, and $\alpha = \tilde{L}_z - a\tilde{E}$, where $J$ is the angular momentum of the central black hole and $\tilde{L}_z$ is the angular momentum per unit $m_{\rm bh}$ of the orbiting black hole. The critical angular momentum for capture is given by that value such that the turning point occurs at the unstable peak of $V(r)$.  Since the orbiting stars and black holes are in non-relativistic orbits, we can set $\tilde{E} \approx 1$. Under these conditions, it is straightforward to show that 
\begin{equation}
(\tilde{L}_z)_c = \pm 2\tilde{m} \left (1 + \sqrt{1 \mp a/\tilde{m}} \right ) \,,
\end{equation}
where the upper (lower) sign corresponds to prograde (retrograde) orbits.  For $a/\tilde{m} =1$, the critical angular momenta are $2\tilde{m}$ and $-2(1+\sqrt{2}) \tilde{m}$.  Converting to the language of orbital elements, where $L_z^2 =m_{\rm bh}^2 Gma(1-e^2)$, we find in the large $e$ limit, $L_z^2 \approx 2m_{\rm bh}^2 Gmr_p$ where $r_p$ is the pericenter distance of the stellar mass black hole orbit. The result is that
\begin{equation}
r_{\rm min}^{\rm bh} \approx 2\tilde{m} \left (1 + \sqrt{1 \mp a/\tilde{m}} \right )^2 \,.
\end{equation}
This ranges from $2Gm$ to $11.6Gm$ for $a/\tilde m=1$ and is $8Gm$ for $a=0$ (Schwarzschild).   We adopt the latter value as a suitable estimate; inserting numbers gives the second of Eqs.\ (\ref{ch2-97}).

\clearpage 


\chapter{Effects of Tidal Deformations} 

\label{AppendixC} 

\thispagestyle{myplain}

\lhead{Appendix C. \emph{Effects of Tidal Deformations}} 

Even if stars survive tidal disruption on passing very close to the MBH at pericenter, they will be tidally distorted, and these distortions can affect their orbits.  However, we argue that, for the stellar orbits of interest, these effects are negligible.  For example, the rate of pericenter advance due to tidal distortions is given by (Eq.\ (12.31) of \cite{Will93})
\begin{equation}
\frac{{\rm d}\omega}{{\rm d}t} = \frac{30 \pi}{P} k_2 \frac{M}{m} \left ( \frac{R}{a} \right )^5 \frac{1 + 3e^2/2 + e^4/8}{(1-e^2)^5} \,,
\end{equation}
where $k_2$ is the so-called ``apsidal constant'' of the star, a dimensionless measure of how centrally condensed it is.  Inserting $R = R_\odot (m/m_\odot)^{0.8}$
, we obtain
\begin{equation}
\frac{{\rm d}\omega}{{\rm d}t} = 0.04 \left( \frac{k_2}{10^{-2}} \right )
\left ( \frac{m}{m_\odot} \right )^{3} 
\left ( \frac{0.1 \, {\rm mpc}}{a} \right )^{13/2} 
\left ( \frac{0.05}{1-e} \right )^5  \, {\rm arcmin/yr} \,.
\end{equation}
The variations in $\imath$ and $\Omega$ scale in exactly the same way, but are further suppressed by the sine of the angle by which the tidal bulge points out of the orbital plane, resulting from the rotation of the star coupled with molecular viscosity, leading to a lag between the radial direction and the tidal bulge.  This angle is expected to be very small.   Thus we can conclude that, as far as perturbations of the orbital planes are concerned, tidal distortions will not be important.

\clearpage 


\chapter{Distribution Function Invariance in Adiabatic Growth of a Point Mass } 

\label{AppendixD} 

\thispagestyle{myplain}

\lhead{Appendix D. \emph{Distribution Function Invariance in Adiabatic Growth of a Point Mass}} 

Young has shown in \cite{Young80} that for the adiabatic growth of a black hole in the center of a star cluster, the conservation of the two adiabatic invariants, namely the angular momentum $L$ and the radial action $I_r$ of each star, leads to the invariance of the distribution function.

In this appendix we first review his argument in our notation for the adiabatic growth of the central black hole in the distribution of dark matter particles and then we show that the result holds in the general relativistic domain too.

As the black hole grows, the gravitational potential evolves from the initial potential $\Phi'$ to a new potential $\Phi$ that includes the point mass and a dark matter particle, initially with conserved quantities $(E', L)$ in $E-L$ space, moves to $(E,L)$ such that $I_r(E,L)=I'_r(E',L)$, therefore:
\be \label{App.D-1}
N'(E',L){\rm d} E' {\rm d}L=N(E,L){\rm d} E {\rm d}L \ .
\ee
where $N(E,L)$ is the density of particles in $E-L$ space.

The number of particles in phase space for a spherically symmetric system is
\begin{eqnarray} \nonumber
f({\bm x},{\bm v}) {\rm d}^3 x {\rm d}^3 v&=&f(r,E,L)(4 \pi r^2 {\rm d}r)\left( \frac{4\pi L}{r^2 |v_r|}{\rm d}E{\rm d}L \right ) \ , \\ \label{App.D-2}
&=&16 \pi^2 f(r,E,L) \frac{L}{|v_r|} {\rm d}r {\rm d}E {\rm d}L \ ,
\end{eqnarray}
where we used the same change of variables that we have in Chapter 3 to get Eq. (\ref{ch3-18}) assuming the distribution function is independent of $L_z$. The corresponding number of dark matter particles in $E-L$ space with energy $E$ in $[E,E+{\rm d}E]$ and angular momentum $L$ in $[L,L+{\rm d}L]$ in the ${\rm d}E{\rm d}L$ volume element is $N(E,L){\rm d}E{\rm d}L$ and to equate this with Eq. (\ref{App.D-2}), we need to integrate Eq. (\ref{App.D-2}) over all values of $r$. Assuming the distribution function is independent of position we have:
\begin{eqnarray} \label {App.D-3}
16 \pi^2L f(E,L) {\rm d}E{\rm d}L \int_{r_-}^{r_+} \frac{{\rm d}r}{|v_r|}&=&N(E,L){\rm d}E{\rm d}L \ , \\  \label {App.D-4}
\Rightarrow \quad 8 \pi^2L f(E,L) \underbrace{\left ( 2\int_{r_-}^{r_+} \frac{{\rm d}r}{|v_r|}\right )}_{P(E,L)}&=&N(E,L) \ ,
\end {eqnarray}
where $r_{\pm}$ are the turning points of the dark matter particles equation of motion and $P(E,L)$ is the orbital period of the dark matter particle. Equation (\ref{App.D-4}) agrees with Eq. (26a) of Young's paper \cite{Young80}. According to the definition of the radial action $I_r(E,L)$ in Eq. (\ref{ch3-28}) we have:
\be \label {App.D-5}
\frac{\partial I_r(E,L)}{\partial E }|_L=\oint \frac{{\rm d}r}{|v_r|}=P(E,L) \ ,
\ee
and using $I_r(E,L)=I'_r(E',L)$ leads to 
\be \label {App.D-6}
\frac{\partial E}{\partial E'}|_L=\frac{P'(E',L)}{P(E,L)} \ ,
\ee
where $P'(E',L)=\oint {\rm d}r / \sqrt{2E'-2\Phi'(r)-L^2/r^2}$. Substituting Eq. (\ref{App.D-4}) for $N(E,L)$ in Eq. (\ref{App.D-1}) gives:
\be \label {App.D-7}
f(E,L)P(E,L){\rm d}E=f'(E',L)P'(E',L){\rm d}E' \ ,
\ee
Now by using Eqs. (App.D-6) and (App.D-7), we get the invariance of the distribution function (Eq. (29) of Young's paper):
\be \label {App.D-8}
f(E,L)=f'(E',L) \ .
\ee
where we used ${\rm d}E'=\left( \partial E' / \partial E |_L \right) {\rm d}E$. So by equating the radial actions and deriving the $E'=E'(E,L)$ relation, we will have the final distribution function.

Now we generalize the derivation of Eq. (\ref{App.D-8}) to the relativistic formalism for the growth of a Schwarzschild black hole. Here we need to use the relativistic radial action given in Eq. (\ref{rel:invar1}). Similar to the non-relativistic case, the conservation of the number of particles in phase space gives:
\be \label {App.D-9}
N({\cal E},L) {\rm d}{\cal E} {\rm d}L=N'({\cal E'},L) {\rm d}{\cal E'} {\rm d}L \ ,
\ee 
To get a similar equation to Eq. (\ref{App.D-2}), we need to use the relativistic Jacobi to change the variables. In spherical symmetry limit, the Jacobi is similar to what we have in Eq. (\ref{J0finalS}):
\begin{eqnarray} \nonumber
f({\bm x},{\bm v}) \ {\rm d}^3 x {\rm d}^3 v&=&f(r,{\cal E},L)(4 \pi r^2 {\rm d}r)\left( \frac{4\pi}{r^2|v_r|} {\cal E} L  {\rm d}{\cal E}{\rm d}L \right ) \ , \\ \label{App.D-10}
&=&16 \pi^2 f(r,{\cal E},L) \frac{L{\cal E}}{|v_r|} {\rm d}r {\rm d}{\cal E} {\rm d}L \ ,
\end{eqnarray}
where $v_r=\sqrt{{\cal E}^2 - (1-2Gm/r)(1 + L^2/r^2)}$. Therefore, if $f(r,{\cal E},L)=f({\cal E},L)$, by integrating Eq. (\ref{App.D-10}) over $r$, for the number of particles in ${\rm d}{\cal E}{\rm d}L$ volume element we get
\begin{eqnarray} \label {App.D-11}
16 \pi^2 {\cal E}L f({\cal E},L) {\rm d}{\cal E}{\rm d}L \int_{r_-}^{r_+} \frac{{\rm d}r}{|v_r|}&=&N({\cal E},L){\rm d}{\cal E}{\rm d}L \ , \\  \label {App.D-12}
\Rightarrow \quad 8 \pi^2 {\cal E}L f({\cal E},L) \underbrace{\left ( 2\int_{r_-}^{r_+} \frac{{\rm d}r}{|v_r|}\right )}_{P({\cal E},L)}&=&N({\cal E},L) \ ,
\end {eqnarray}
Note that the differences of Eq. (\ref{App.D-12}) with the non-relativistic case (Eq. (\ref{App.D-4})), are an extra factor of $\cal E$ and the definition of $v_r$. Also the $P({\cal E},L)$ in Eq. (\ref{App.D-12}) is not the orbital period of the dark matter particle's orbit measured by an observer sitting at infinity. In fact, since $v_r={\rm d}r/{\rm d}\tau$, $P({\cal E},L)$ is the orbital period measured by the clock moving with the particle.

Using the definition of $I_r({\cal E},L)$ in Eq. (\ref{rel:invar1}) we have
\begin{eqnarray} \nonumber
\frac{\partial I_r({\cal E},L)}{\partial {\cal E}}|_L&=& \frac{\partial}{\partial {\cal E}}\oint \sqrt{{\cal E}^2 - (1-2Gm/r)(1 + L^2/r^2)} {\rm d}r \ , \\  \nonumber
&=&\oint \frac{{\cal E} {\rm d}r}{\sqrt{{\cal E}^2 - (1-2Gm/r)(1 + L^2/r^2)}} \ , \\  \nonumber
&=&{\cal E} \oint \frac{{\rm d}r}{|v_r|} \ , \\ \label{App.D-13}
&=&{\cal E}P({\cal E},L) \ .
\end{eqnarray}
Assuming $I_r({\cal E},L)=I'_r({\cal E}',L)$, Eq. (\ref{App.D-13}) results in
\be \label{App.D-14}
\frac{\partial {\cal E}}{\partial {\cal E}'}|_L=\frac{{\cal E'} P'({\cal E'},L)}{{\cal E}P({\cal E},L)} \ .
\ee

Substituting Eq. (\ref{App.D-12}) in Eq. (\ref{App.D-9}) gives
\be \label{App.D-15}
{\cal E} f({\cal E},L)P({\cal E},L) {\rm d} {\cal E}={\cal E'} f({\cal E'},L)P({\cal E'},L) {\rm d} {\cal E'} \ ,
\ee 
again since ${\rm d}{\cal E'}=\left( \partial {\cal E}'/\partial {\cal E} \right)|_L {\rm d}{\cal E}$, using Eqs. (\ref{App.D-14}) and (\ref{App.D-15}) leads to the invariance of the distribution function in the relativistic formalism:
\be \label{App.D-16}
f({\cal E},L)=f'({\cal E'},L) \ .
\ee

\clearpage 

\addtocontents{toc}{} 

\backmatter


\pagestyle{myfancy}
\label{Bibliography}
\lhead{\emph{Bibliography}} 
\footnotesize{
\bibliographystyle{unsrtnat}
}
\bibliography{Bibliography} 

\end{document}